\documentclass[twocolumn]{aastex631}
\received{2021 September 10}
\revised{2022 May 17}
\accepted{2022 May 27}
\submitjournal{the Astronomical Journal}
\newcommand{\cii}{[C\,\textsc{ii}]}
\shorttitle{\cii\ in Virgo Dwarfs}
\shortauthors{Minchin R. et al.}
\begin{document}
\title{Environmental effects in {\it Herschel} observations of the ionized carbon content of star forming dwarf galaxies in the Virgo cluster\footnote{{\it Herschel} is an ESA space observatory with science instruments provided by European-led Principal Investigator consortia and with important participation from NASA.}}
\author[0000-0002-1261-6641]{Robert Minchin}\affiliation{SOFIA Science Center, Universities Space Research Association, MS 232-11, Moffett Field, CA 94035, USA}
\author[0000-0002-3698-7076]{Dario Fadda}\affiliation{SOFIA Science Center, Universities Space Research Association, MS 232-11, Moffett Field, CA 94035, USA}
\author[0000-0002-3782-1457]{Rhys Taylor}\affiliation{Astronomical Institute of the Czech Academy of Sciences, Bocni II 1401/1a, 141 00 Praha 4, Czech Republic}
\author[0000-0002-7898-5490]{Boris Deshev}\affiliation{Astronomical Institute of the Czech Academy of Sciences, Bocni II 1401/1a, 141 00 Praha 4, Czech Republic}
\author{Jonathan Davies}\affiliation{School of Physics \& Astronomy, Cardiff University, Queen's Buildings, The Parade, Cardiff, CF24 3AA, UK}
\correspondingauthor{Robert Minchin}\email{rminchin@sofia.usra.edu}
\begin{abstract}
We use archival {\it Herschel} data to examine the singly ionized carbon (\cii) content of 14 star forming dwarf galaxies in the Virgo cluster. We use spectral energy distribution (SED) fits to far infrared, mid infrared, near infrared, optical and ultraviolet data to derive the total infrared continuum (TIR) for these galaxies. We compare the \cii/TIR ratio for dwarf galaxies in the central part of Virgo to those in the southern part of the cluster and to galaxies with similar TIR luminosities and metallicities in the {\it Herschel} Dwarf Galaxy Survey (DGS) sample of field dwarf galaxies to look for signs of \cii\ formation independent of star formation. Our analysis indicates that the sample of Virgo dwarfs in the central part of the cluster has significantly higher values of \cii/TIR than the sample from the southern part of the cluster and the sample from the DGS, while the southern sample is consistent with the DGS. This \cii/TIR excess implies that a significant fraction of the \cii\ in the dwarf galaxies in the cluster center has an origin other than star formation and is likely to be due to environmental processes in the central part of the cluster. We also find a surprisingly strong correlation between \cii/TIR and the local ram pressure felt by the dwarf galaxies in the cluster. In this respect, we claim that the excess \cii\ we see in these galaxies is likely to be due to formation in ram pressure shocks.
\end{abstract}

\section{Introduction}\label{intro}
The \cii\ line at 158\micron\ is commonly used in local galaxies (including our own Milky Way) as a measure of their star formation rate (SFR), e.g. \citet{1991ApJ...373..423S,2001ApJ...561..766M,2011MNRAS.416.2712D,2014AA...568A..62D,2014ApJ...788L..17D,2014AA...570A.121P,2015ApJ...800....1H,2019ApJ...886...60S}. There has also been interest in its use as a SFR tracer at high redshifts, where it is accessible to ALMA, e.g. \citet{2015Natur.522..455C,2018MNRAS.478.1170C,2019MNRAS.489....1F,2020AA...643A...1L,2020AA...643A...3S,2022AA...660A..14R}. This has revived interest in determining circumstances where environmental factors might cause departures from the \cii--SFR correlation. Recent studies with the FIFI-LS instrument onboard the Stratospheric Observatory for Infrared Astronomy (SOFIA) have shown that turbulence in the interstellar medium (ISM) associated with interactions with jets can lead to the formation of \cii\ distinct from star formation processes. As the infrared continuum also traces star formation (albeit only obscured star formation, while the \cii\ can, in principle, reflect both the obscured star formation and the unobscured star formation traced by ultra-violet emission), and \cii\ associated with star formation is excited via gas heated by the photoelectric effect of UV photons on small dust grains and polycyclic aromatic hydrocarbons \citep[e.g.,][]{2015ApJ...800....1H}, \cii\ from other sources is identifiable as an excess in the \cii/infrared continuum ratio, as seen in the host galaxies of active galactic nuclei (AGN) where the jet is interacting with the disk \citep{2018ApJ...869...61A,2019AA...626L...3S,2021ApJ...909..204F}. Similarly, {\it Herschel} observations of gas in the collisionally-formed bridge between the Taffy Galaxies (UGC 12914/12915) found enhanced \cii/infrared continuum ratios that were attributed to turbulently heated H$_2$ and high column-density H\,{\sc i} resulting from the collision of the two galaxies \citep{2018ApJ...855..141P}, and models of warm molecular gas shocks in Stephan's Quintet point to collisional heating from the warm H$_2$ being responsible for boosting the \cii\ emission in that system \citep{2013ApJ...777...66A,2017ApJ...836...76A}.

Galaxies in clusters are affected by environmental processes including interactions with other galaxies and with the intra-cluster medium (ICM). The interaction of the ISM of galaxies with the ICM – ram pressure – can cause shocks at the `bows' of the galaxies as well as stripping of their gas, a process known as ram pressure stripping \citep{1972ApJ...176....1G}. As \cii\ can be formed in shocks, it has long been speculated that it could be formed by ram pressure  processes, e.g. by \citet{1999MNRAS.303L..29P} who found \citep[using the Infrared Space Observatory, ISO;][]{1996AA...315L..27K} that NGC 4522 had an `exceptional' value of \cii/TIR, which they ascribed to it probably experiencing ram pressure stripping from the Virgo cluster ICM \citep[NGC 4522 has since been confirmed as being subject to ram pressure stripping, e.g.][]{2004AJ....127.3361K,2007ApJ...659L.115C}. However, observations of 19 late-type galaxies in Virgo by \citet{1999MNRAS.310..317L}, also using ISO, found that the influence of the cluster environment on the \cii\ emission was small compared to its dependence on other factors. This seemed to rule out any large-scale environmental effect in Virgo, despite the anomalously high \cii\ value found for NGC 4522. 

In this study we look at 14 dwarf galaxies in the Virgo cluster that have been observed in \cii\ with the PACS instrument on the {\it Herschel } space telescope \citep{2010AA...518L...2P,2010AA...518L...1P} to see whether there is any excess in \cii/infrared continuum that can be attributed to environmental effects injecting energy into the ISM. These observations are more sensitive than the earlier ISO observations,\footnote{Errors on the \cii\ fluxes here range from 0.3 to $1.8\times 10^{-18}$ W\,m$^{-2}$, compared to 2 to $7\times 10^{-17}$ W\,m$^{-2}$ in \citet{1999MNRAS.310..317L}.} while the galaxies studied here are smaller than those observed by \citet{1999MNRAS.310..317L}; as dwarf galaxies are known to be more affected by ram pressure than larger galaxies \citep{2014AA...570A..69B} they may thus be more prone to environmental influences in their integrated \cii\ signal.

\section{Sample and analysis}\label{sample}

The sample analyzed here was extracted from the {\it Herschel} Science Archive by searching the archive for dwarf galaxies observed in the \cii\ line in or near the Virgo cluster. The observations that were found were all from a single project, drawn from the star-forming dwarf galaxies in Virgo studied by \citet{2015AA...574A.126G} and \citet{2016AA...590A..27G}. This gives a sample for which far infrared (FIR) continuum data, which are essential for the modeling used to determine the total infrared flux (TIR), and metallicities, which are needed for identifying suitable comparator galaxies, are available. Basic information from the literature for the galaxies in the sample is given in Table~\ref{sampletable}.

\begin{deluxetable*}{lCCCCR@{ $\pm$ }L}
\tablecaption{Literature data on the sample: position and distance from GOLDMine \citep{2014arXiv1401.8123G}, metallicity from  \citet{2016AA...590A..27G}, and systemic velocity from \citet{2018ApJ...861...49H}.
\label{sampletable}}
\tablehead{\colhead{Galaxy ID}&\colhead{R.A.}&\colhead{Decl.}&\colhead{Distance}&\colhead{Metallicity}&\multicolumn2c{Velocity}\\
&(J2000)&(J2000)&\colhead{(Mpc)}&\colhead{(12 + log(O/H))}&\multicolumn2c{(km/s)}
}
\startdata
\\[-2ex]
VCC 144  &12:15:18.3&+05:45:39.8&32&8.21\pm 0.10&2016&32\\[1ex]
VCC 213  &12:16:56.0&+13:37:31.5&17&8.77\pm 0.12& -164&61\\[1ex]
VCC 324  &12:19:09.9&+03:51:23.4&17&8.14\pm 0.10&1525&23\\[1ex]
VCC 334  &12:19:14.2&+13:52:55.9&17&8.22\pm 0.10& -252&21\\[1ex]
VCC 340  &12:19:22.1&+05:54:37.7&32&8.26\pm 0.10&1510&28\\[1ex]
VCC 562  &12:22:35.9&+12:09:29.2&17&8.10\pm 0.10&      9&22\\[1ex]
VCC 693  &12:24:03.2&+05:10:50.2&17&8.43\pm 0.10&2051&50\\[1ex]
VCC 699  &12:24:07.4&+06:36:26.9&23&8.30\pm 0.10&  727&43\\[1ex]
VCC 737  &12:24:39.5&+03:59:43.8&17&8.28\pm 0.10&1725&78\\[1ex]
VCC 841  &12:25:47.5&+14:57:06.8&17&8.33\pm 0.10&  499&20\\[1ex]
VCC 1437&12:33:15.4&+09:10:25.2&17&8.38\pm 0.10&1155&29\\[1ex]
VCC 1575&12:34:39.5&+07:09:36.7&17&8.76\pm 0.10&  593&44\\[1ex]
VCC 1686&12:24:43.4&+13:15:33.6&17&8.33\pm 0.15&1120&53\\[1ex]
VCC 1725&12:37:41.5&+08:33:31.1&17&8.25\pm 0.10&1076&38\\[1ex]
\enddata
\end{deluxetable*}

\subsection{{\it Herschel} \cii\ intensity maps and spectra}\label{observations}

Data were downloaded from the {\it Herschel} Science Archive in the form of fully-calibrated and flat-fielded Level 2 products output by the PACS pipeline reduction.\footnote{{\it Herschel} Science Archive data products are described at \url{https://www.cosmos.esa.int/web/herschel/data-products-overview}}

\begin{figure*}[h]
\gridline{
\fig{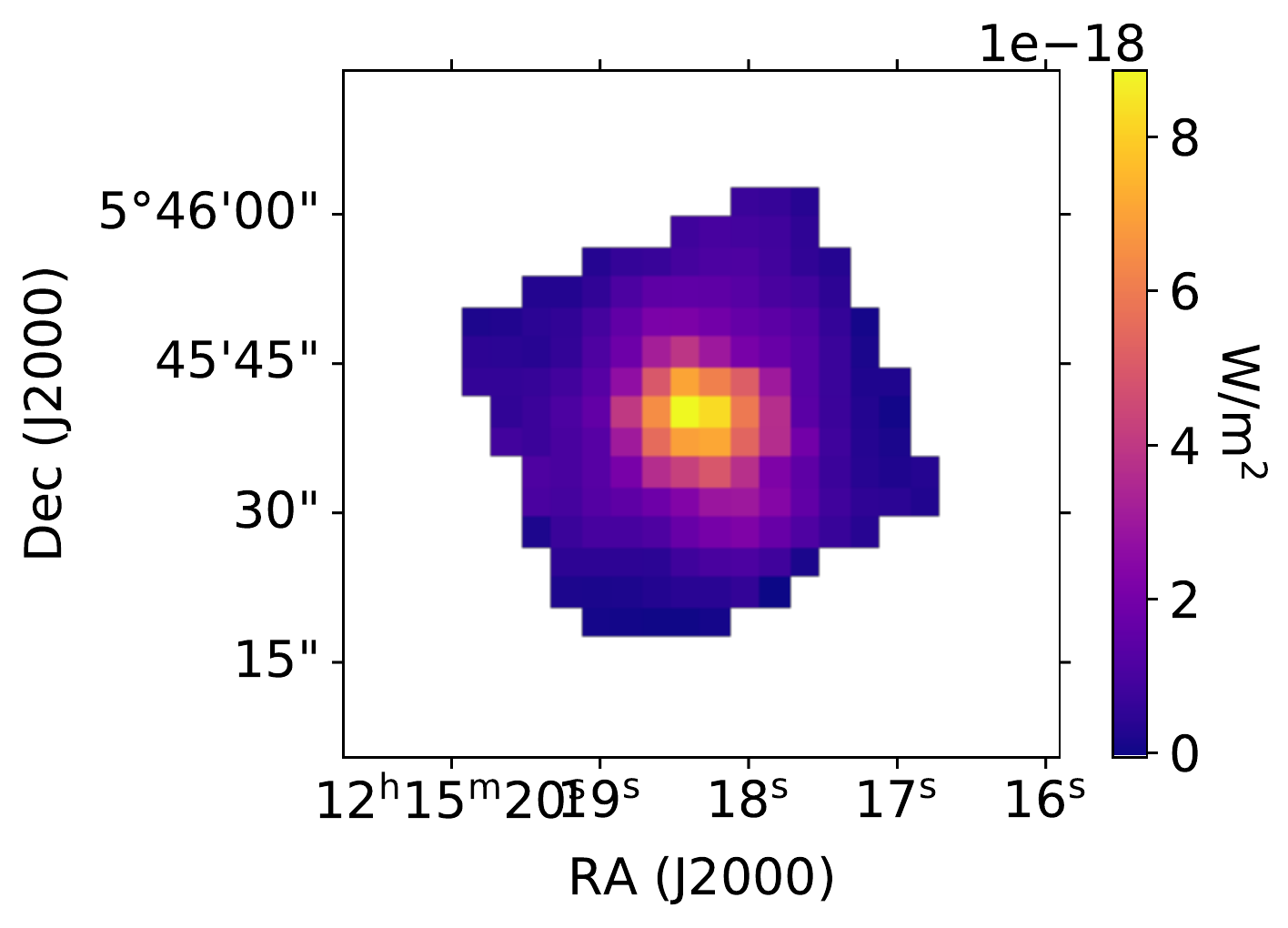}{0.25\textwidth}{VCC 144}\fig{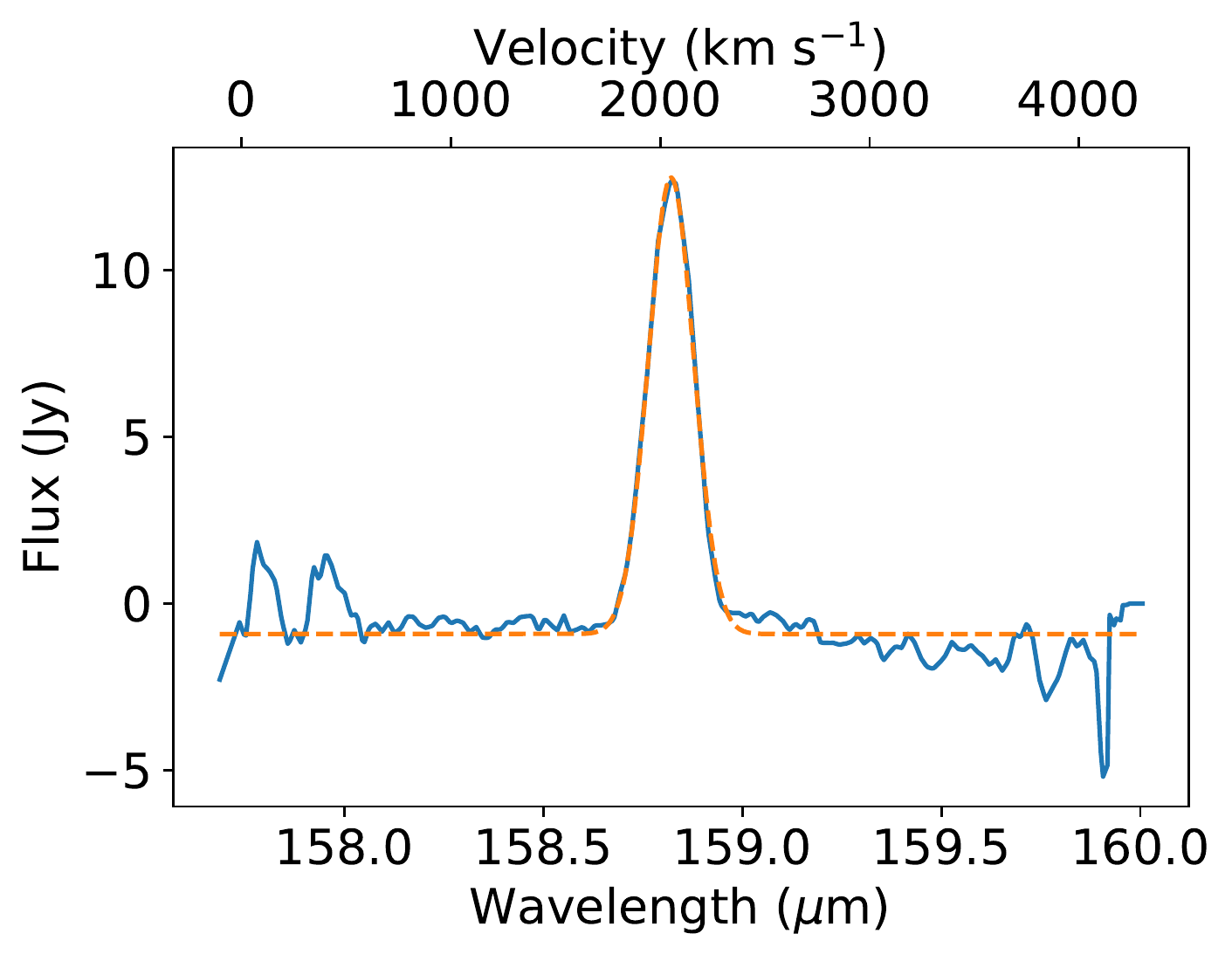}{0.25\textwidth}{VCC 144}
\fig{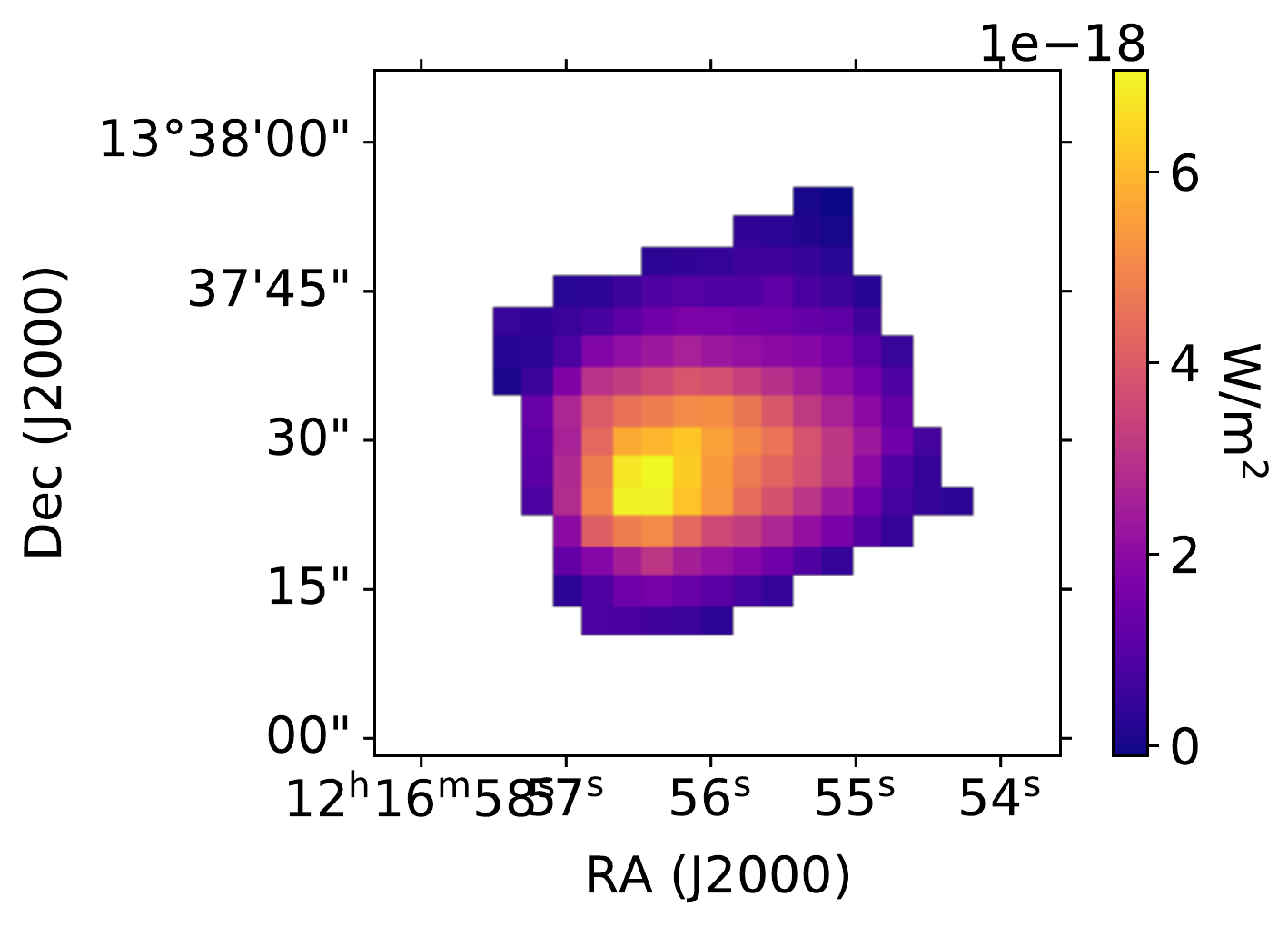}{0.25\textwidth}{VCC 213}\fig{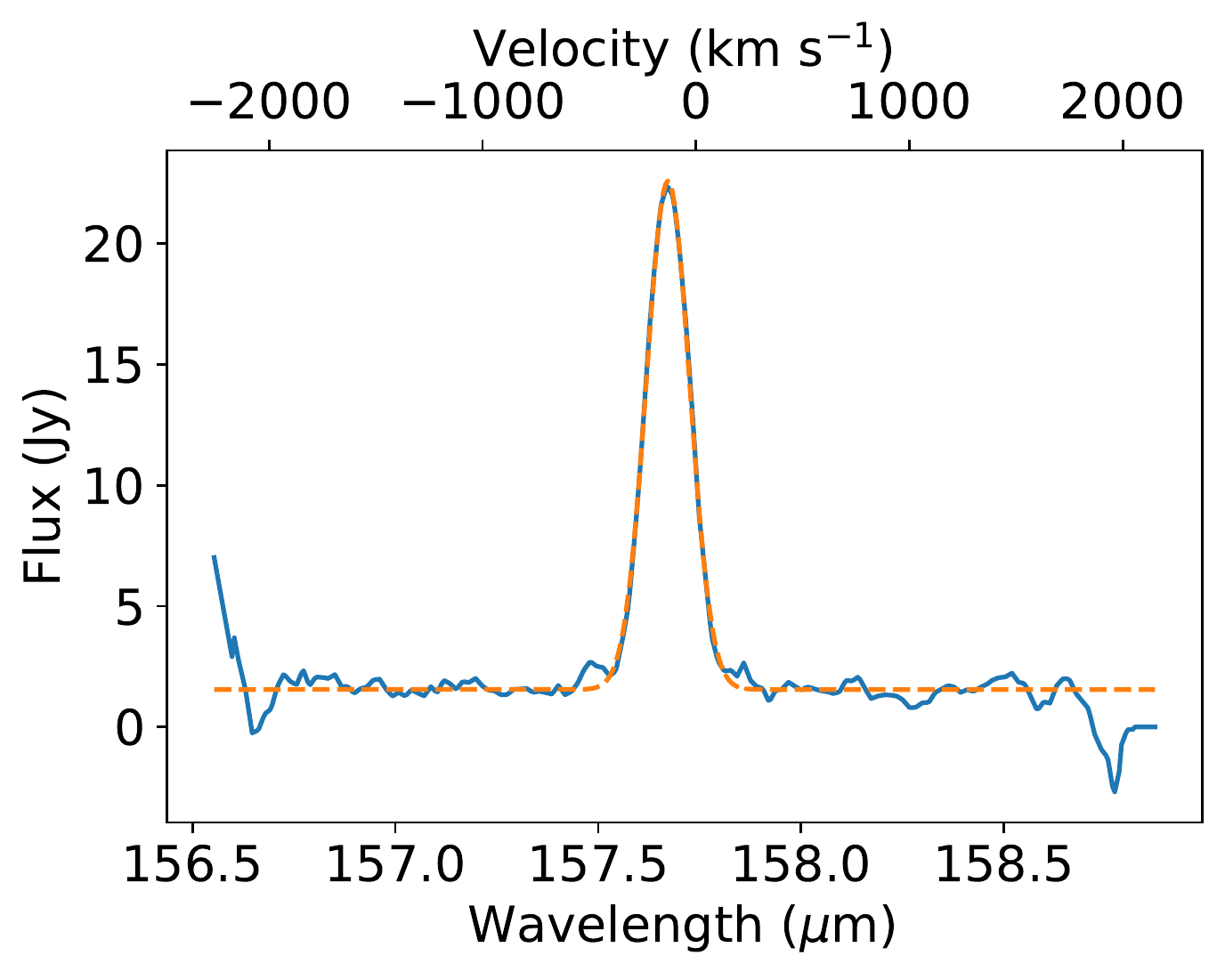}{0.25\textwidth}{VCC 213}
}
\gridline{
\fig{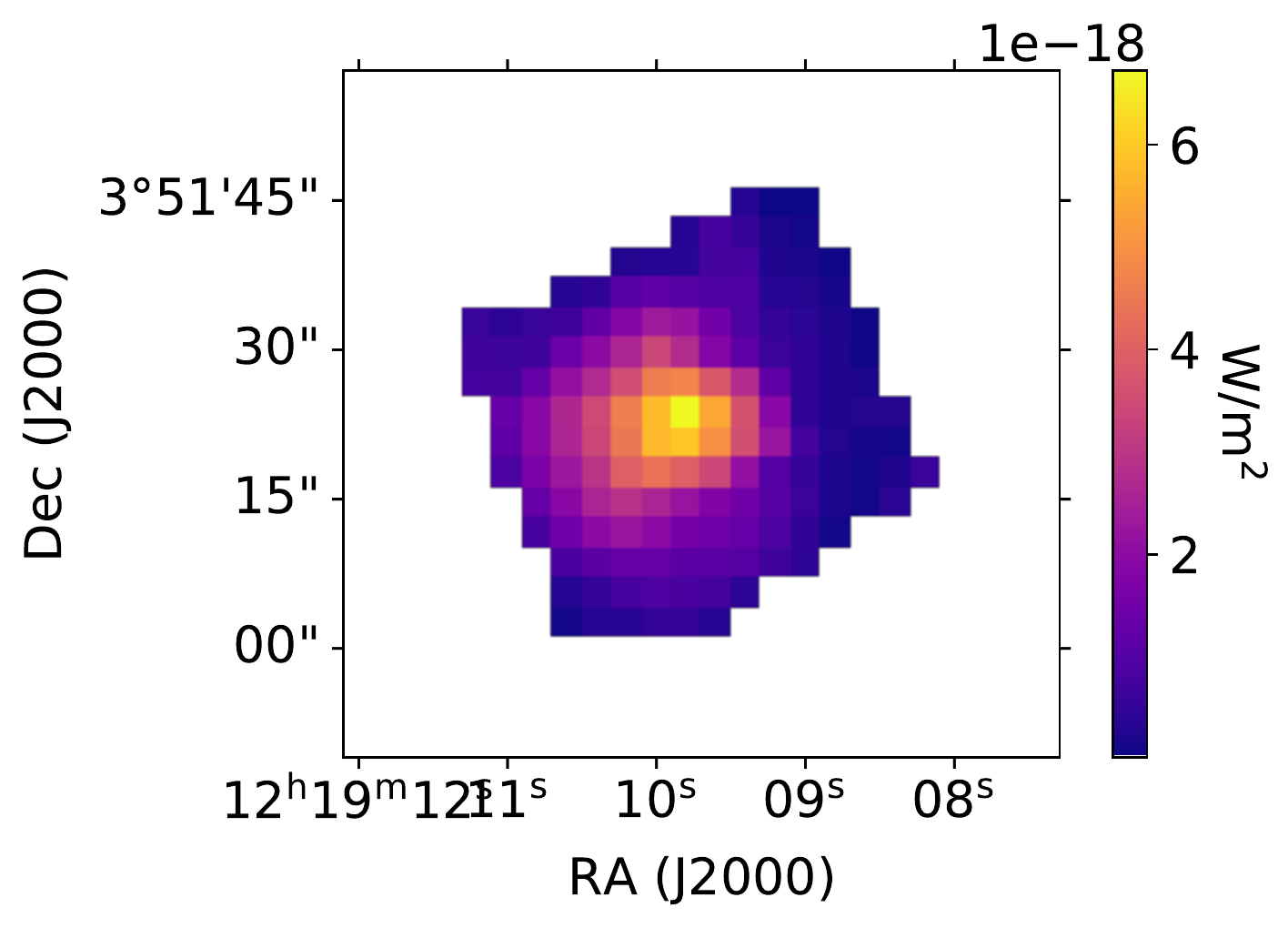}{0.25\textwidth}{VCC 324}\fig{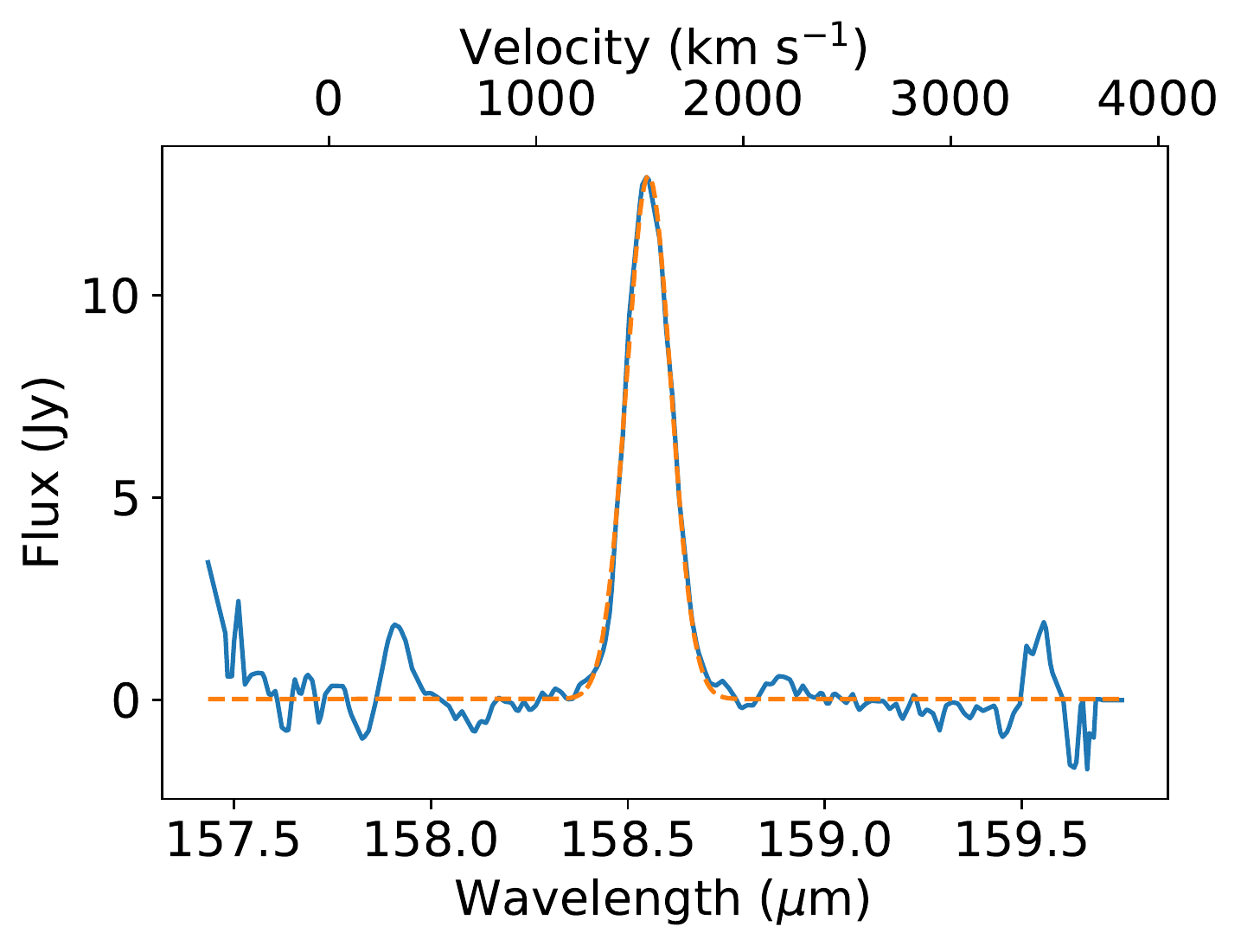}{0.25\textwidth}{VCC 324}
\fig{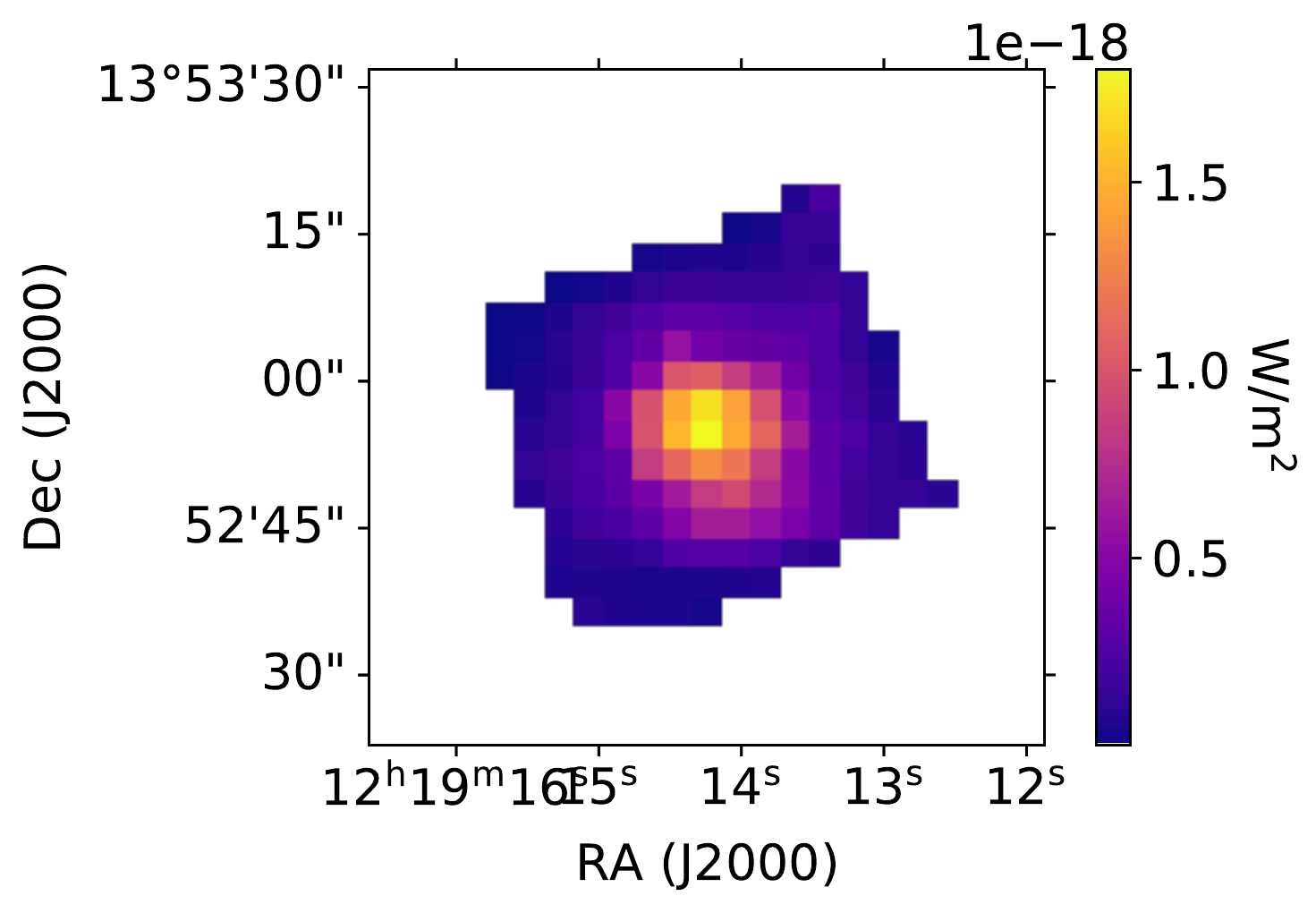}{0.25\textwidth}{VCC 334}\fig{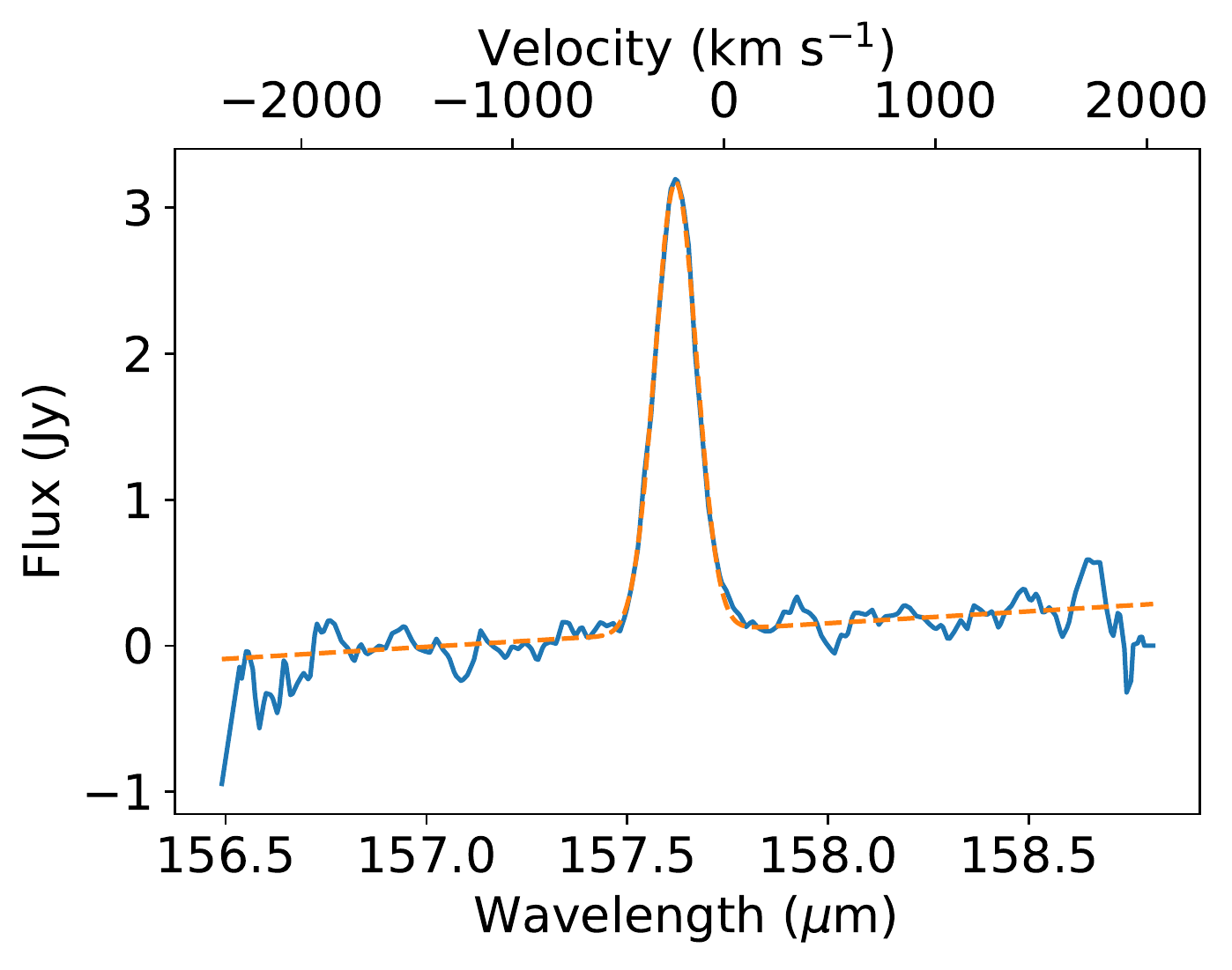}{0.25\textwidth}{VCC 334}
}
\gridline{
\fig{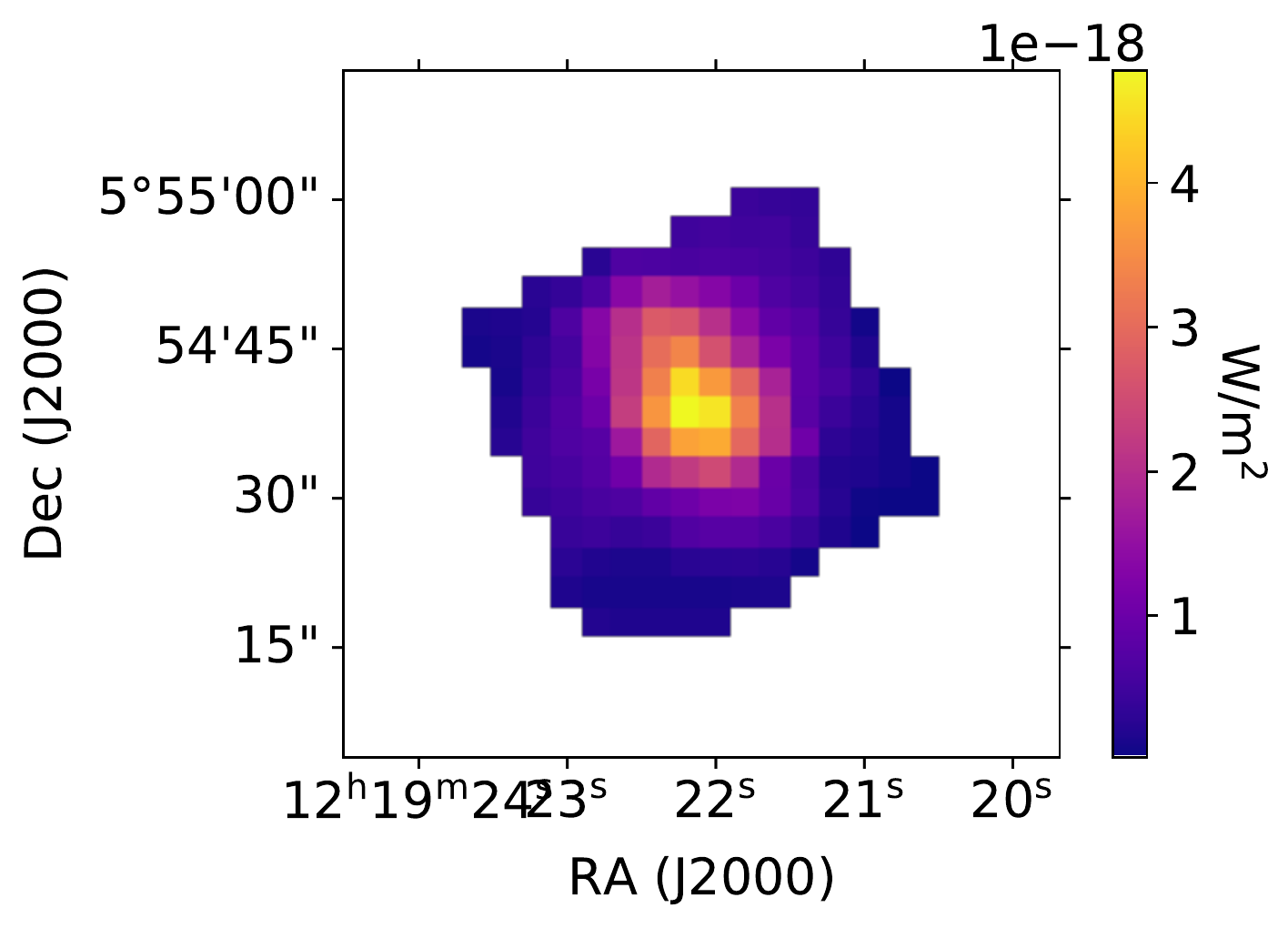}{0.25\textwidth}{VCC 340}\fig{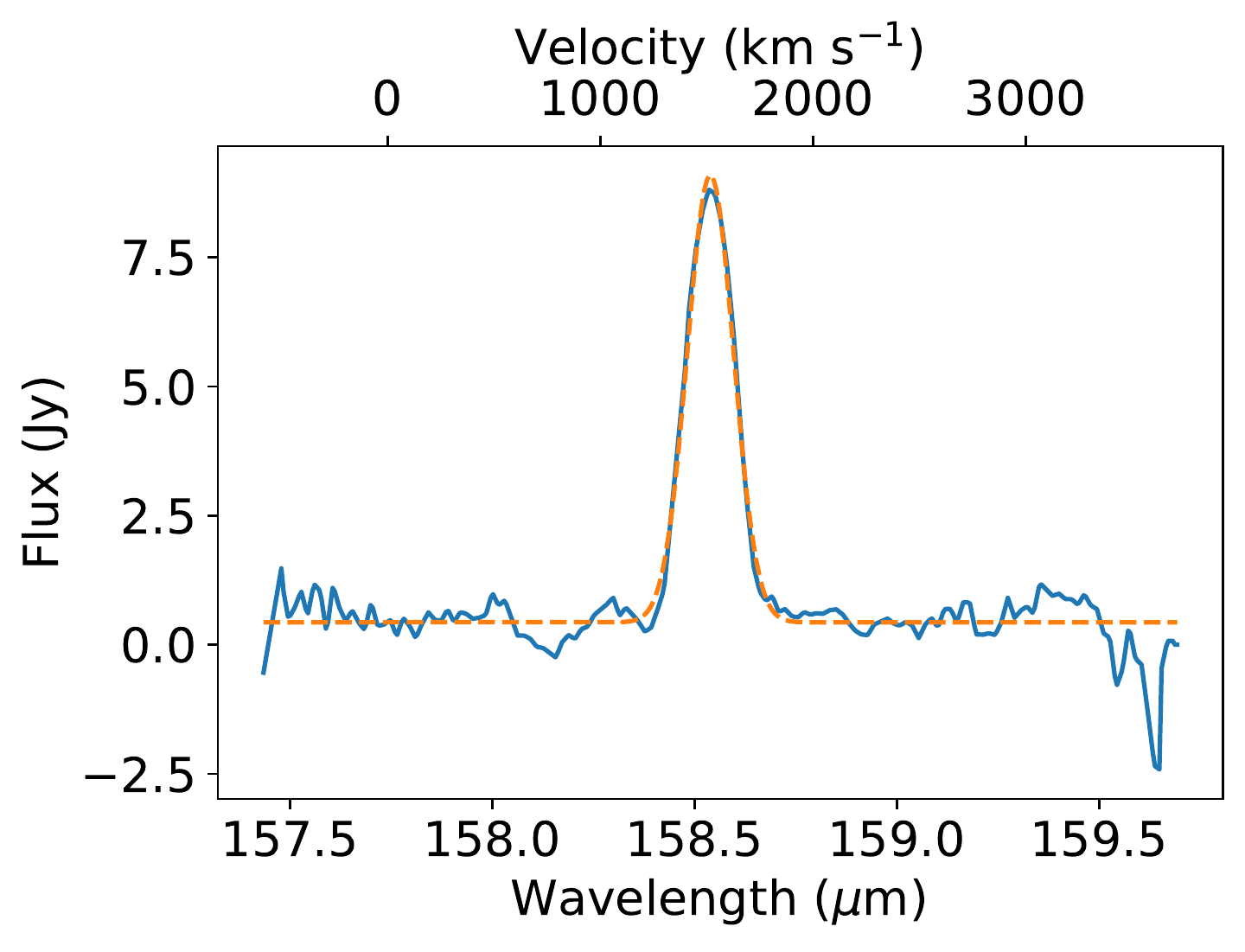}{0.25\textwidth}{VCC 340}
\fig{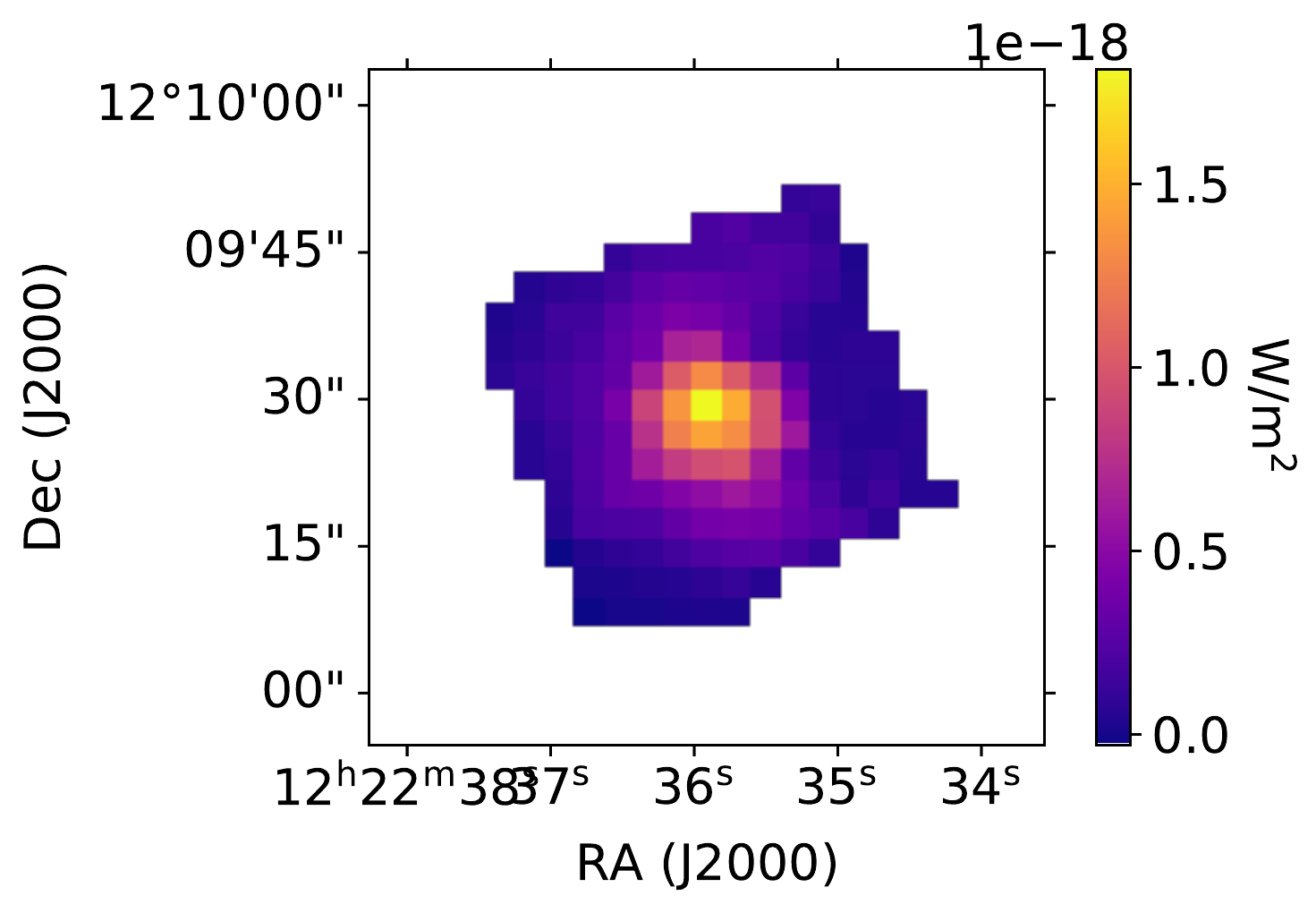}{0.25\textwidth}{VCC 562}\fig{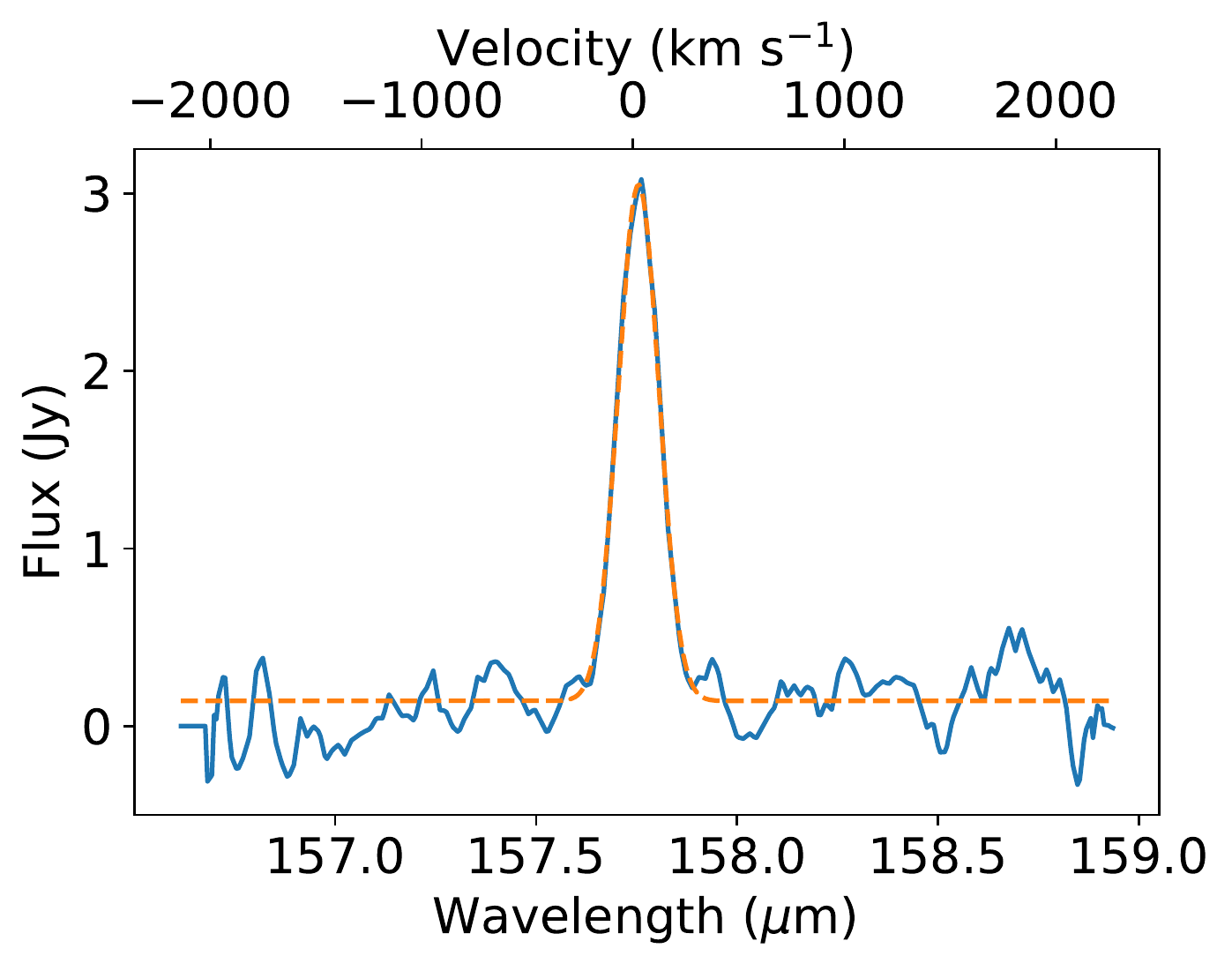}{0.25\textwidth}{VCC 562}
}
\gridline{
\fig{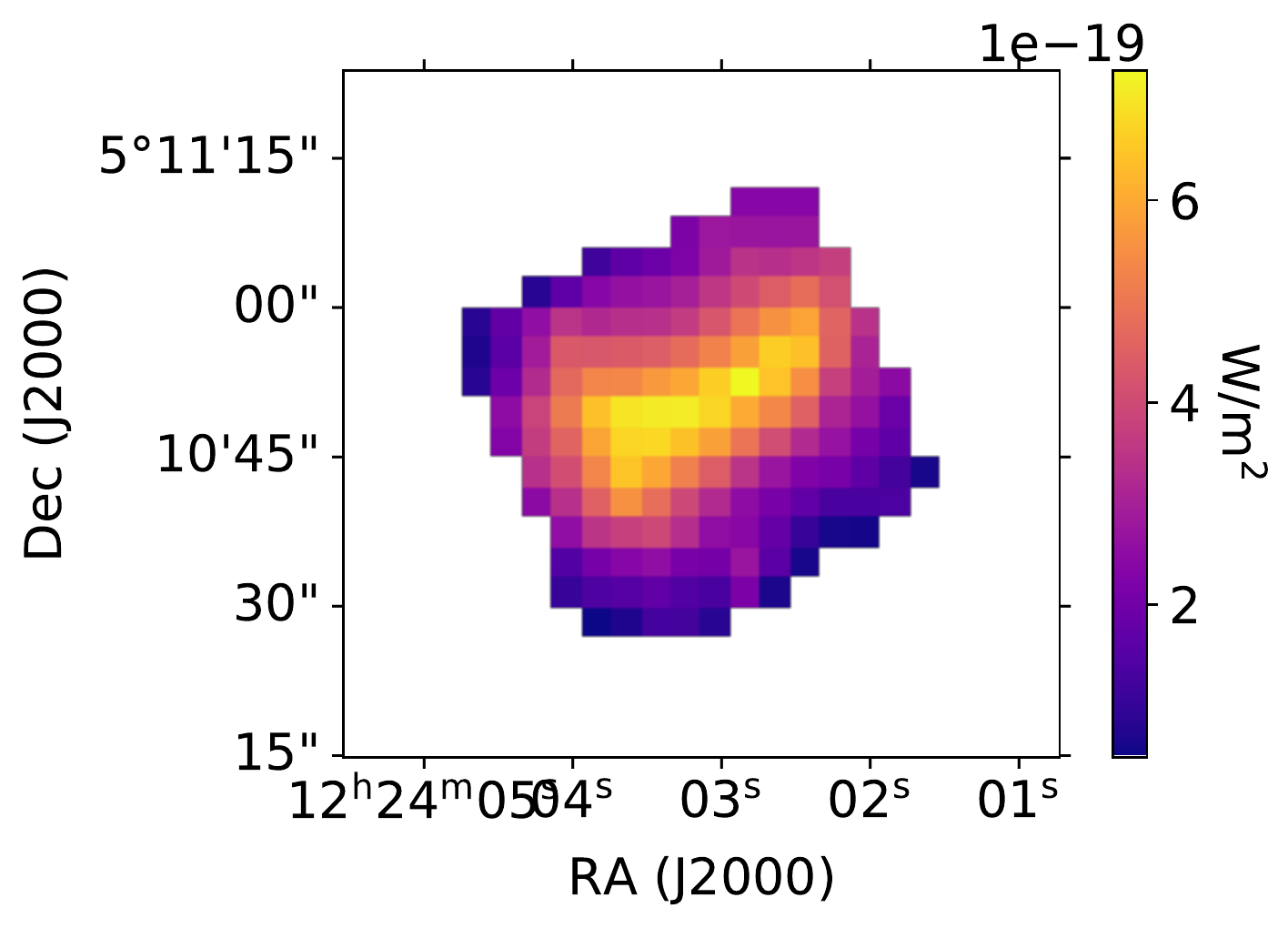}{0.25\textwidth}{VCC 693}\fig{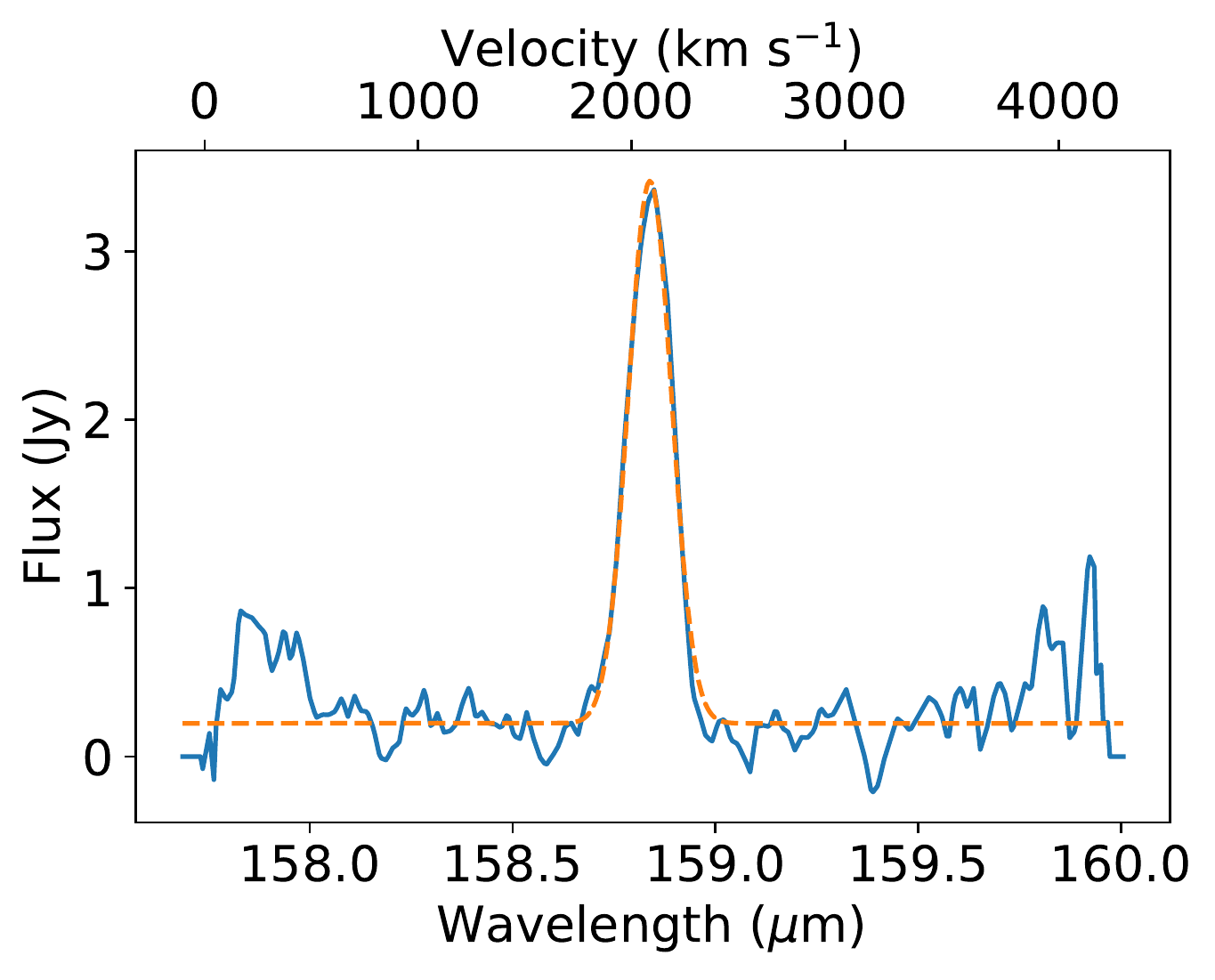}{0.25\textwidth}{VCC 693}
\fig{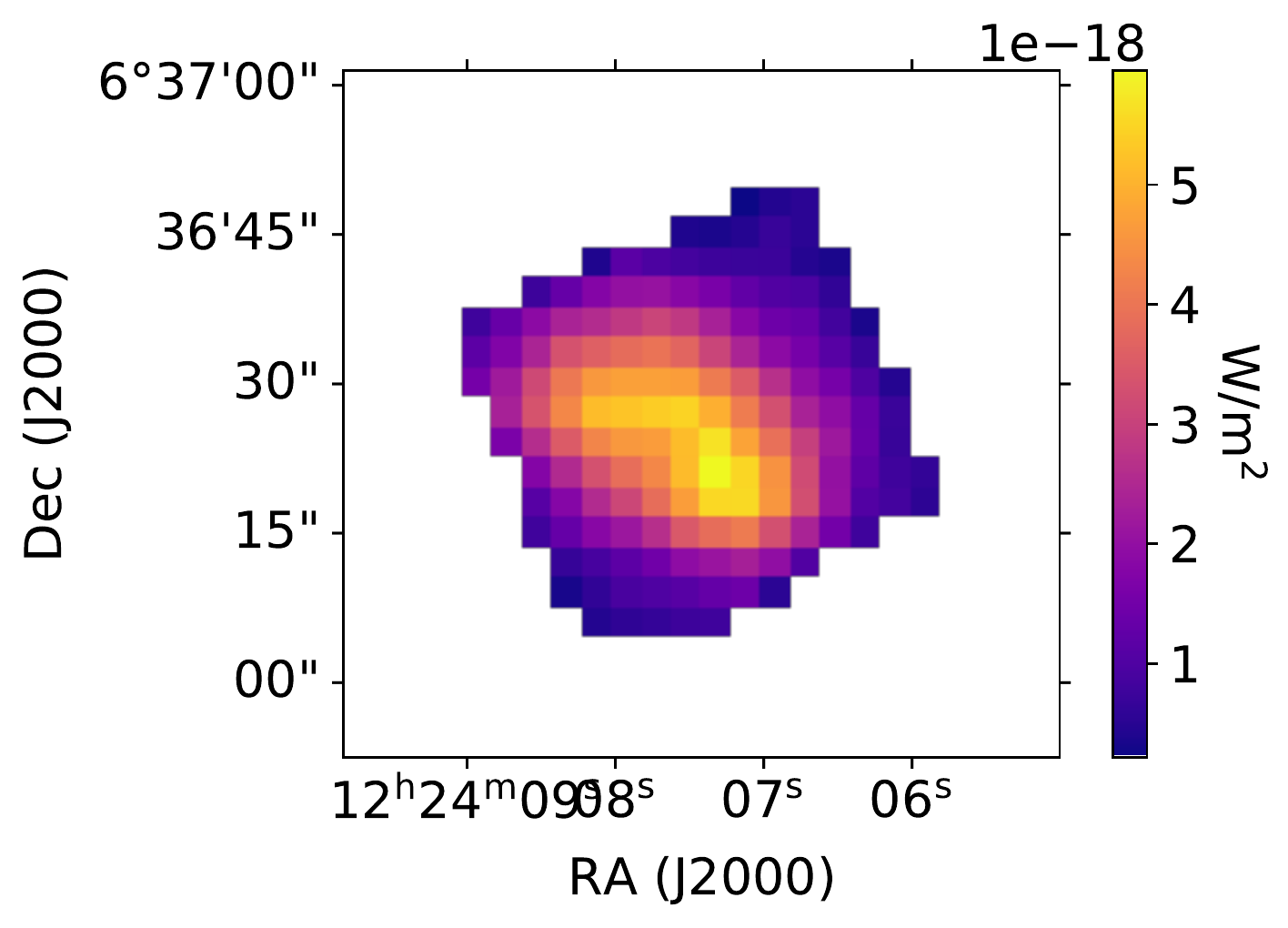}{0.25\textwidth}{VCC 699}\fig{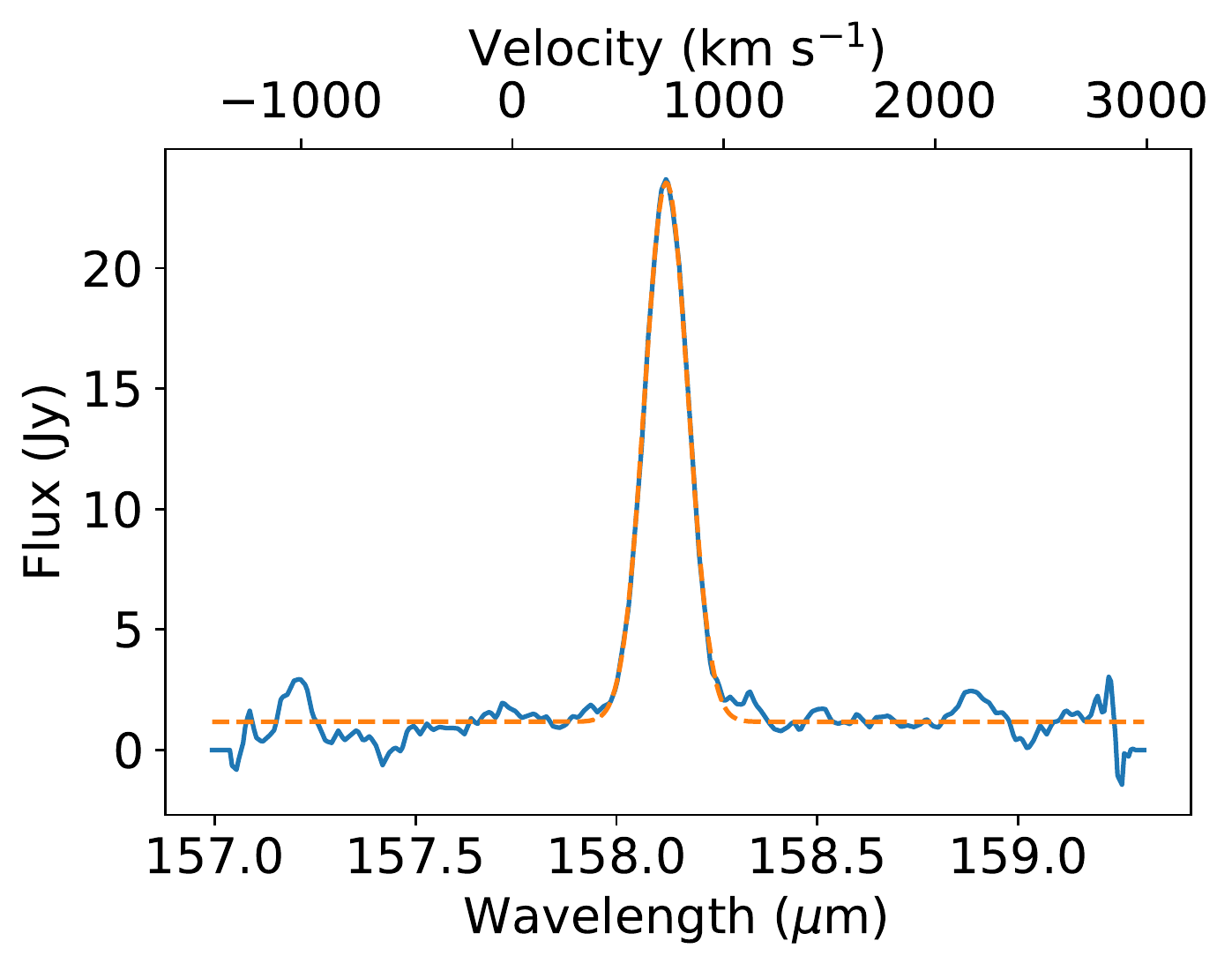}{0.25\textwidth}{VCC 699}
}
\caption{Herschel \cii\ line intensity maps and spatially integrated spectra for the fourteen galaxies in the sample. Spectra show the {\it Herschel} spectrum (solid blue line) and the fitted line profile from SOSPEX (orange dashed line).\label{HerschelCII}}
\end{figure*}

\begin{figure*}
\figurenum{\ref{HerschelCII}}
\gridline{
\fig{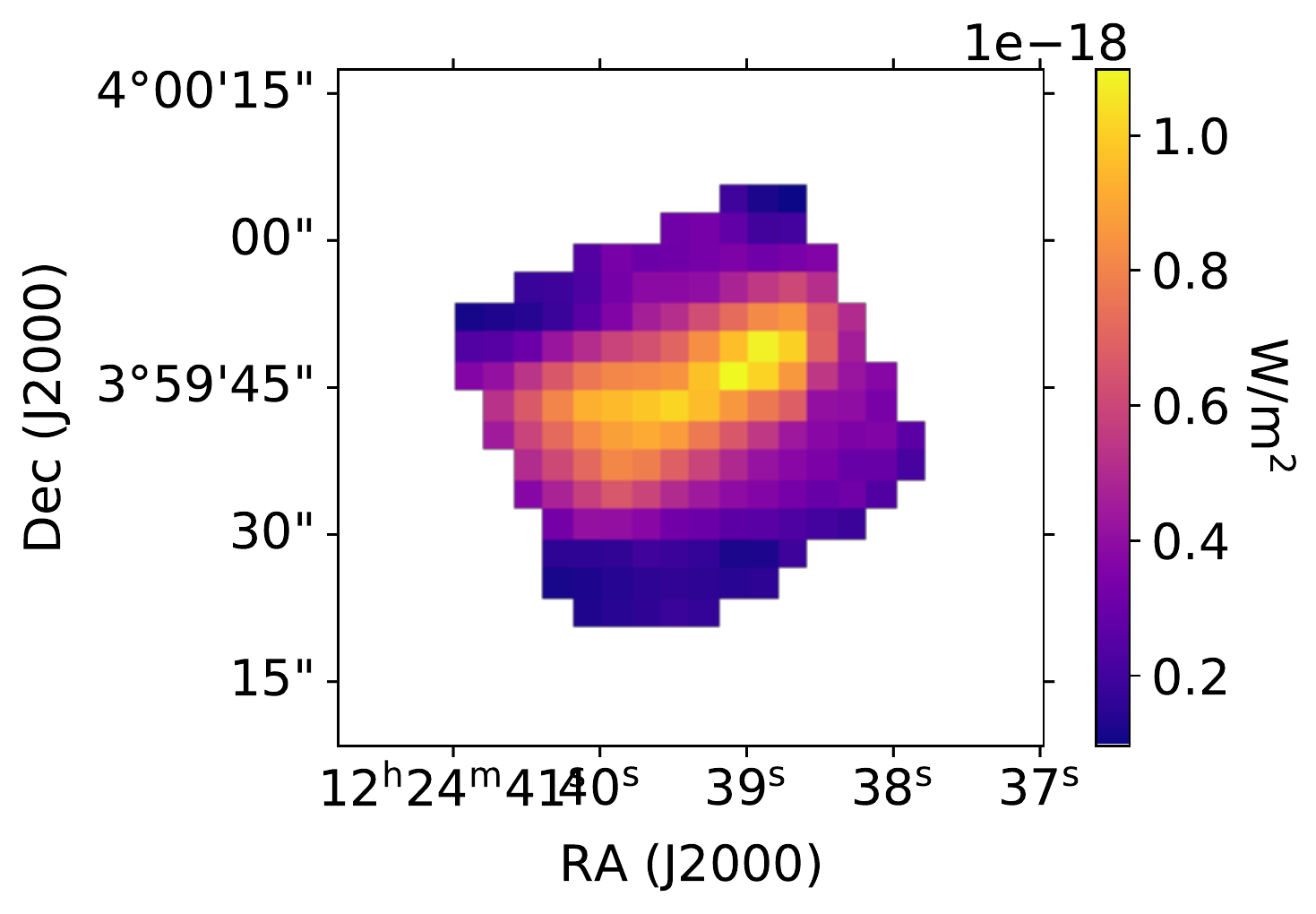}{0.25\textwidth}{VCC 737}\fig{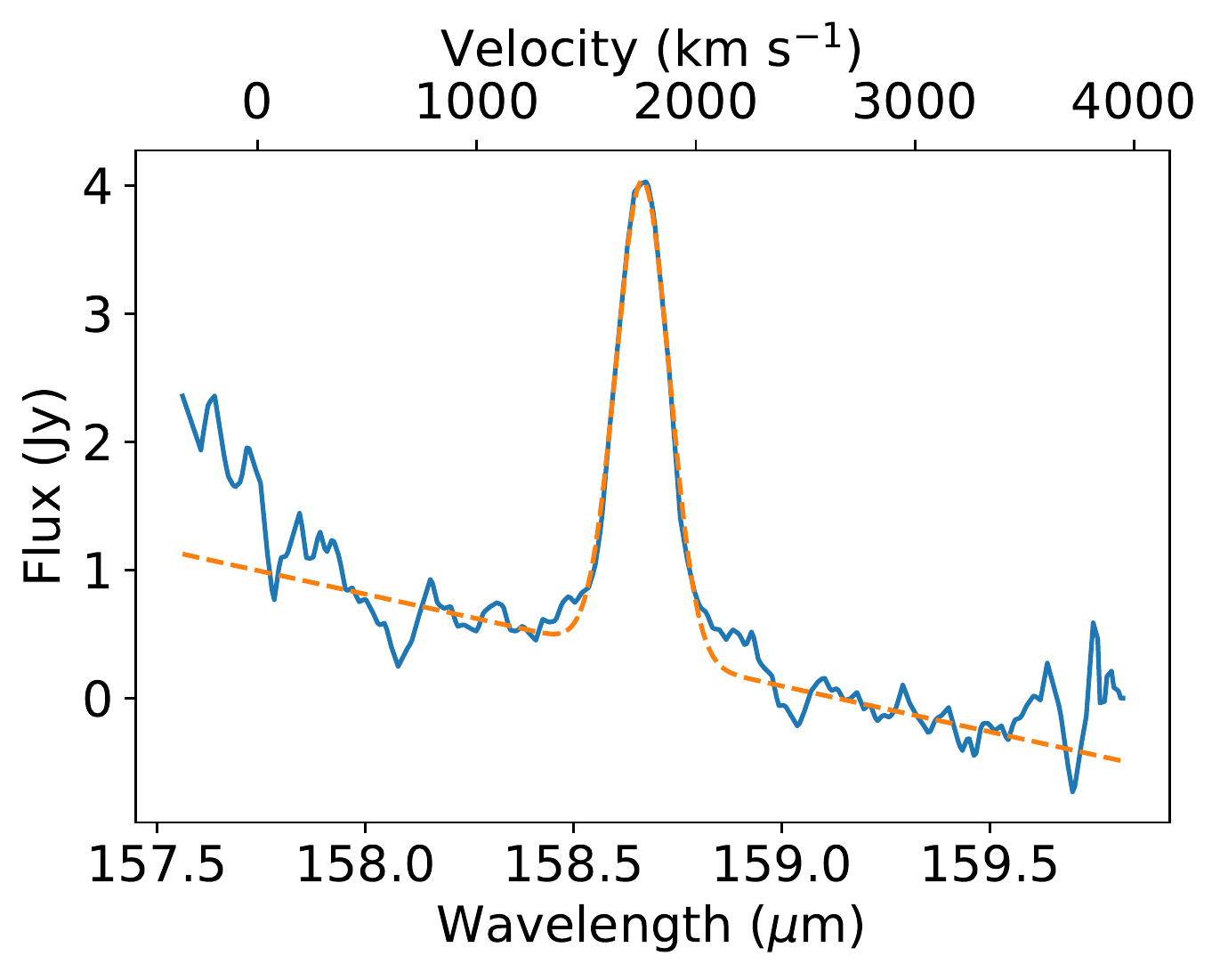}{0.25\textwidth}{VCC 737}
\fig{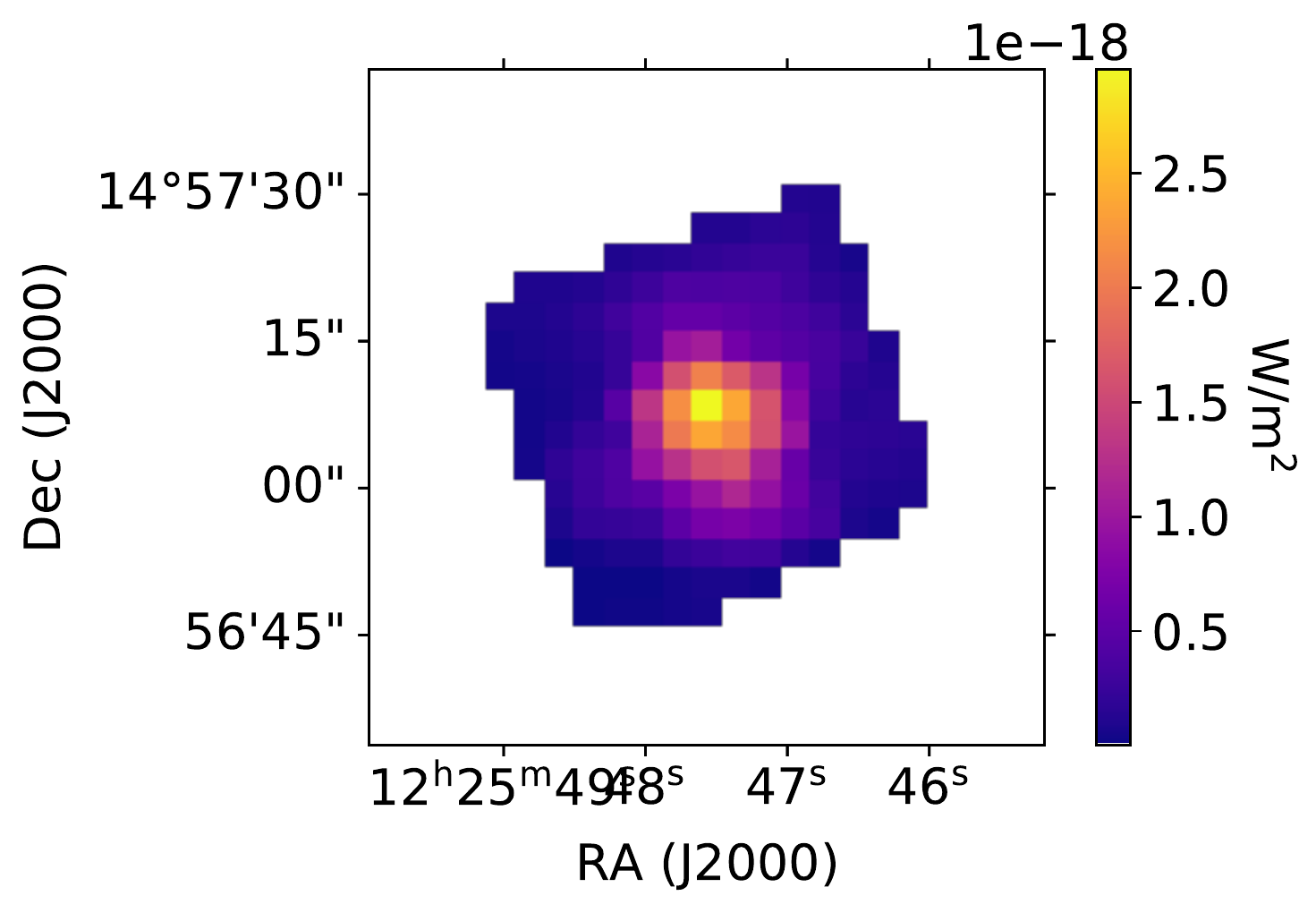}{0.25\textwidth}{VCC 841}\fig{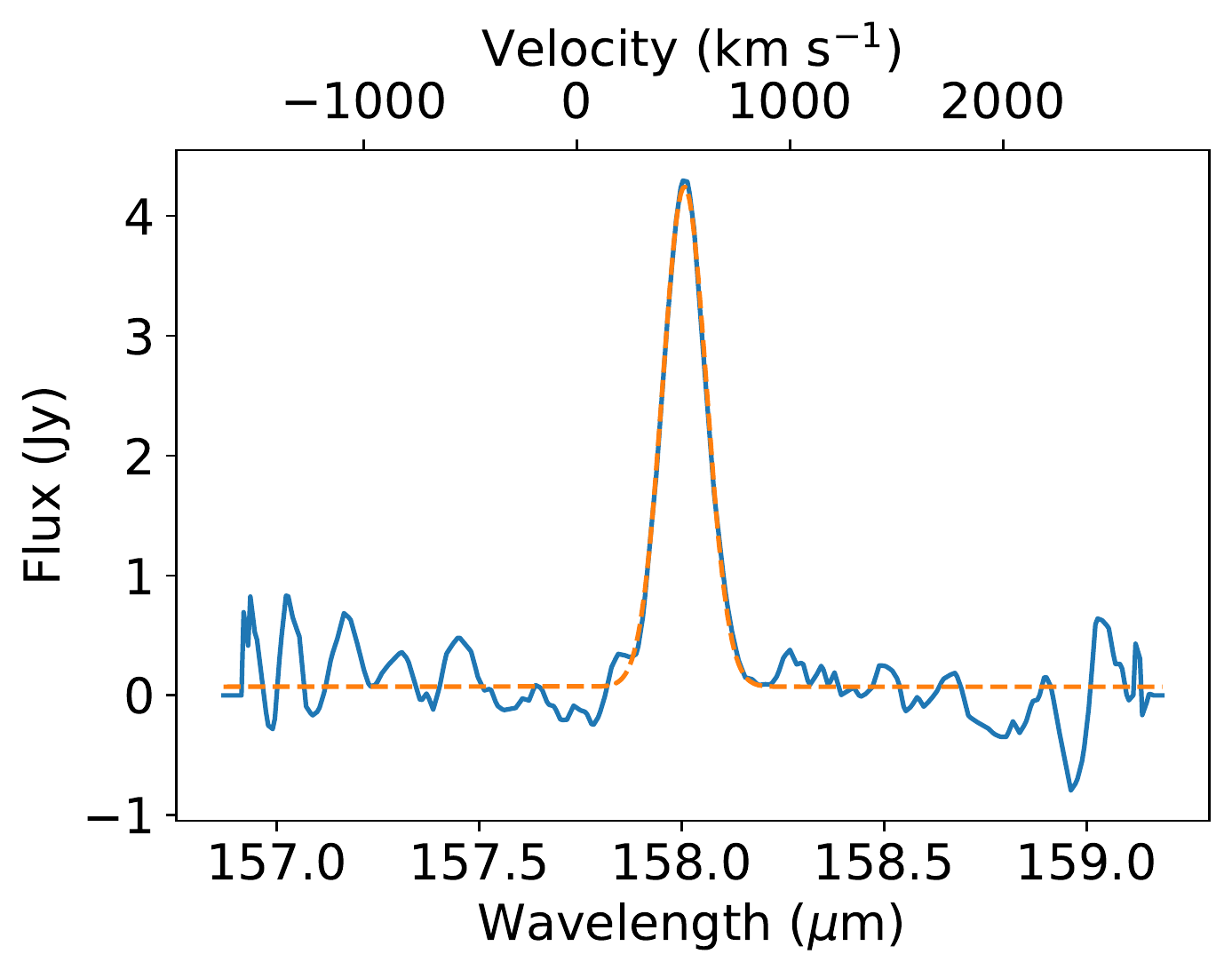}{0.25\textwidth}{VCC 841}
}
\gridline{
\fig{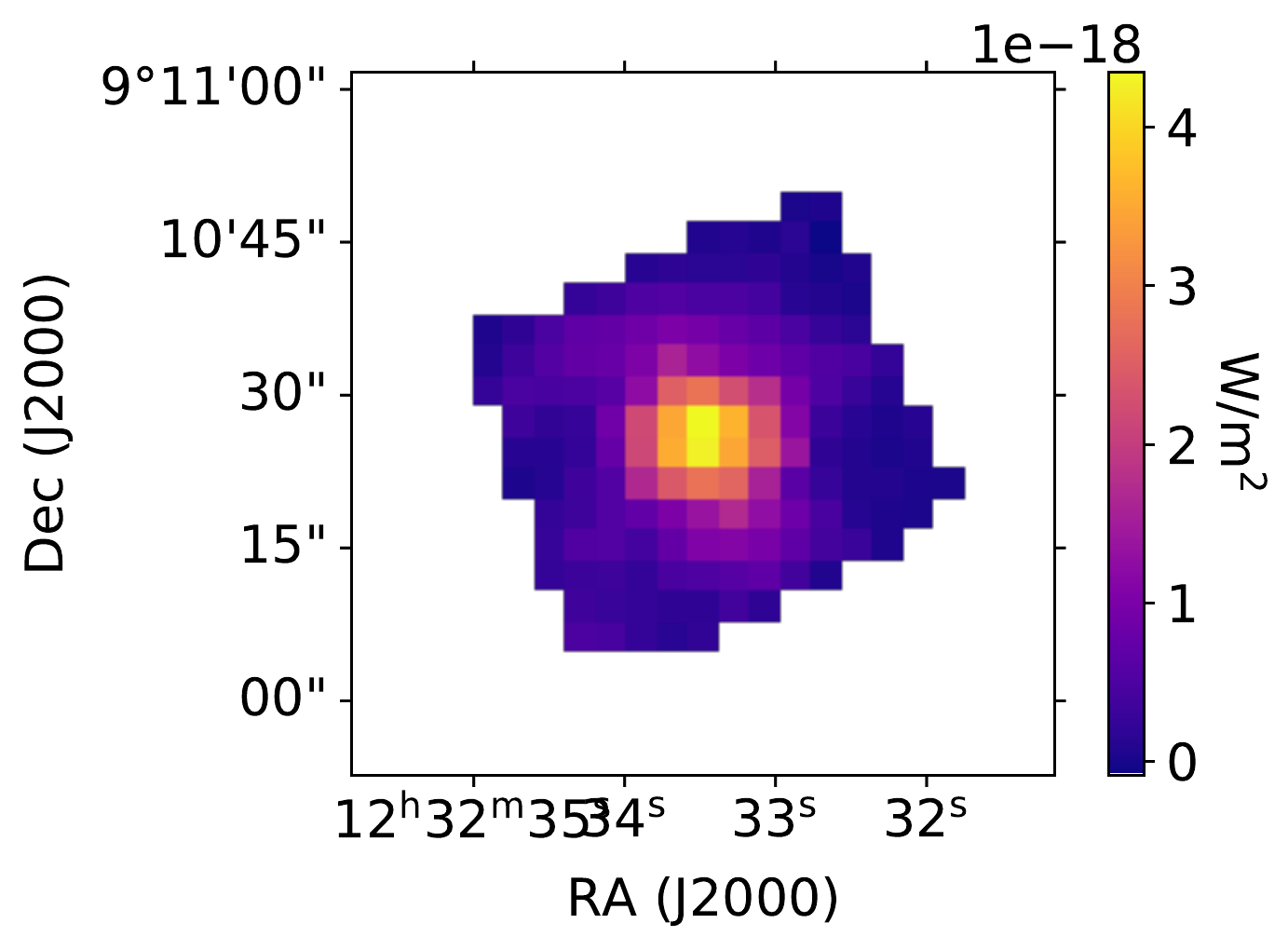}{0.25\textwidth}{VCC 1437}\fig{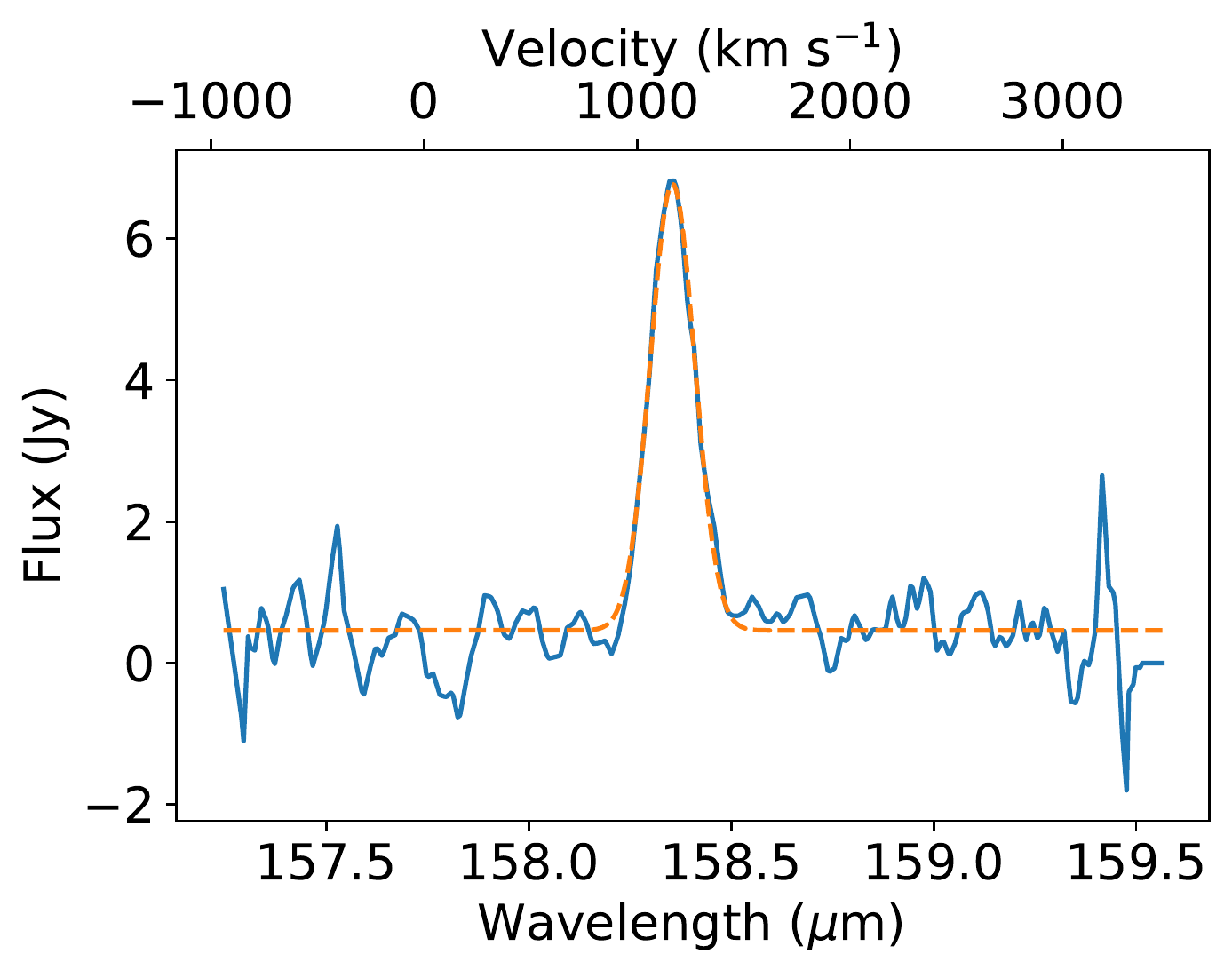}{0.25\textwidth}{VCC 1437}
\fig{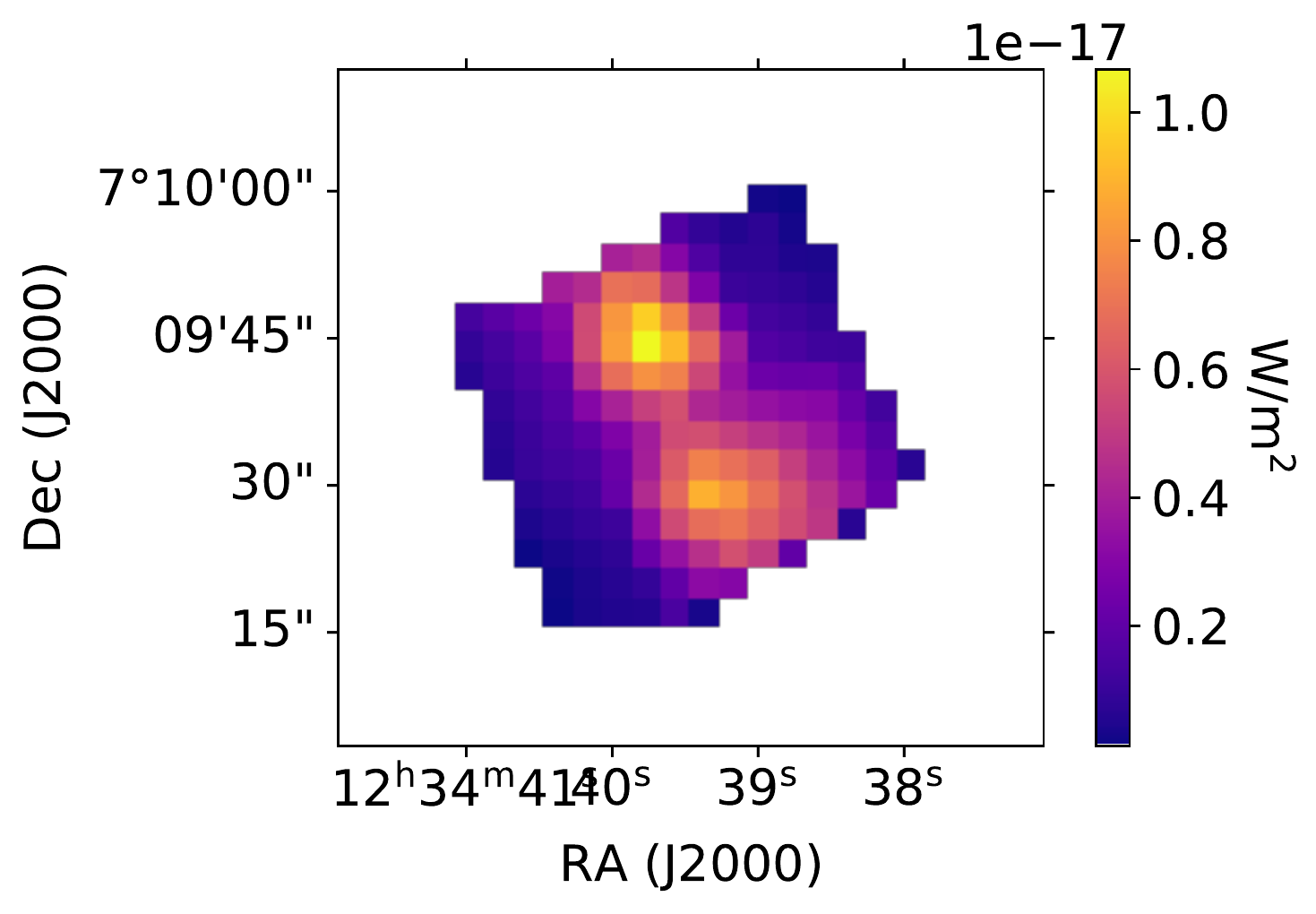}{0.25\textwidth}{VCC 1575}\fig{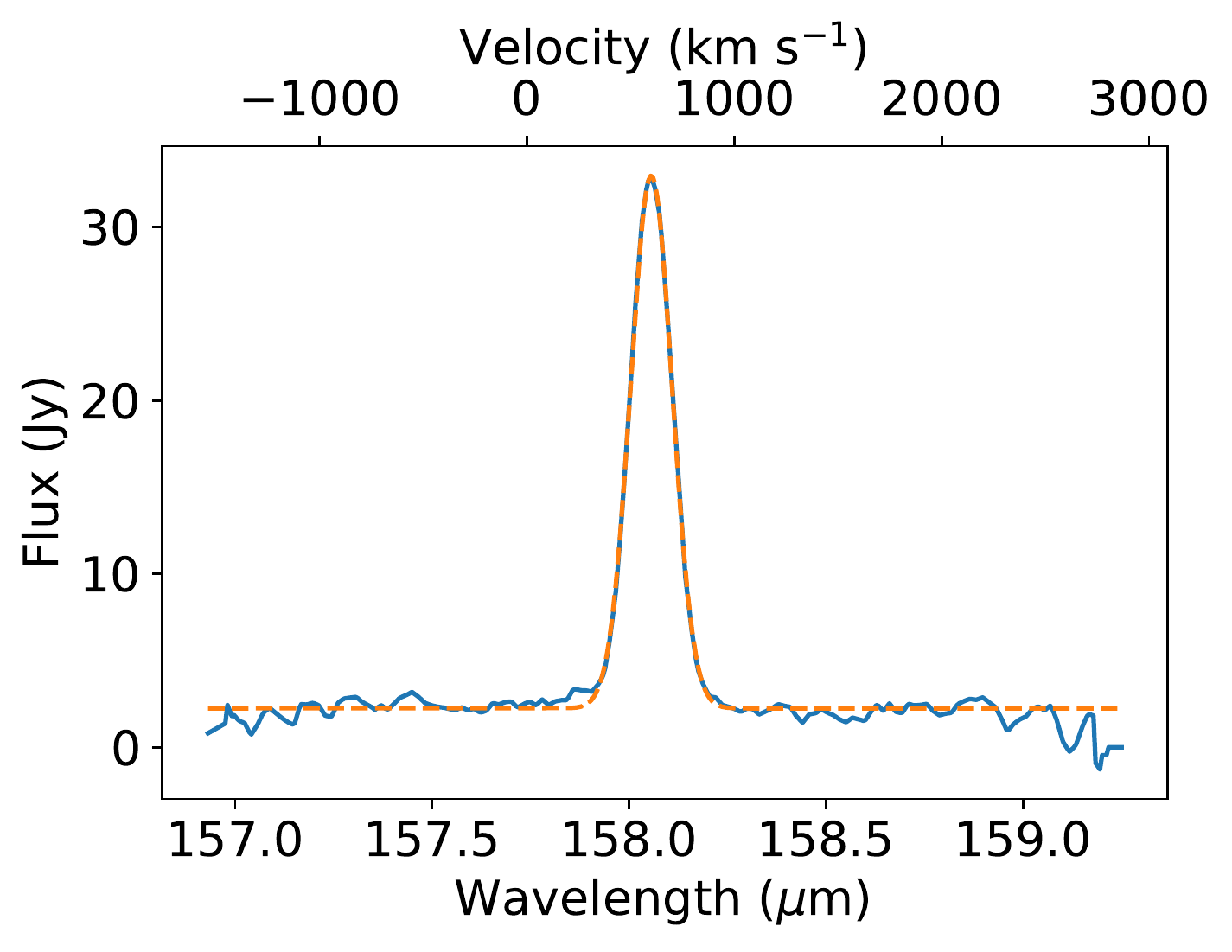}{0.25\textwidth}{VCC 1575}
}
\gridline{
\fig{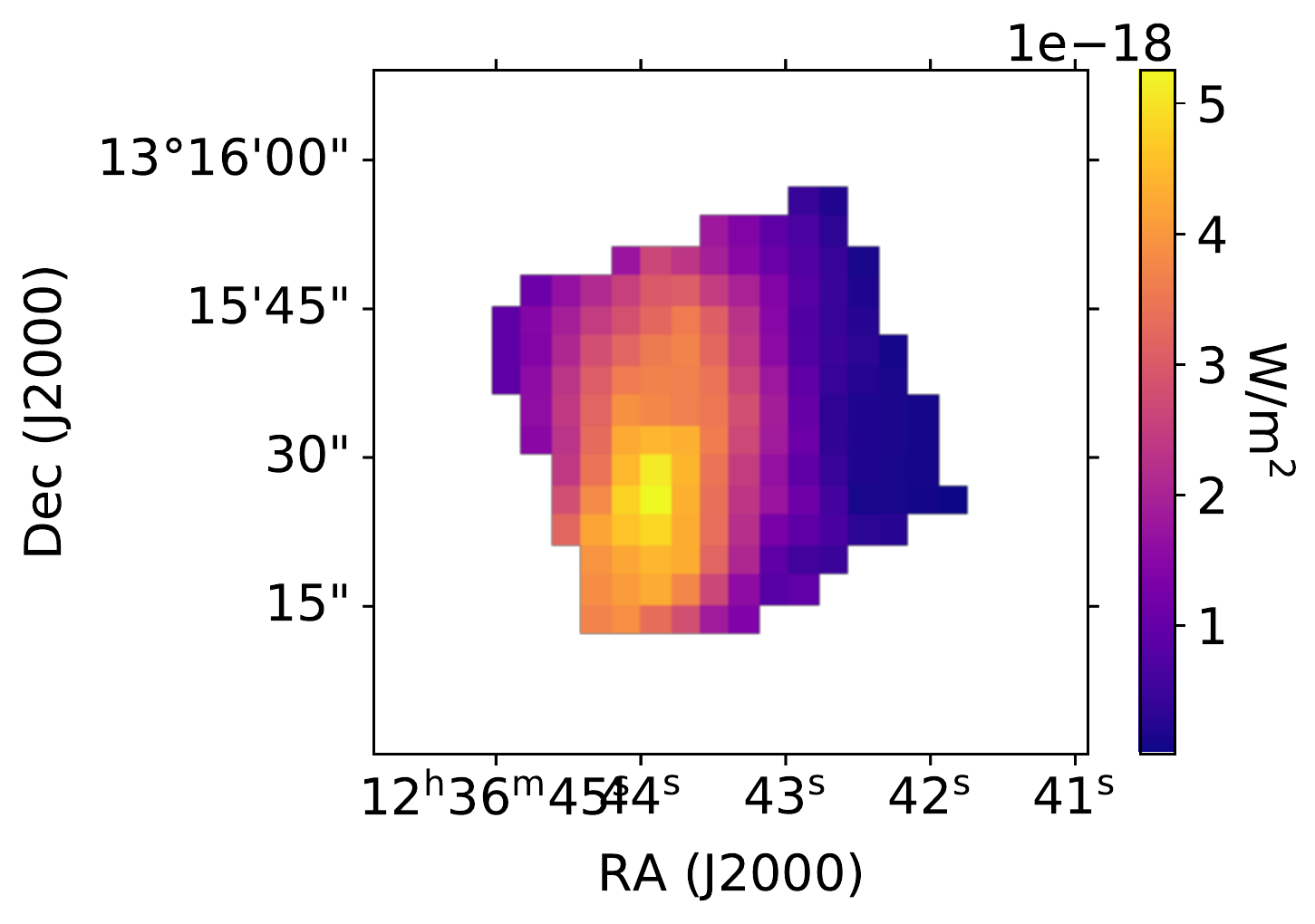}{0.25\textwidth}{VCC 1686}\fig{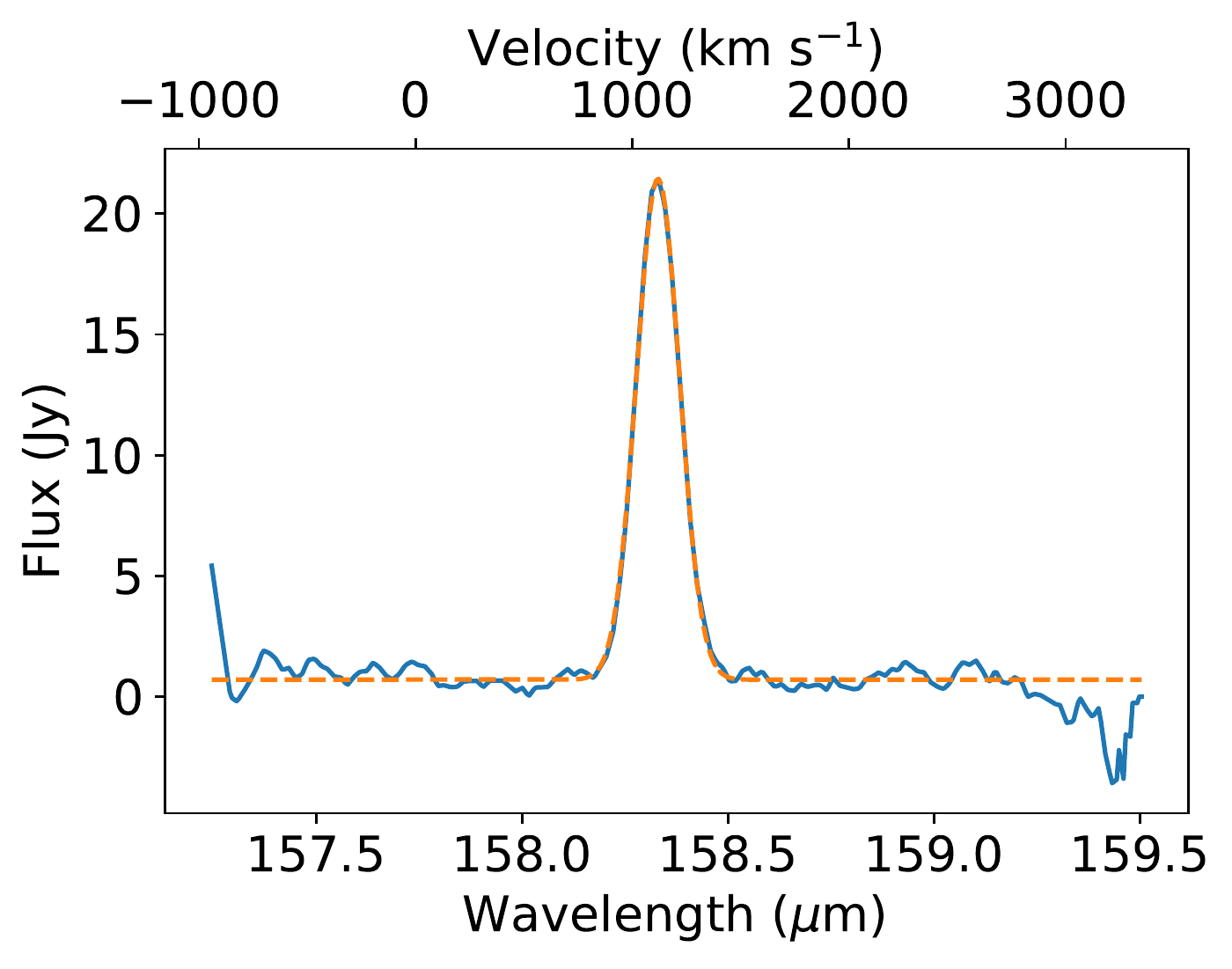}{0.25\textwidth}{VCC 1686}
\fig{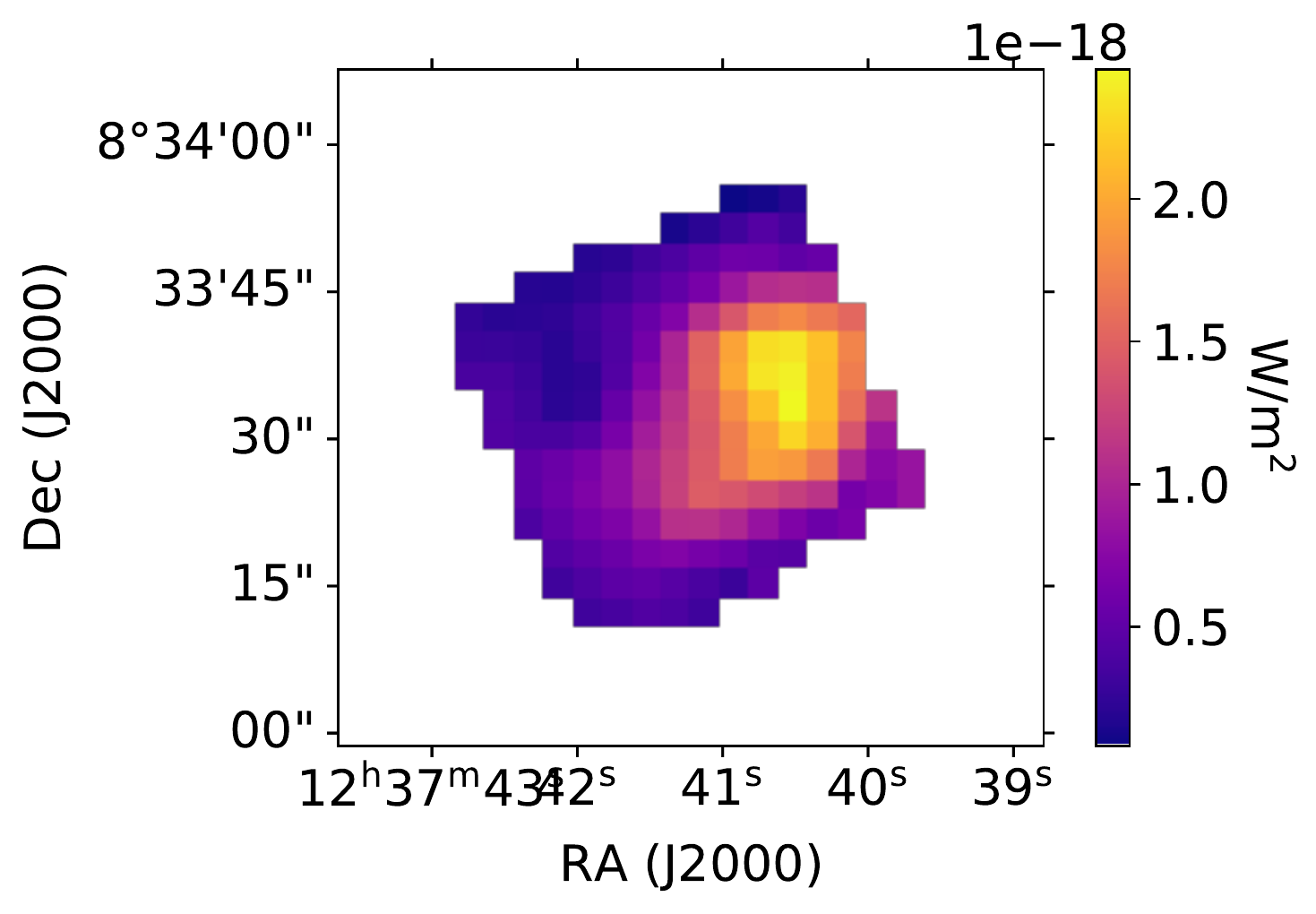}{0.25\textwidth}{VCC 1725}\fig{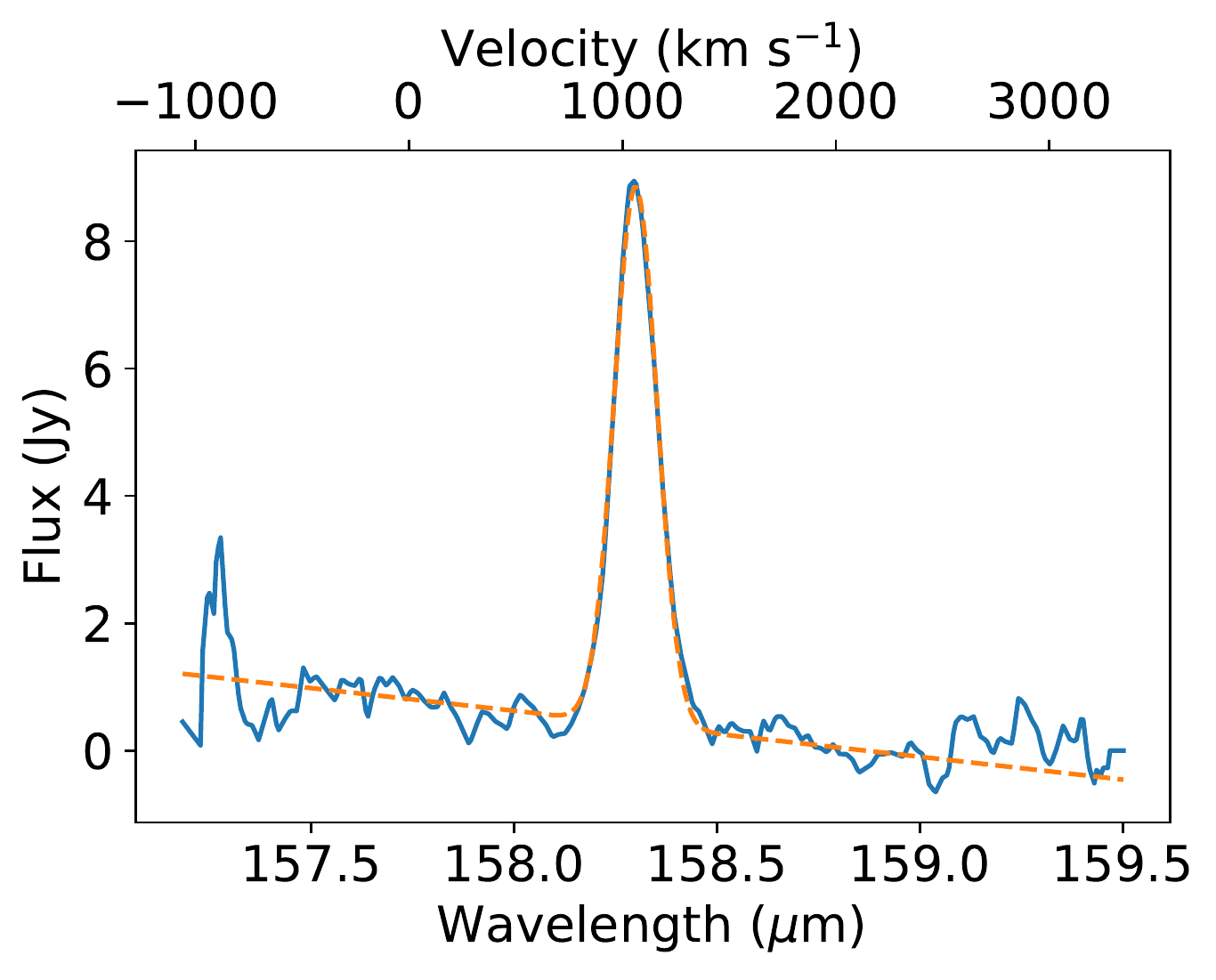}{0.25\textwidth}{VCC 1725}
}
\caption{\it(continued)}
\end{figure*}

The {\it Herschel} \cii\ data cubes were analyzed using SOSPEX \citep{2018AAS...23115011F}. Spatially integrated spectra were created in SOSPEX using an aperture that took in the whole map and line profiles were fitted to these spectra in SOSPEX using Gaussian profiles; baseline regions were defined visually to be clear of the emission lines and of the noisy end regions of the spectra. These fits were used to obtain the integrated \cii\ flux and its associated error. With the exceptions of VCC 334, VCC 737 and VCC 1725, baselines were assumed to be flat and were set to the median of the data (excluding the line region and the noisy ends of the spectra). For the three galaxies mentioned, visual inspection showed that the flat baseline initially fitted was not a good match for the actual spectral baseline. These three galaxies were fitted with a sloped baseline: for VCC 737 this reduced the line flux by 4\%, while for VCC 334 and VCC 1725 the reduction was less than 1\% and was within the measurement errors. The fitted line profile is shown together with the spectrum in Figure \ref{HerschelCII}. The values of the \cii\ intensity derived from these fitted line profiles are used for the rest of the analysis presented here (see Table \ref{resultstable}). Line maps were created using line fitting to each pixel in SOSPEX. The line maps are used to illustrate and examine the extent of the \cii\ but measurements taken from these maps were not used in the analysis presented here. 

\subsection{Spectral Energy Distribution fitting}\label{sedfit}

\begin{figure*}
\gridline{
\fig{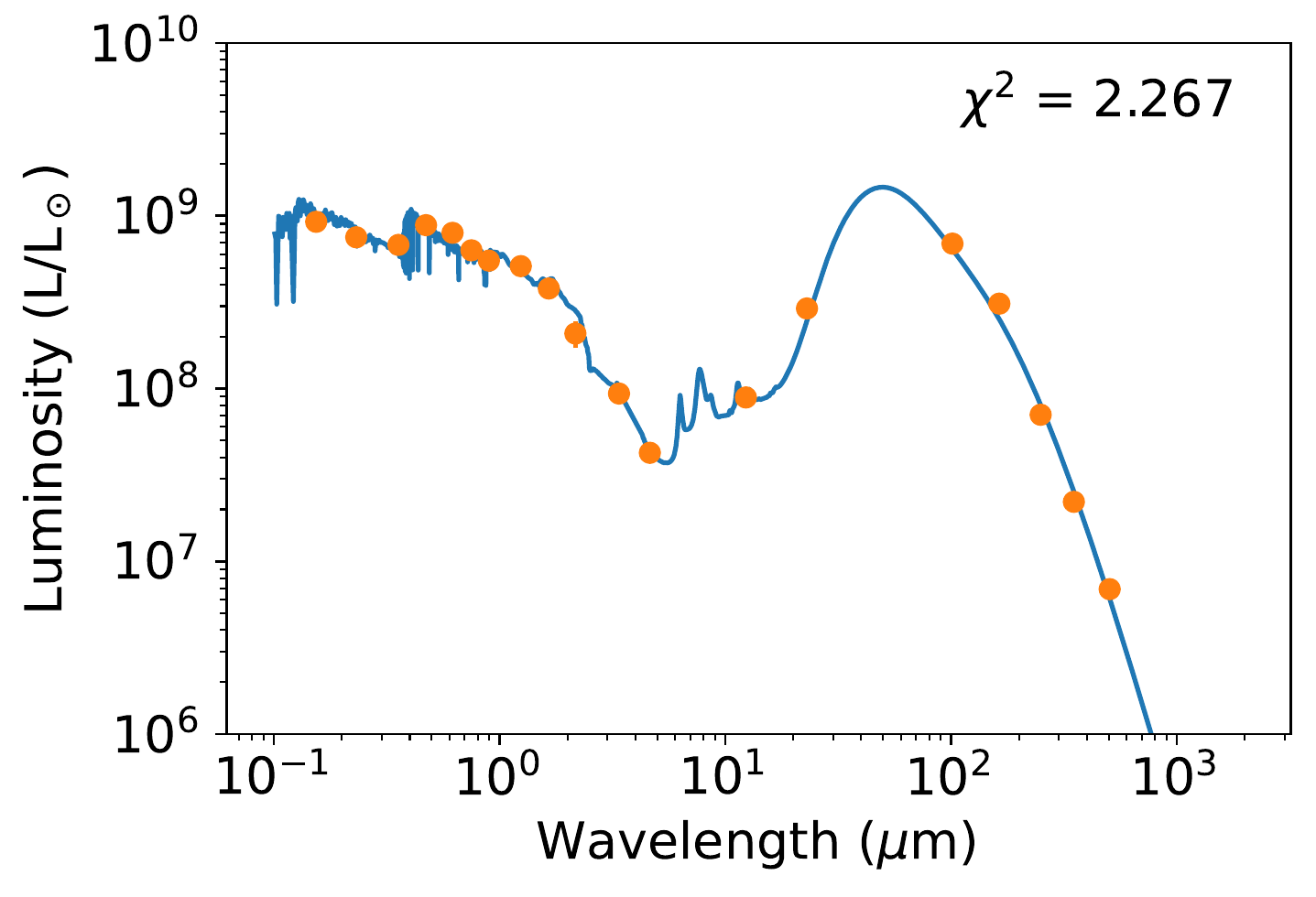}{0.3\textwidth}{VCC 144}
\fig{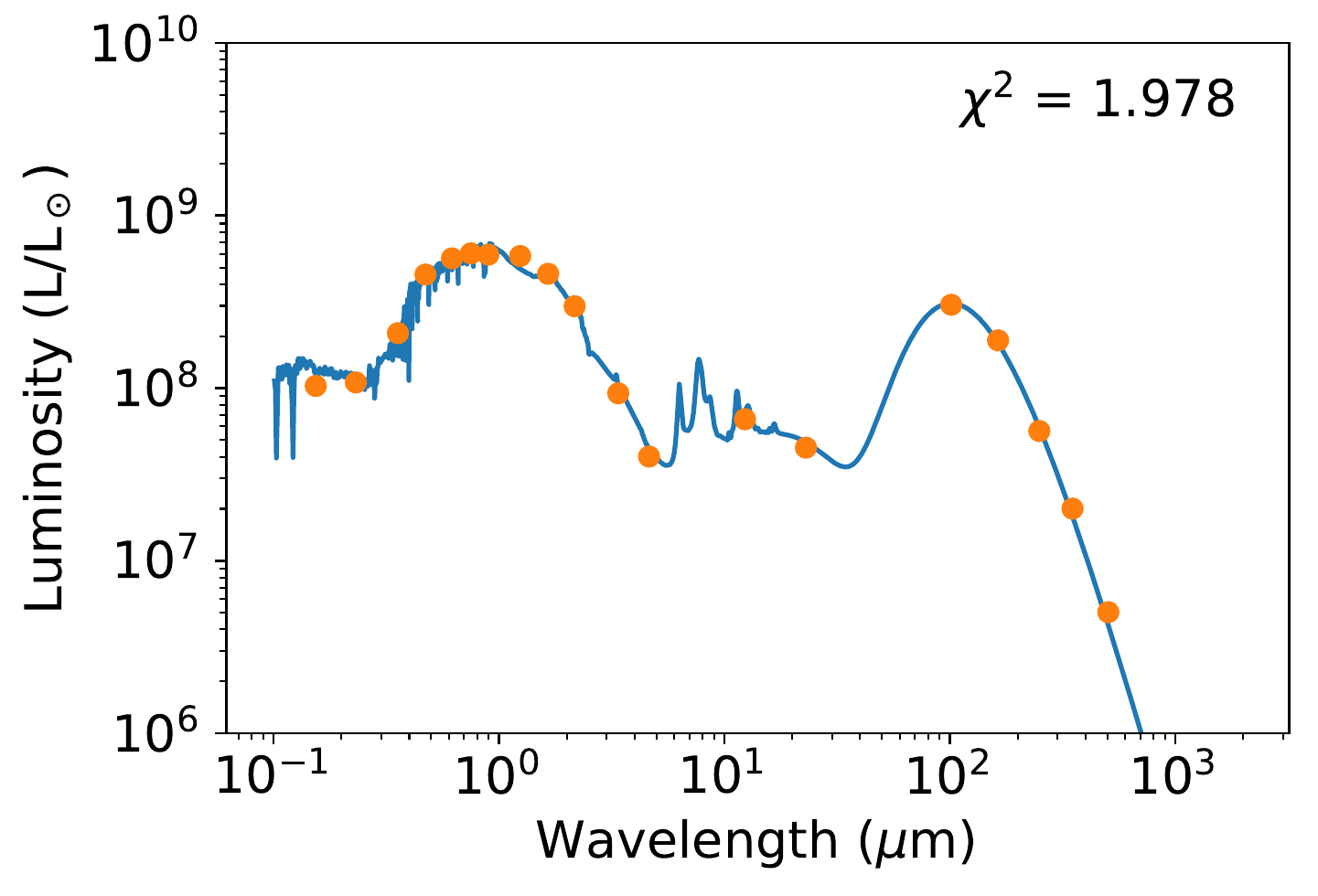}{0.3\textwidth}{VCC 213}
\fig{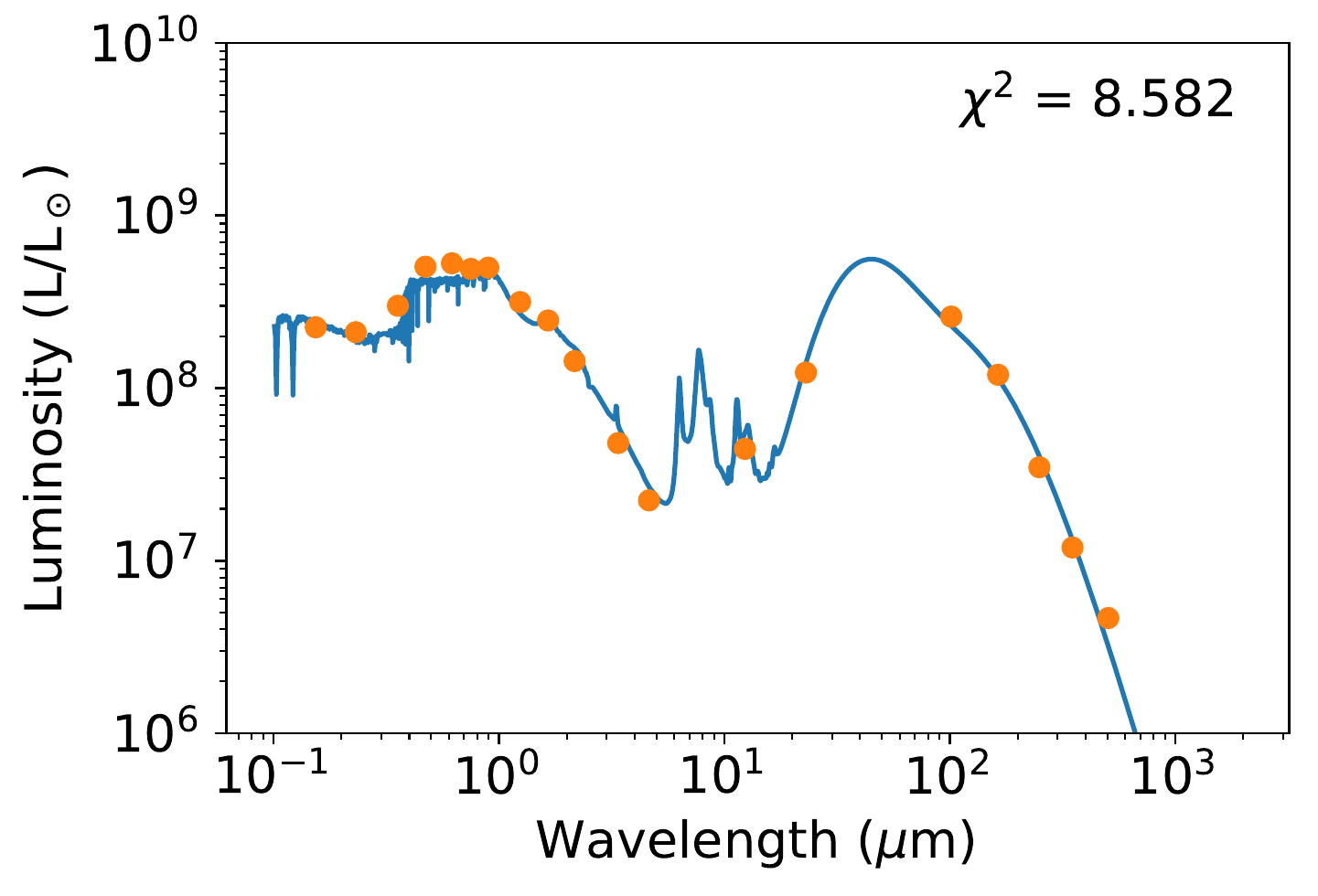}{0.3\textwidth}{VCC 324}
}
\gridline{
\fig{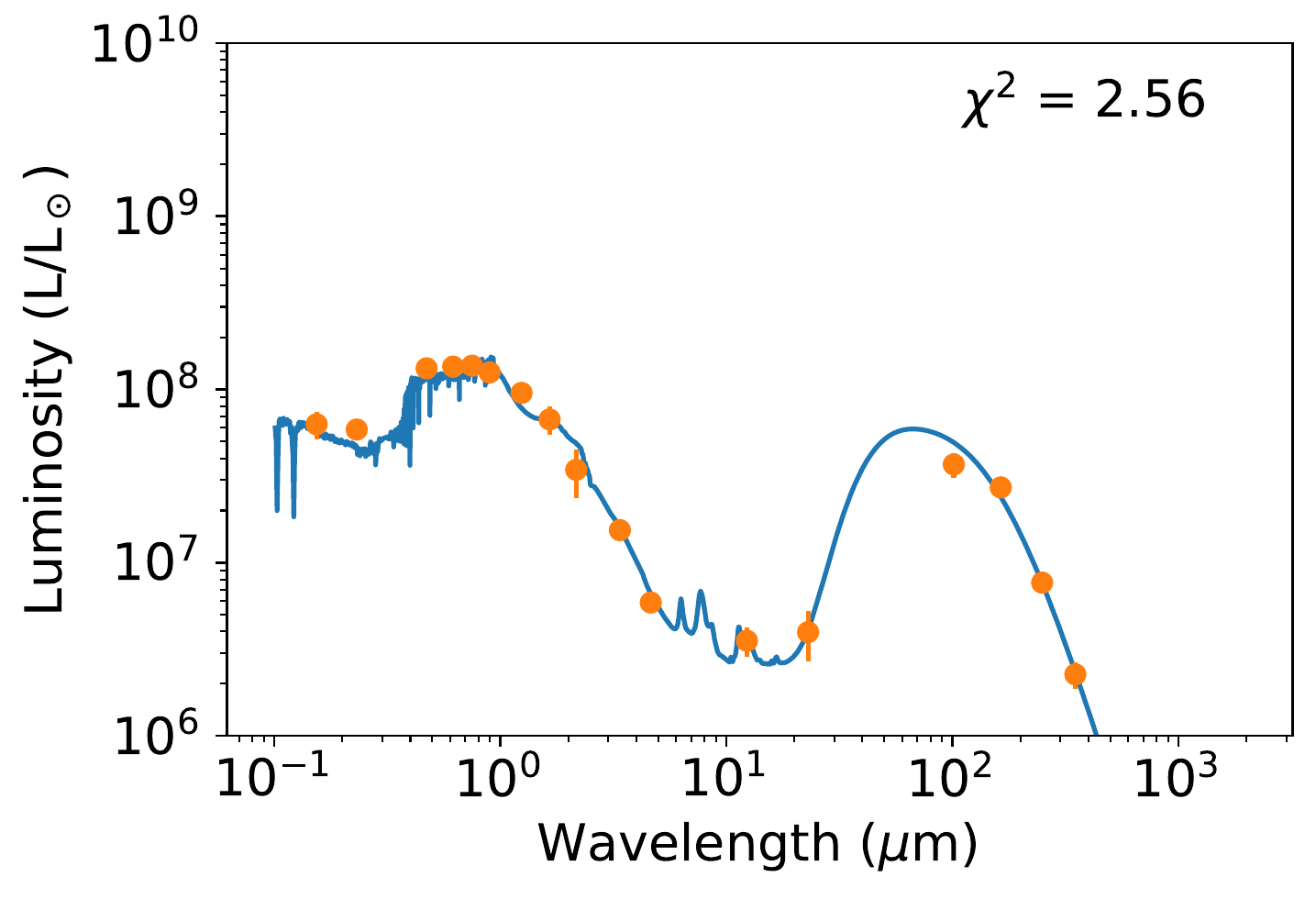}{0.3\textwidth}{VCC 334}
\fig{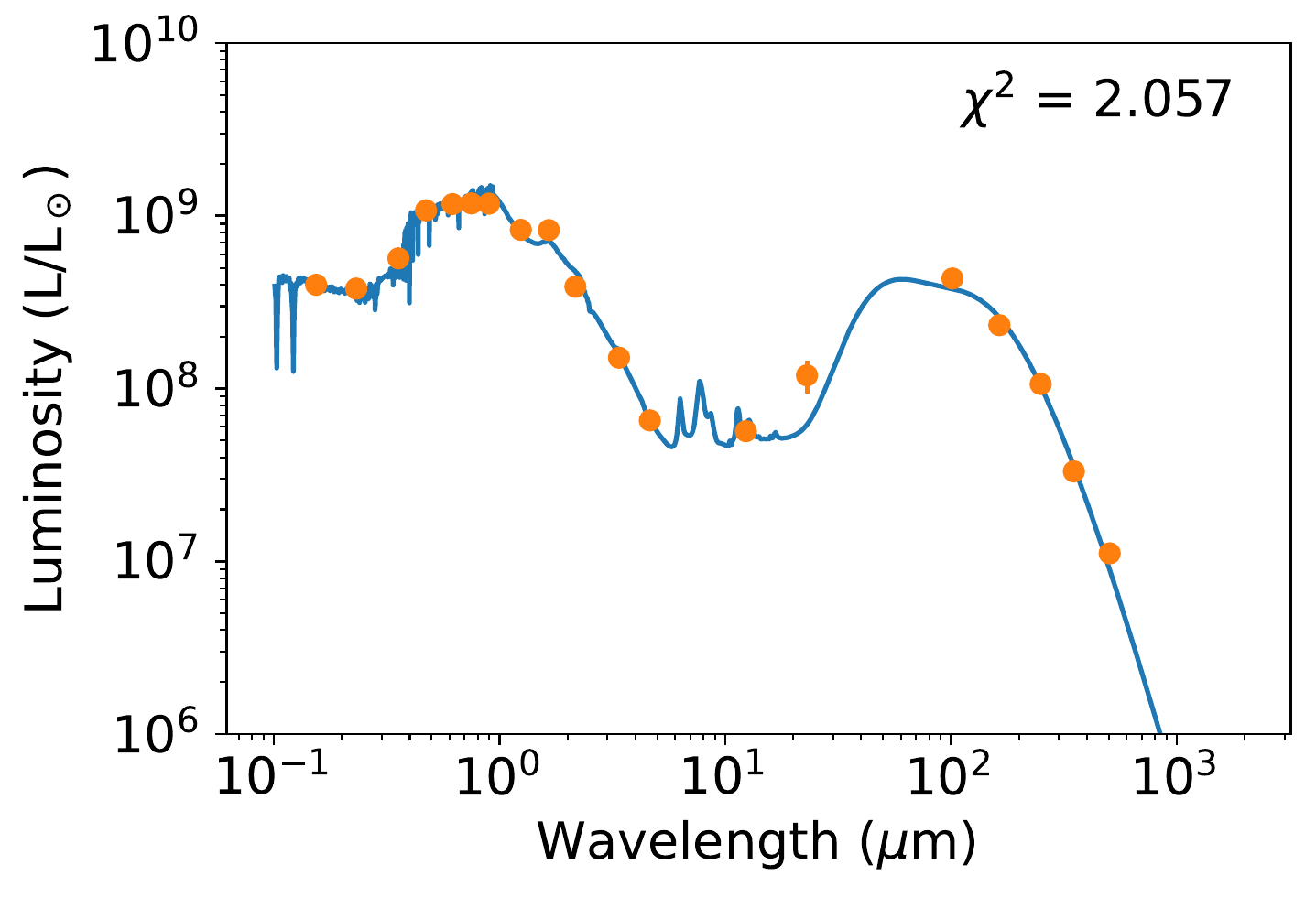}{0.3\textwidth}{VCC 340}
\fig{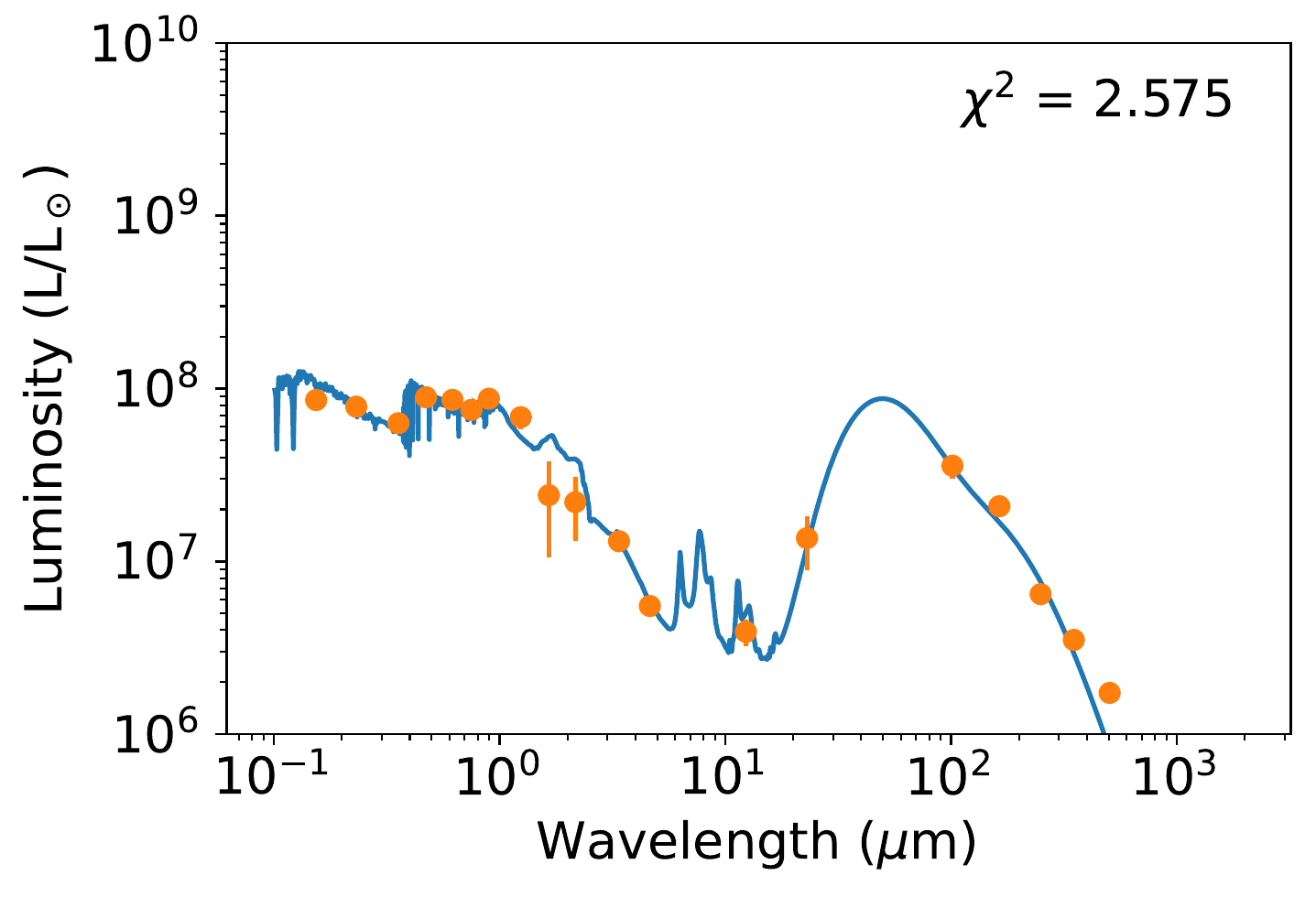}{0.3\textwidth}{VCC 562}
}
\gridline{
\fig{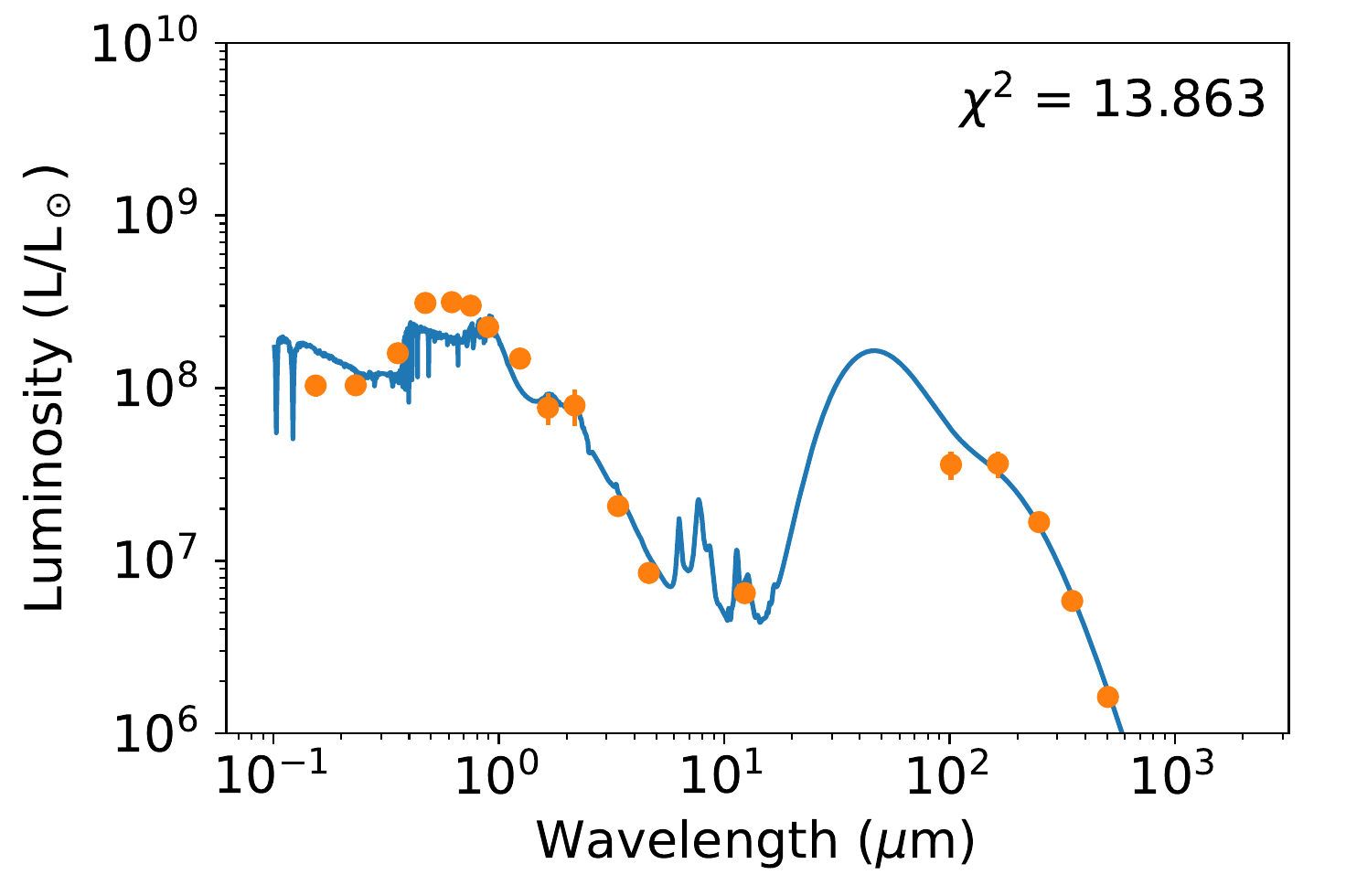}{0.3\textwidth}{VCC 693}
\fig{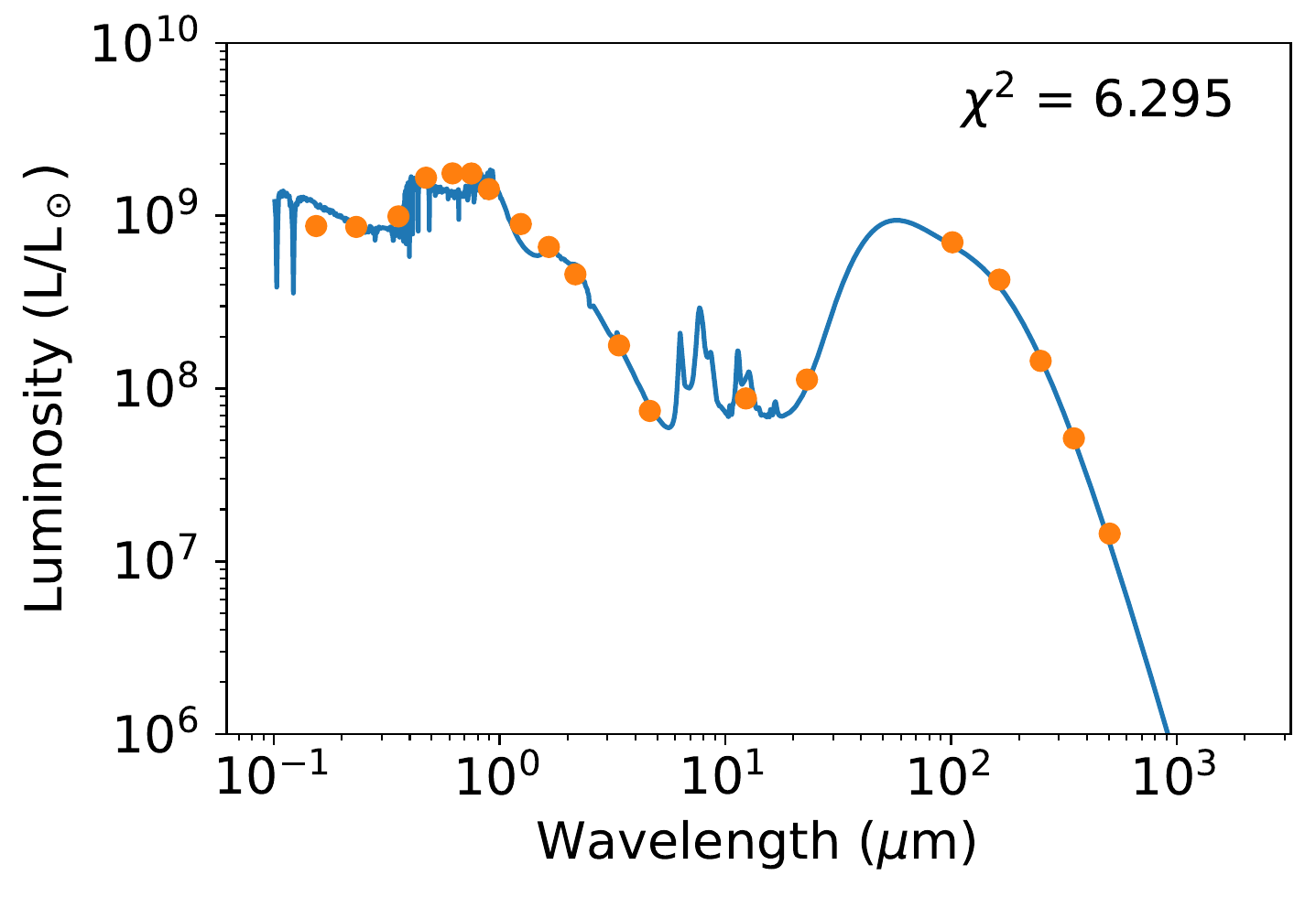}{0.3\textwidth}{VCC 699}
\fig{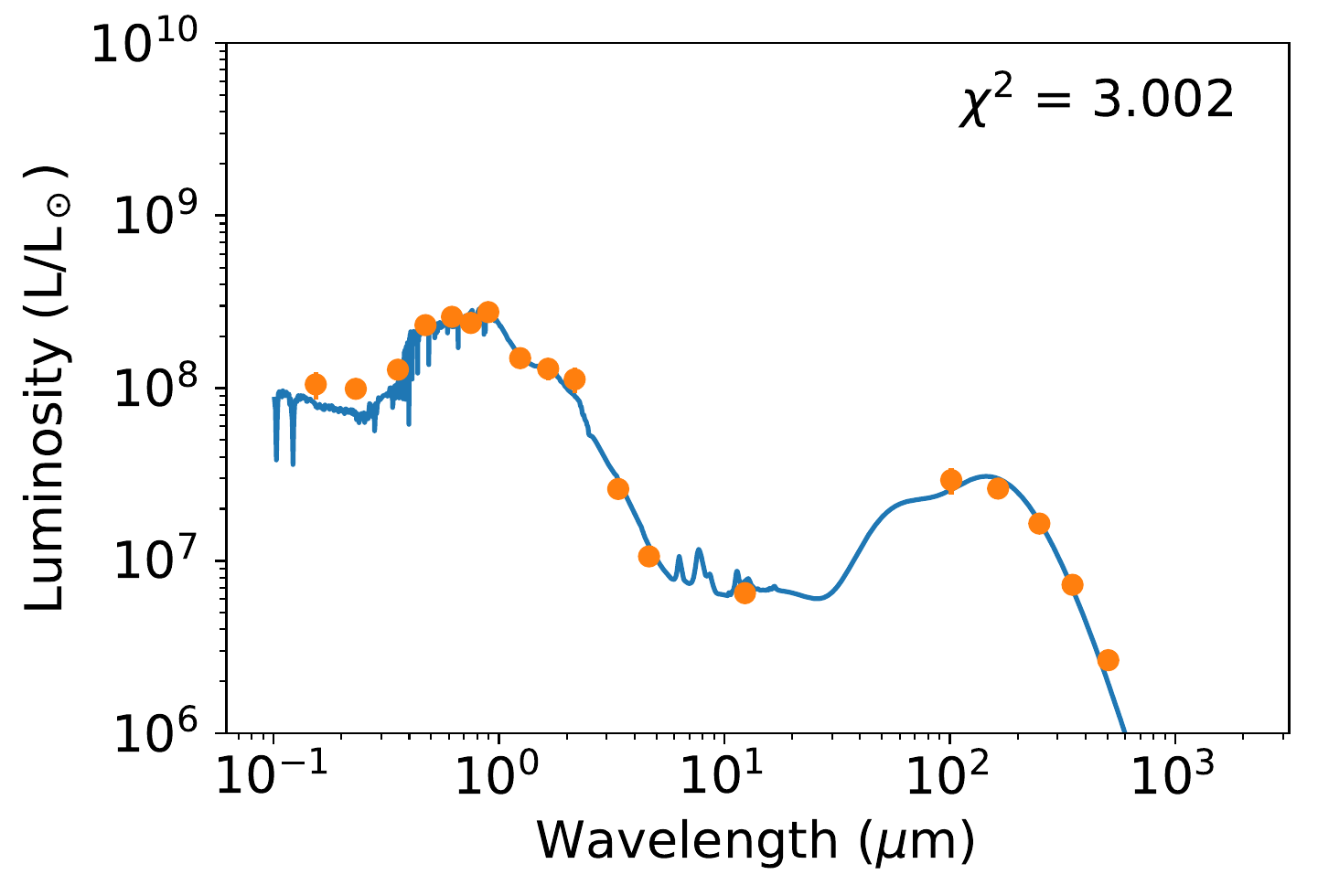}{0.3\textwidth}{VCC 737}
}
\gridline{
\fig{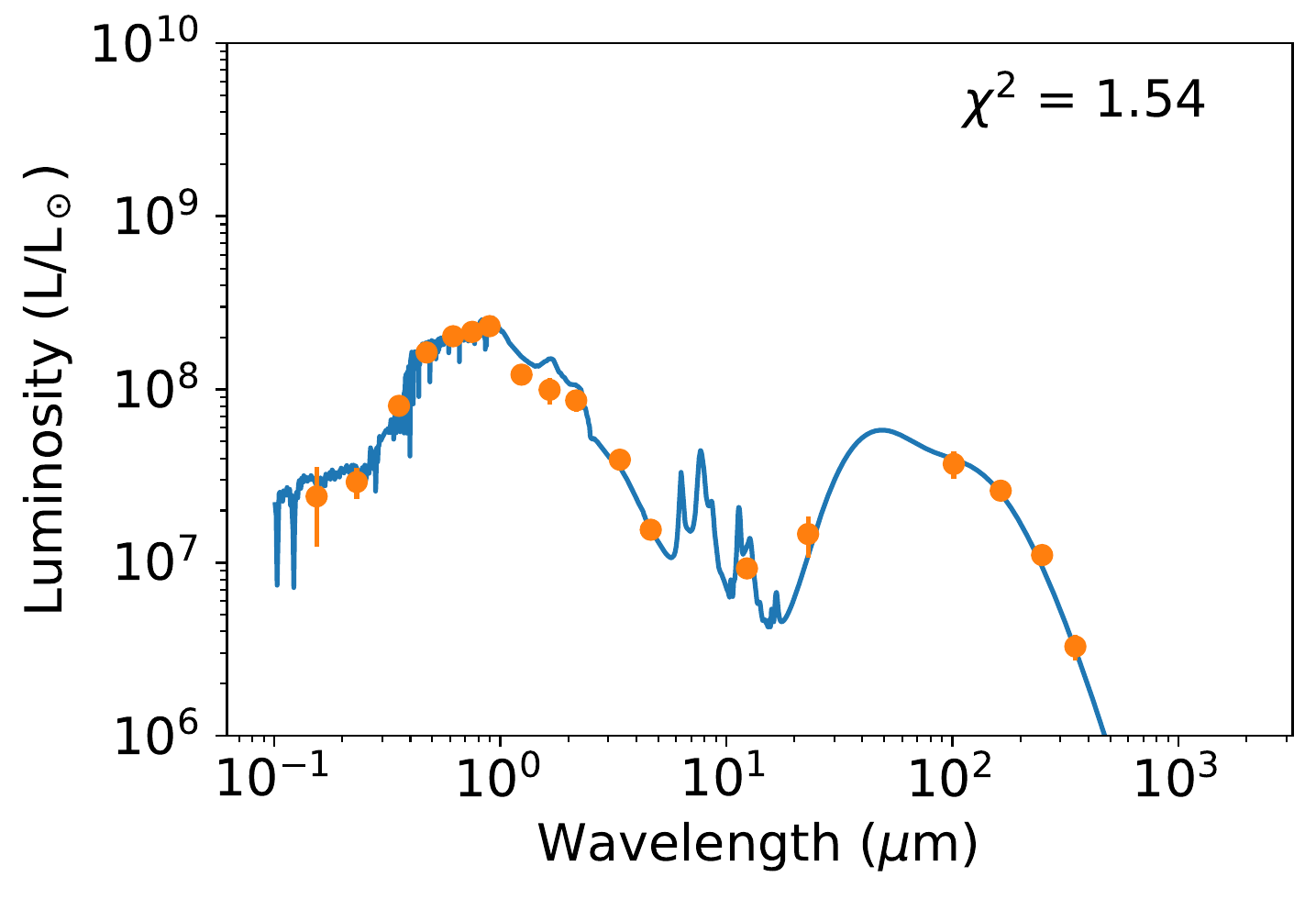}{0.3\textwidth}{VCC 841}
\fig{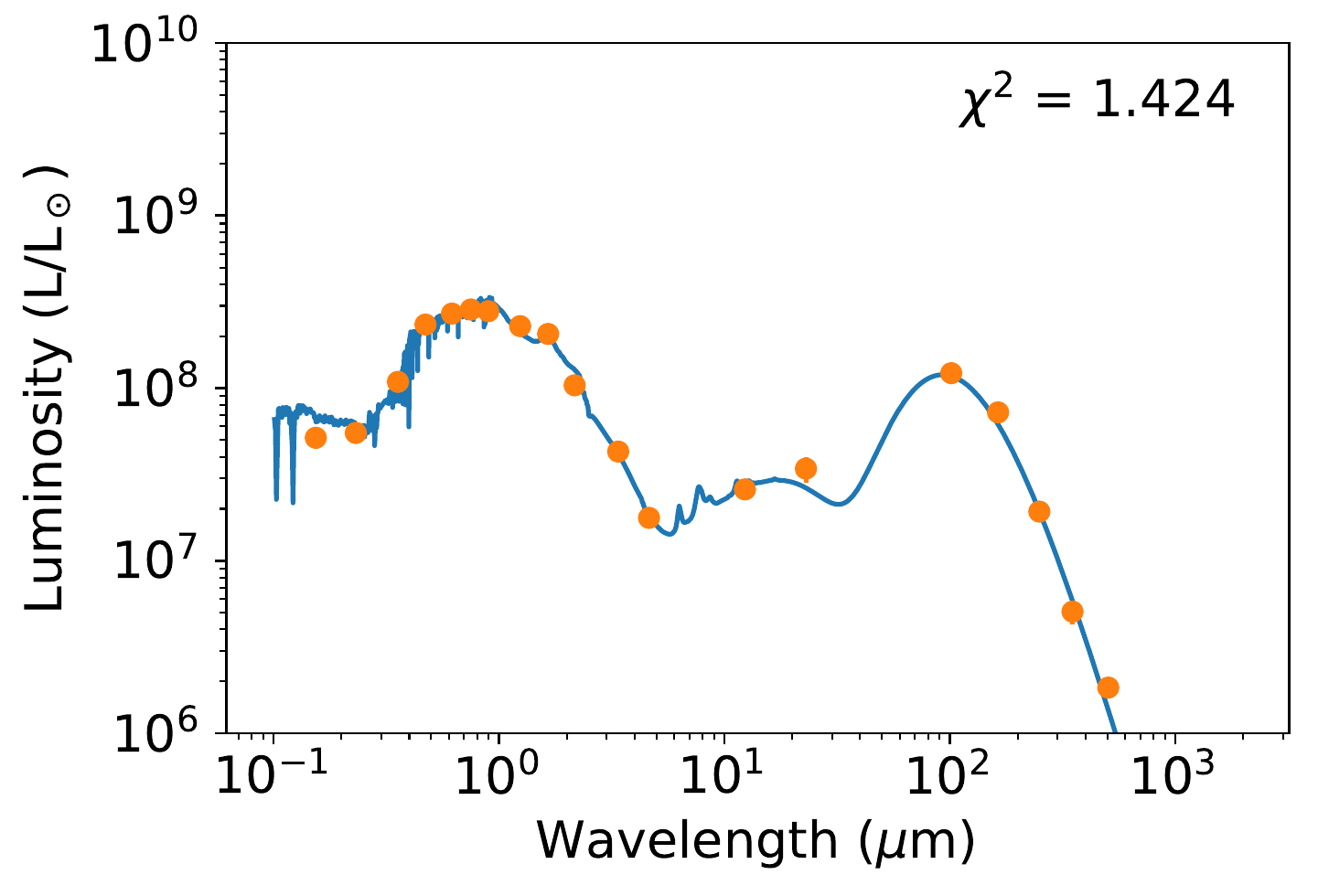}{0.3\textwidth}{VCC 1437}
\fig{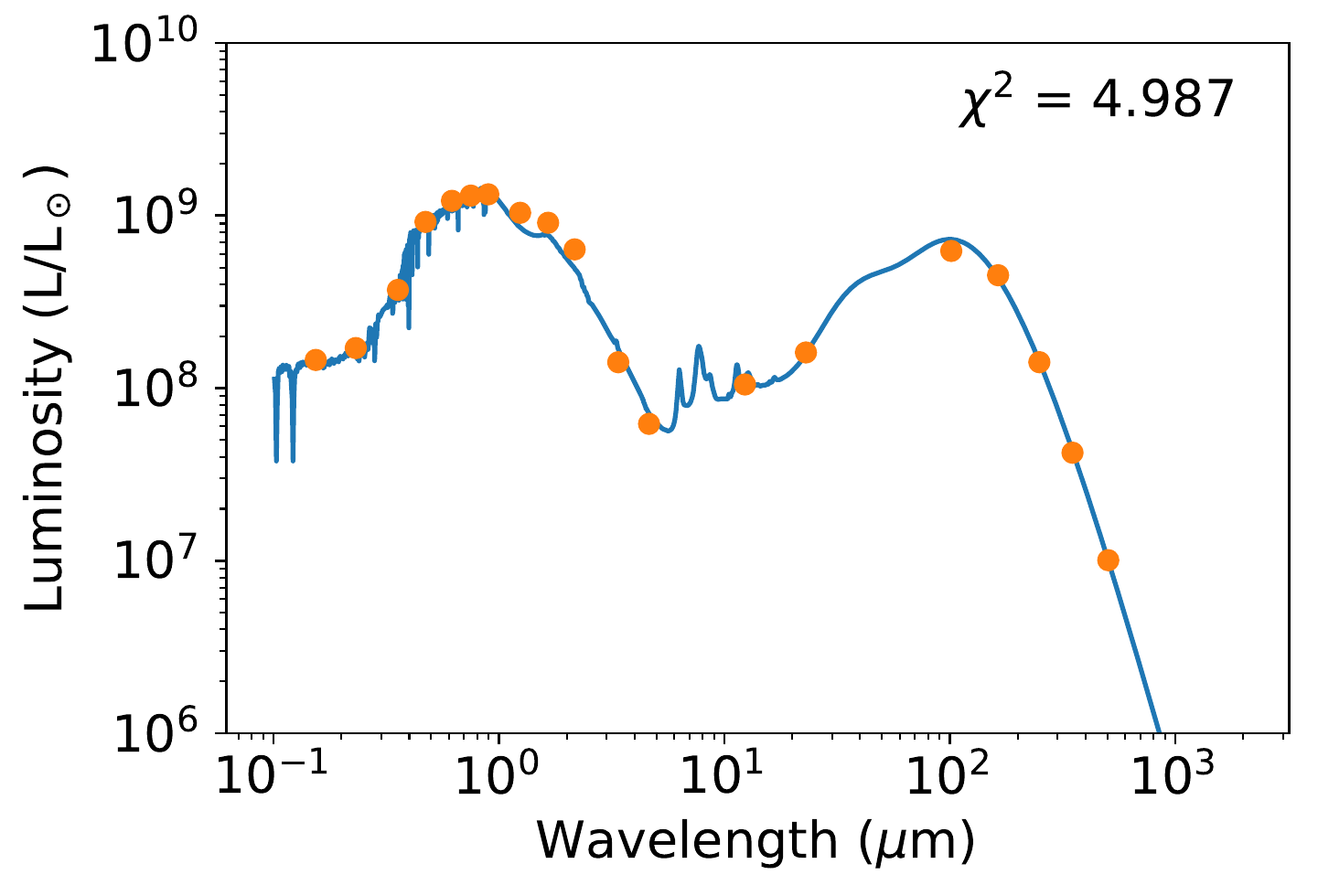}{0.3\textwidth}{VCC 1575}
}
\gridline{\hfill
\fig{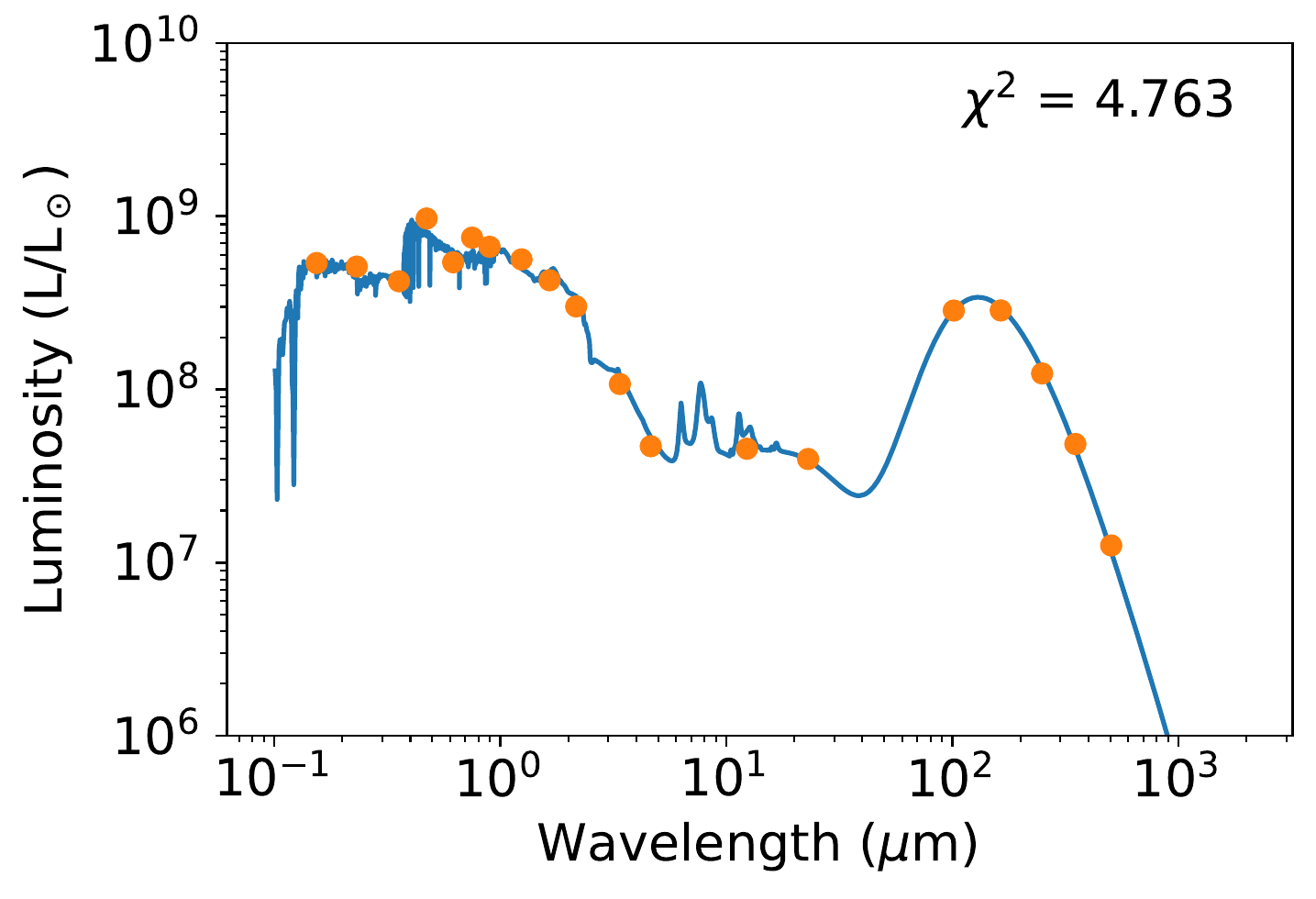}{0.3\textwidth}{VCC 1686}
\fig{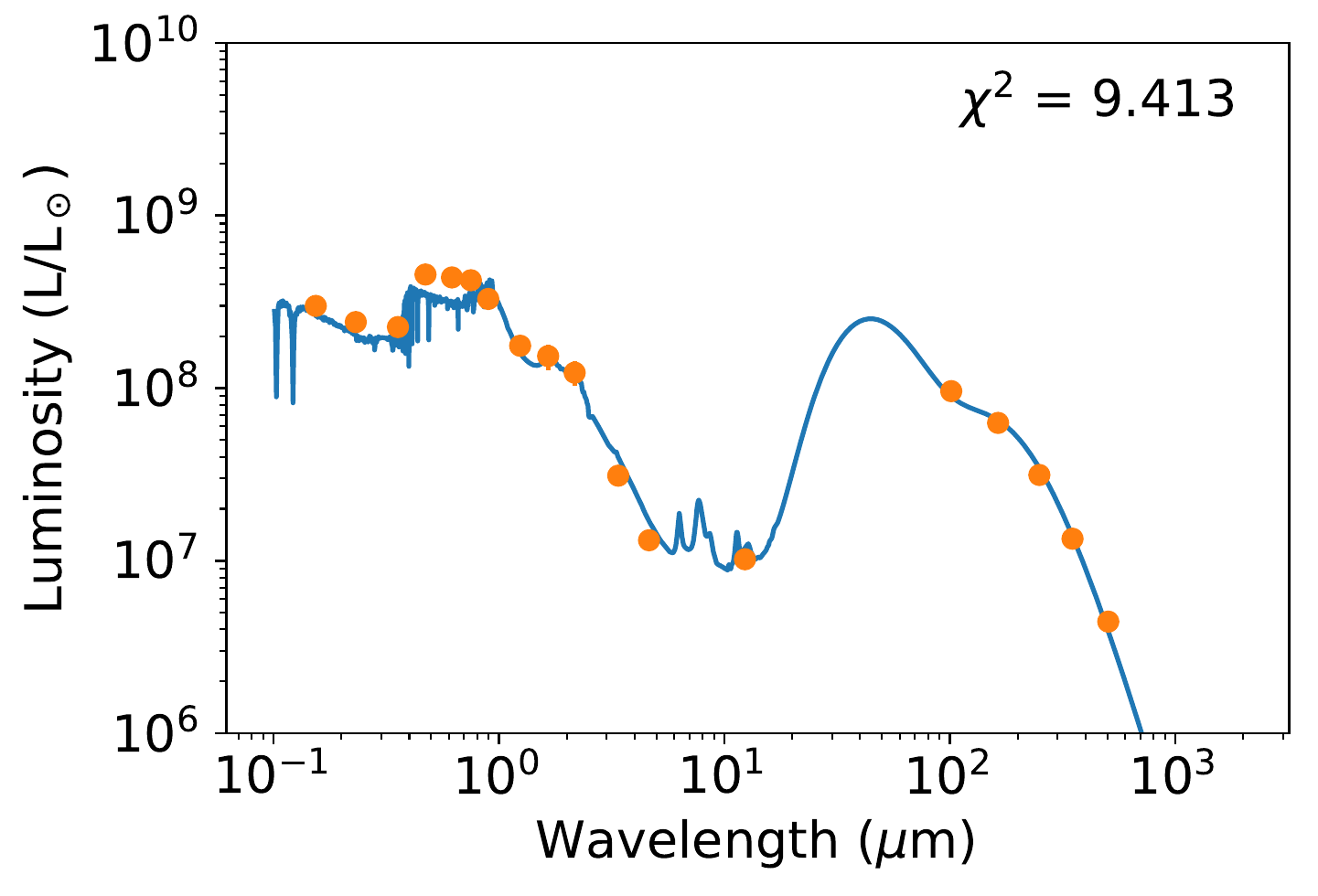}{0.3\textwidth}{VCC 1725}\hfill
}
\caption{Spectral Energy Distributions of the fourteen galaxies in the sample. Measured datapoints are shown as orange circles; errorbars are shown on these datapoints but may be smaller than the symbol used. The best-fit SED from MAGPHYS is shown as a solid blue line with its $\chi^2$ given in the upper-right.\label{MagphysSEDs}}
\end{figure*}

Spectral Energy Distributions (SEDs) were created using FIR data from the {\it Herschel} Virgo Cluster Survey \citep[HeVICS;][]{2013MNRAS.428.1880A}, as reanalyzed by \citet{2015AA...574A.126G}, mid infrared data from AllWISE \citep{2013wise.rept....1C, 2010AJ....140.1868W} and optical data from the SDSS/Extended Virgo Cluster Catalogue \citep[EVCC;][]{2014ApJS..215...22K}, supplemented with our own analysis of near infrared data from 2MASS \citep{2006AJ....131.1163S} and UV data from GALEX \citep{2007ApJS..173..682M}. For VCC 1686, which has a foreground star superposed on the galaxy, we used our own measurement from the AllWISE data and the SDSS fluxes, masking out the star and patching the region with a similar region from elsewhere in the galaxy. We also use our own measurement of the SDSS {\it z} band flux of VCC 693, as the literature value from \citet{2014ApJS..215...22K} was highly discrepant from the other fluxes (and our measurement) for unknown reasons. Errors were calculated following the prescriptions given in the documentation in the archives, or following \citet{2014ApJS..215...22K} for the SDSS. Absorption corrections to the UV, optical and NIR fluxes were made using the dust measurements of \citet{2011ApJ...737..103S}, via the IRSA Galactic Dust Reddening and Extinction service.\footnote{\url{https://irsa.ipac.caltech.edu/applications/DUST/}} As the galaxies in this sample are similar in size to or smaller than the beam (i.e. are point sources) for the crucial {\it Herschel} measurements of the FIR fluxes, it is not reliable to use a single aperture to measure the SED. We therefore use the integrated whole-galaxy flux at all wavelengths rather than attempting to make measurements within a defined aperture.

SED fitting was carried out using MAGPHYS \citep[Multi-wavelength Analysis of Galaxy Physical Properties;][]{2008MNRAS.388.1595D} which returns both an overall best-fit SED and marginalized probability distributions for the individual parameters. Fluxes and errors, both converted to Jy, were supplied as inputs to the fitting, either from the literature or based on our own measurements as described above (see Table \ref{SEDtable}). The plots in Figure 2 show the overall best-fit SED output by MAGPHYS; for our analysis we use the marginalized probabilities for $L_d^{tot}$ (the MAGPHYS parameter that gives the TIR) using the 50\% point of the probability distribution as the central estimator and the 16\% and 84\% points as the estimators for the 1$\sigma$ error. The difference between the value of $L_d^{tot}$  returned for the best-fit model and the 50\% point of the probability distribution of $L_d^{tot}$  is less than 1$\sigma$ (or less than 1\% in the cases where the error estimate is zero) except for VCC 562 (best-fit $L_d^{tot}$ 0.05 dex higher) and VCC 841 (best-fit $L_d^{tot}$ 0.21 dex higher).

\begin{deluxetable*}{lcr@{ $\pm$ }lCr@{}Lr@{}L}
\tablecaption{Measured values for the galaxies from the \cii\ observations and the SED fitting.\label{resultstable}}
\tablehead{\colhead{Galaxy ID}&{\it Herschel} \cii&\multicolumn2c{\cii\ flux} &\colhead{TIR Luminosity}&\multicolumn2c{TIR flux}&\multicolumn2c{\cii/TIR}\\&OBSID&\multicolumn2c{(10$^{-18}$ W\,m$^{-2}$)}&\colhead{log(L$_{\rm TIR}$/\nom{L})}&\multicolumn2c{(10$^{-15}$ W\,m$^{-2}$)}&\multicolumn2c{(\%)}}
\startdata
\\[-2ex]
 VCC 144 & 1342224404 &\phn\phn 232.8 & 1.6 & 9.28^{+0.05}_{-0.01} &\phn\phn\phn  60.2&^{+8.1}_{-0.7}  & 0.387&^{+0.005}_{-0.052} \\[1ex]
 VCC 213 & 1342224405 & 338.4 & 1.4 & 8.68^{+0.04}_{-0.02} &  53.0&^{+5.8}_{-2.4}  & 0.638&^{+0.029}_{-0.070} \\[1ex]
 VCC 324 & 1342225152 & 217.9 & 1.8 & 8.90^{+0.00}_{-0.01} &  88.0&^{+0.0}_{-1.0}  & 0.248&^{+0.003}_{-0.002} \\[1ex]
 VCC 334 & 1342225154 &  48.6 & 0.3 & 7.94^{+0.00}_{-0.00} &   9.6&^{+0.1}_{-0.0}  & 0.504&^{+0.003}_{-0.007} \\[1ex]
 VCC 340 & 1342225156 & 147.4 & 1.1 & 8.89^{+0.02}_{-0.00} &  24.2&^{+0.9}_{-0.3}  & 0.608&^{+0.008}_{-0.022} \\[1ex]
 VCC 562 & 1342225158 &  44.5 & 0.5 & 8.01^{+0.04}_{-0.06} &  11.2&^{+1.2}_{-1.4}  & 0.397&^{+0.051}_{-0.044} \\[1ex]
 VCC 693 & 1342225160 &  50.9 & 0.5 & 8.33^{+0.01}_{-0.00} &  23.7&^{+0.6}_{-0.0}  & 0.215&^{+0.002}_{-0.005} \\[1ex]
 VCC 699 & 1342225162 & 363.6 & 1.9 & 9.18^{+0.00}_{-0.00} &  91.1&^{+0.0}_{-0.0}  & 0.399&^{+0.002}_{-0.002} \\[1ex]
 VCC 737 & 1342225165 &  70.6 & 0.7 & 7.79^{+0.00}_{-0.00} &   6.8&^{+0.0}_{-0.1}  & 1.046&^{+0.016}_{-0.011} \\[1ex]
 VCC 841 & 1342225166 &  67.6 & 0.7 & 7.82^{+0.04}_{-0.01} &   7.2&^{+0.7}_{-0.2}  & 0.934&^{+0.023}_{-0.091} \\[1ex]
VCC 1437 & 1342225221 & 103.7 & 1.5 & 8.35^{+0.01}_{-0.08} &  24.8&^{+0.6}_{-4.4}  & 0.418&^{+0.075}_{-0.011} \\[1ex]
VCC 1575 & 1342225223 & 486.5 & 1.5 & 9.11^{+0.00}_{-0.00} & 141.1&^{+0.0}_{-0.0}  & 0.345&^{+0.001}_{-0.001} \\[1ex]
VCC 1686 & 1342225225 & 324.8 & 1.3 & 8.67^{+0.01}_{-0.00} &  51.8&^{+0.6}_{-0.6}  & 0.627&^{+0.008}_{-0.008} \\[1ex]
VCC 1725 & 1342225226 & 133.5 & 0.8 & 8.54^{+0.00}_{-0.00} &  38.4&^{+0.0}_{-0.0}  & 0.348&^{+0.002}_{-0.002} \\[1ex]
\enddata
\end{deluxetable*}

\subsection{Results}\label{results}

\begin{figure}
\plotone{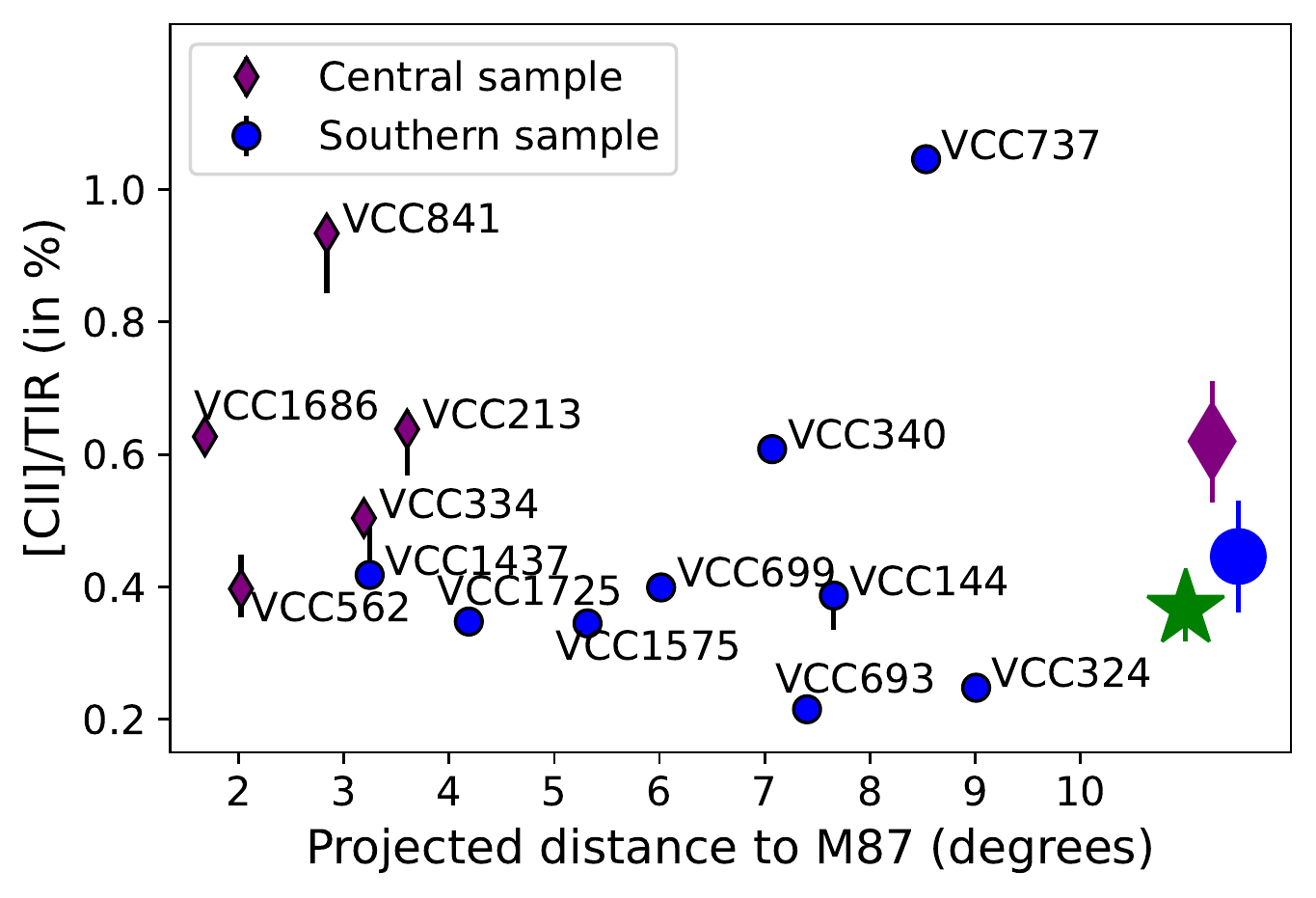}
\caption{\cii/TIR versus projected distance to M87, the central elliptical galaxy in the Virgo cluster. Error bars indicate 1$\sigma$ errors based on the {\it Herschel} \cii\ error budget and the 16--84 percent range of the marginalized probability distribution of the TIR flux; larger errors are dominated by the uncertainty in the TIR flux. Points indicate the central value of \cii/TIR, based on the 50 percent point of the marginalized probability distribution; shape and color indicate the sample to which each galaxy is assigned: central (purple lozenges) or southern (blue circles). The large symbols (shown for illustrative purposes on the right hand side) indicate the arithmetic mean and error of \cii/TIR for each sample, with the large green star showing the mean and error for the comparator sample drawn from the {\it Herschel} Dwarf Galaxy Survey (see Section \ref{differences}).\label{clusterdistance}}
\end{figure}

\begin{figure}
\plotone{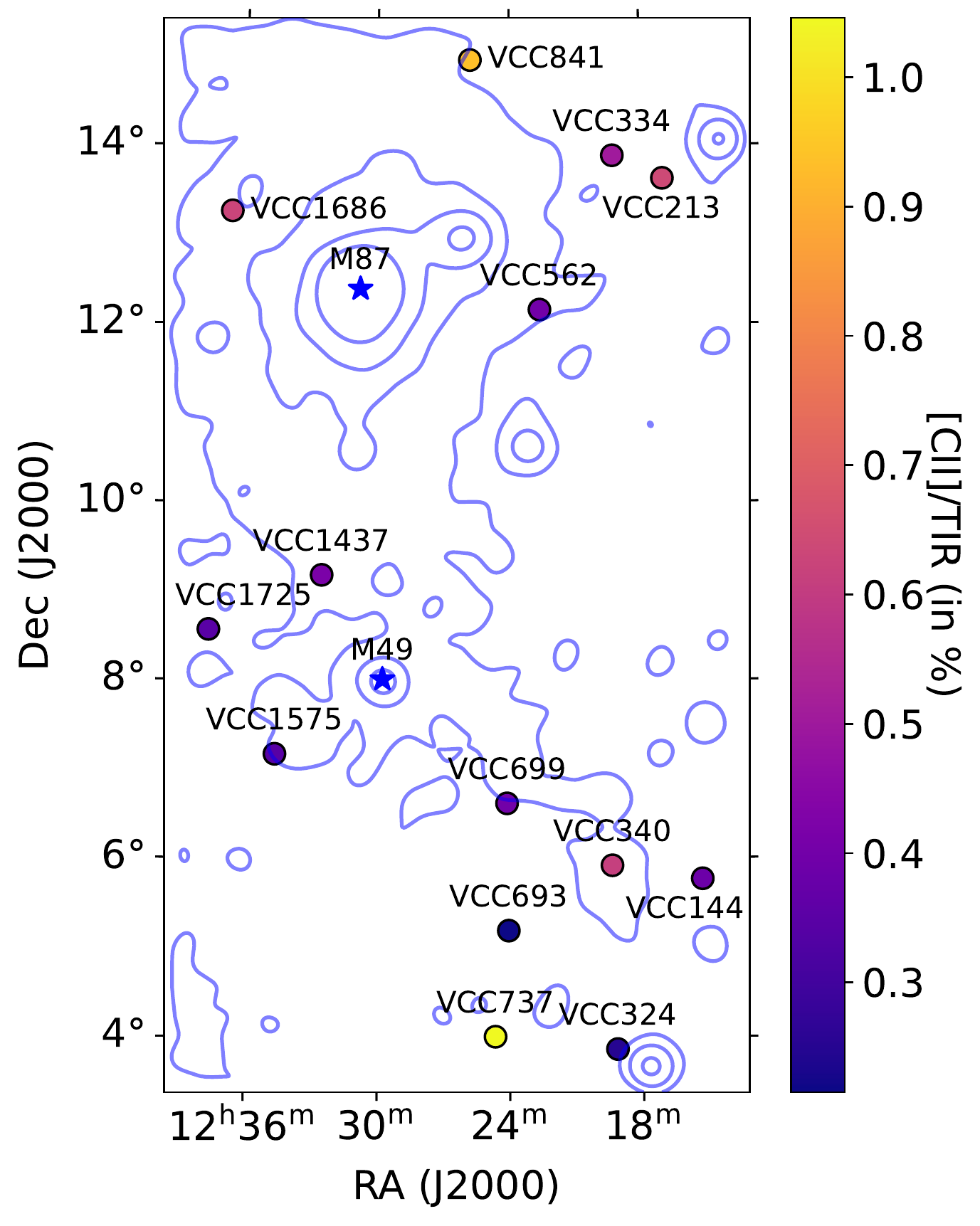}
\caption{Positions of sample galaxies in the cluster with the central value for \cii/TIR indicated by the color. The positions of M87 and M49 are indicated by blue stars. Contours (at 0.1, 0.2, 0.4 and 0.8 counts\,s$^{-1}$\,pixel$^{-1}$) indicate the smoothed hard X-ray counts from the ROSAT All Sky Survey  \citep{1999AA...349..389V}.\label{clustermap}}
\end{figure}

The results of the \cii\ observations and the SED fitting are given in Table \ref{resultstable}. Figure \ref{clusterdistance} shows how the \cii/TIR ratio varies with angular distance from the giant elliptical M87 (taken to represent the cluster center), while Figure \ref{clustermap} shows the positions of the sources within the cluster. The values of L$_{\rm TIR}$ are calculated from the MAGPHYS output of $L_d^{tot}$ assuming a distance of 17 Mpc except for three galaxies that, as given in Table \ref{sampletable}, are assigned to more distant subclusters in Virgo at 23 Mpc (VCC 699) and 32 Mpc (VCC 144 and VCC 340); no error on the distance is assumed.

For the purpose of our analysis, we divide the sample into two groups: the central galaxies (VCC 213, VCC 334, VCC 562, VCC 841 and VCC 1686), which lie around M87 and are all north of declination $+12\deg$, and the southern galaxies (VCC 144, VCC 324, VCC 340, VCC 693, VCC 699, VCC 737, VCC 1437, VCC 1575 and VCC 1725), which are all south of declination $+9.5\deg$ and lie around or south of M49. The central galaxies thus correspond to subcluster A and the southern galaxies to subcluster B, with the exception of VCC 699 (W$^\prime$ cloud) and VCC 144 and 340 (W cloud), according to the standard subdivision of Virgo \citep[e.g.][]{2014AA...570A..69B}. The southern galaxies thus combine three different environments, but as these form our control sample of galaxies outside of the center of the Virgo cluster this is not expected to affect our analysis.

\section{Discussion}\label{discussion}
\subsection{Differences between the samples}\label{differences}

The `central' galaxies are those closest to the center of the Virgo cluster while the `southern' galaxies form a control sample away from the cluster core, thus if there is any environmental effect it should manifest as a difference between these two samples. Comparator galaxies from the {\it Herschel} Dwarf Galaxies Survey \citep[DGS;][]{2015AA...578A..53C} are also used as to form a second control sample. The TIR for the DGS galaxies is, as with our galaxies, defined using the modeled dust SED and their measurement over 1--1000\micron\ can be expected to give an equivalent measurement of the total infrared flux to the MAGPHYS measurement over 3--1000\micron\ \citep[particularly Section 4.2 and Table 3]{2008MNRAS.388.1595D,2015AA...582A.121R}. As the \cii/TIR ratio is influenced by both metallicity and FIR luminosity \citep[e.g.,][figure 5]{2015AA...578A..53C}, we take a sub-sample of DGS galaxies that have similar metallicities (12 + log(O/H) = 8.0 -- 8.8) and luminosities (L$_{\rm TIR}/\nom{L} = 5 \times 10^7$ -- $2.5 \times 10^9$) to those of the Virgo dwarfs.

\begin{figure}
\plotone{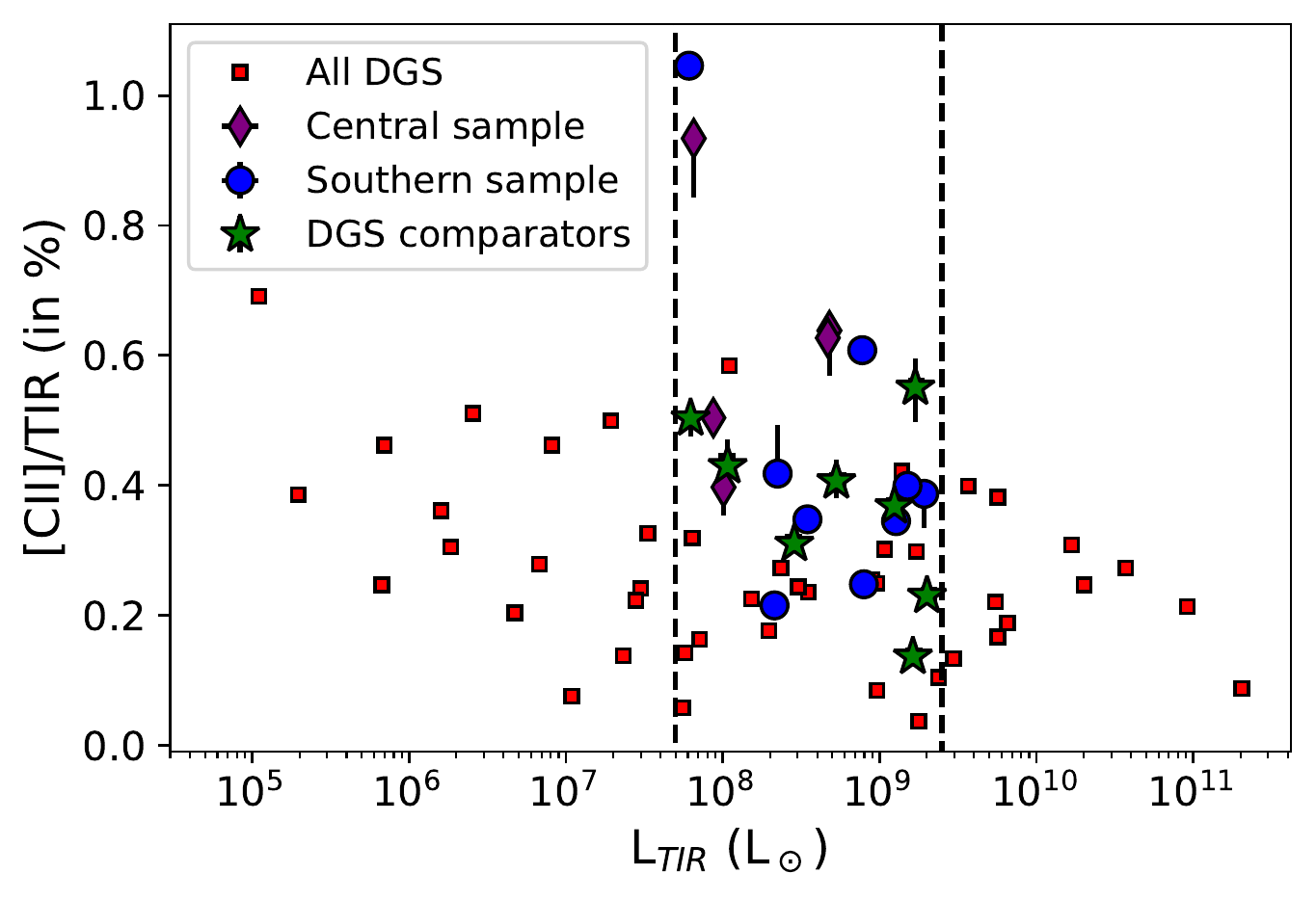}
\plotone{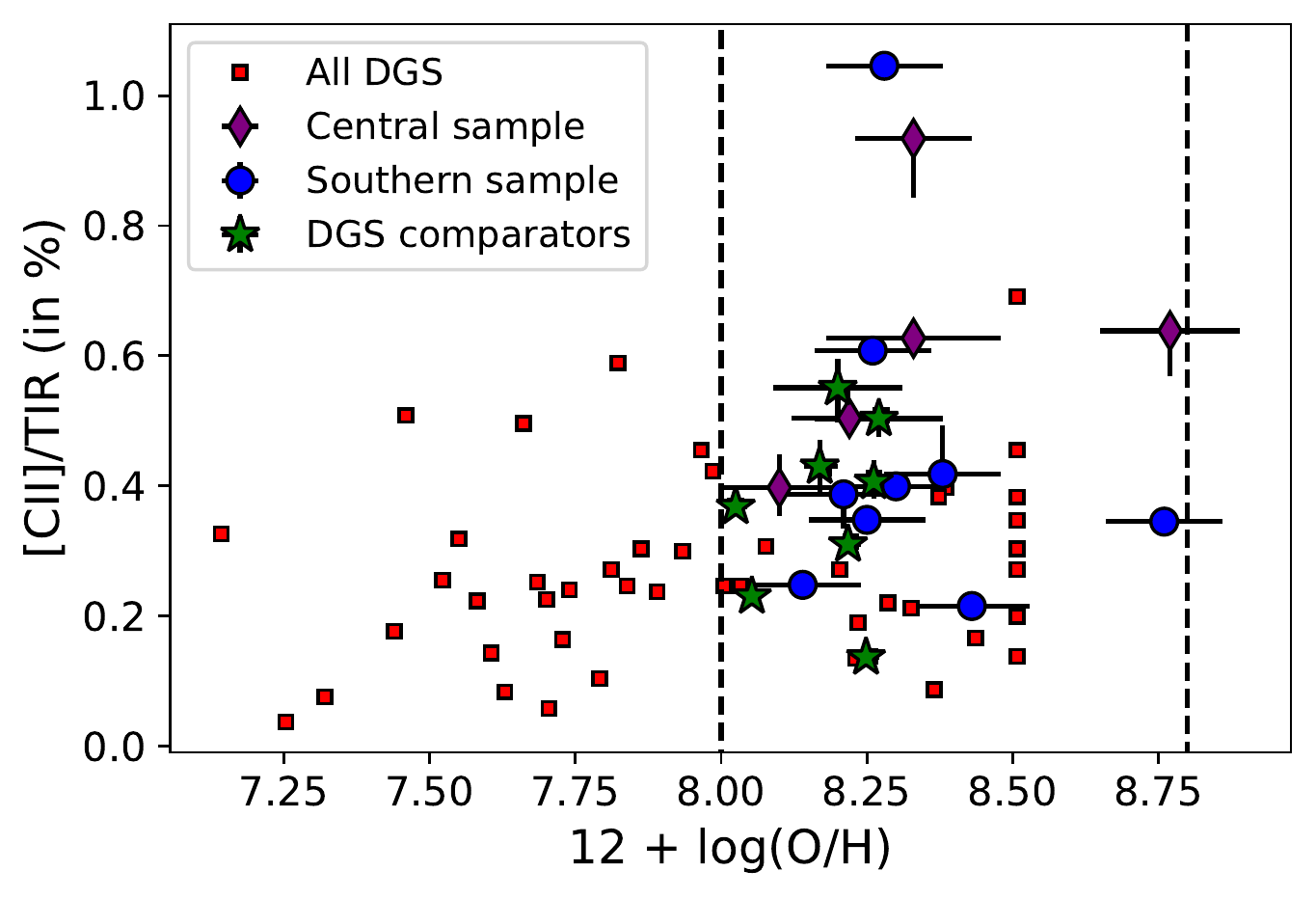}
\caption{\cii/TIR vs L$_{\rm TIR}$ (upper) and metallicity (lower) for our central galaxies (purple lozenges) and southern galaxies (blue circles) along with the comparator sub-sample from the DGS (green stars). Red squares indicate the other DGS galaxies and the black dashed lines indicate the limits for our selection in $L_{\rm TIR}$ and metallicity.\label{samples}}
\end{figure}

Four of the five central galaxies have \cii/TIR $>$ 0.5\% compared to two out of nine in the southern region and two out of eight in the DGS comparator sub-sample. We compare the various samples statistically both by looking at their averages and by using the Kolmogorov-Smirnov test; as we are testing the hypothesis that the central galaxies have a higher \cii/TIR than those in the control samples we use the single-sided two-sample Kolmogorov-Smirnov test. The distribution of the samples in \cii/TIR, L$_{\rm TIR}$ and metallicity is shown in Figure \ref{samples}. Including the full DGS sample (shown in Figure \ref{samples} by red squares), with a mean \cii/TIR of $0.29\pm 0.02$, leads to a very significant difference between our central sample and the DGS (3.5$\sigma$; p = 0.001 from the Kolmogorov-Smirnov test), but also to a significant difference appearing between our southern sample and the DGS (1.9$\sigma$; p = 0.02 from the Komogorov-Smirnov test) due to the inclusion of a large number of galaxies that are not similar to the Virgo dwarfs, motivating us to use only a sub-sample of DGS galaxies with similar luminosities and metallicities for our comparison sample.
	
The central galaxies have a mean \cii/TIR of $0.62\pm0.09$ percent and the southern galaxies have a mean \cii/TIR of $0.45\pm0.08$ percent, indicating a 1.4$\sigma$ difference between these two samples (errors on the means in both cases estimated by propagating the errors on the individual measurements in the samples and combining these in quadrature with the uncertainty in the estimates of the means due to the scatter in the samples). This is confirmed by the one-sided Kolmogorov-Smirnov test, which, despite the small sizes of the samples, gives a likelihood of getting this distribution if both were drawn from the same parent population of p = 0.086. Similarly, the DGS sub-sample has a mean \cii/TIR of $0.37\pm 0.05$ percent, indicating a 2.4$\sigma$ difference from the central sample, and the Kolmogorov-Smirnov test gives p = 0.040. The difference between the means of the DGS sub-sample and the southern sample is 0.8$\sigma$ and the Kolmogorov-Smirnov test gives p = 0.57, both consistent with the DGS sub-sample and the southern Virgo dwarfs being drawn from the same parent population. We therefore combine these two control samples, getting a mean of $0.41\pm0.05$, giving a difference of 2.0$\sigma$ from the mean of the central sample, and a result from the Kolmogorov-Smirnov test of p = 0.032. We conclude that both parametric and non-parametric tests show that there is a statistically significant difference between \cii/TIR in the central sample and the control samples, with the caveat that the numbers in the samples remain small. The results of these statistical tests for difference are summarized in Table \ref{statstable}.

\begin{deluxetable*}{lCr@{\ }LCCC}
\tablecaption{Summary of the samples: number, mean, one-sided Kolmogorov-Smirnov p-value and statistic for difference from the central galaxies, and the significance of the difference in the mean from the central galaxies.\label{statstable}}
\tablehead{\colhead{Sample}&\colhead{N}&\multicolumn2c{Mean}&\colhead{KS p-value}&\colhead{KS statistic}&\colhead{$\sigma$}}
\startdata
Central&5&0.62&\pm\ 0.09&\nodata&\nodata&\nodata\\
Southern&9&0.45&\pm\ 0.08&0.086&0.58&1.4\\
DGS &8&0.37&\pm\ 0.05&0.040&0.68&2.4\\
Southern+DGS &17&0.41&\pm\ 0.05&0.032&0.62&2.0\\
\enddata
\end{deluxetable*}

\subsection{Relationship to ram pressure stripping}\label{rampressure}

We compare our sample to the work of \citet{2018MNRAS.479.4367K} to see whether galaxies with high values of \cii/TIR correspond to those identified there as likely to be undergoing ram pressure stripping. \citet{2018MNRAS.479.4367K} categorize galaxies as ``active strippers'' (likely to be currently undergoing ram pressure stripping) and ``past strippers'' (showing evidence of past gas loss, but not currently undergoing ram pressure stripping) based on an analysis of how tightly bound their H\,{\sc i} disk is and whether the local ram pressure would be sufficient to strip this. The local ram pressure, $p_{loc}$, is expected to decrease with increasing distance from the cluster center \citep[see Figure 14 in][]{2018MNRAS.479.4367K}. This pressure is fundamentally related to two parameters: the density of the ICM and the speed with which a galaxy is moving through that ICM, with $p_{loc} = \rho_{ICM} v_{gal}^2$. 

\citet{2018MNRAS.479.4367K} use a $\beta$ model of the cluster \citep{1976AA....49..137C, 1999AA...343..420S, 2001ApJ...561..708V} to estimate the ICM density and the cluster mass distribution:

\begin{equation}
\rho = \rho_0 \left(1 + \frac{r^2}{r^2_c}\right)^{-(3/2)\beta}
\label{betafn}
\end{equation}

The values adopted for the $\beta$ model parameters in \citet{2018MNRAS.479.4367K} are $\beta = 0.47$, $r_c = 13.4$ kpc and $\rho_0 = 4 \times 10^{-2}$ cm$^{-3}$ for the ICM, and $\beta = 1$, $r_c = 0.32$ Mpc, and $\rho_0 = 3.8 \times 10^{-4}$ \nom{M} pc$^{-3}$ for the dark matter cluster halo. The velocities of the galaxies at a given radius are estimated as being the local escape velocity using the $\beta$ model with the parameters for the dark matter halo given above to estimate the mass distribution. There are limitations to this approach: the $\beta$ model assumes a smooth, symmetrical relationship with radius, not taking into account structures and variations in the density, and sub-clusters, while the assumption that galaxies are moving at their local escape velocity may not be true for individual galaxies; however, it provides a reasonable description overall. They also define $p_{def}$ as the pressure required to strip the galaxy to its current H\,{\sc i} deficiency, with the ratio $p_{loc}/p_{def}$ then giving whether a galaxy is currently being stripped (an ``active stripper'', with $p_{loc}/p_{def} > 0.5$) or was stripped in the past (a ``past stripper'', with $p_{loc}/p_{def} < 0.5$). We look here firstly at how their analysis compares to our data and then at how $p_{loc}$, calculated for our galaxies, corresponds to \cii/TIR.

\begin{figure}
\plotone{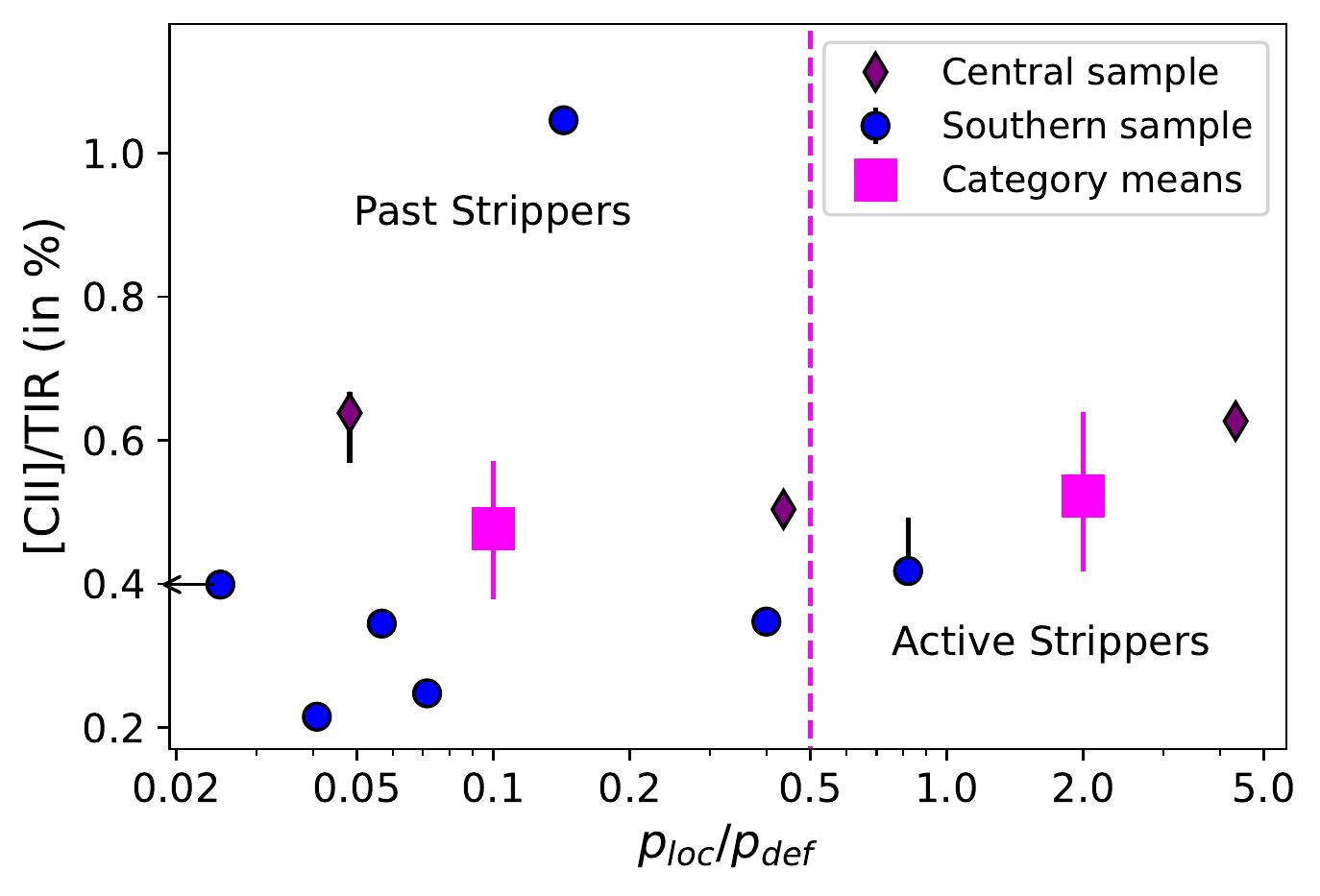}
\caption{\cii/TIR vs $p_{loc}/p_{def}$ with ram pressure stripping category \citep[both from][]{2018MNRAS.479.4367K} for our central galaxies (purple lozenges) and southern galaxies (blue circles). Magenta squares with colored error bars indicate the mean of each category (errors on the means, as before, estimated by propagating the errors on the individual measurements in the samples and combining these in quadrature with the uncertainty in the estimates of the means due to the scatter in the samples) and the dashed magenta line indicates where \citet{2018MNRAS.479.4367K} split their categories. VCC 699, which has $p_{loc}/p_{def} = 0$ in \citet{2018MNRAS.479.4367K}, is artificially placed at $p_{loc}/p_{def} = 0$.025 with an arrow to the left. \label{rps}}
\end{figure}

A comparison of their analysis with our data does not reveal any firm correlation between high values of \cii/TIR and whether a galaxy is identified as likely to be undergoing ram pressure stripping: of the galaxies in the central sample with values for \cii/TIR $> 0.5$\%, only VCC 1686 is an ``active stripper'' while VCC 213 and VCC 334 are both given as ``past strippers'' and VCC 841 is not listed in their sample. The only galaxy in the central sample with a value for \cii/TIR $< 0.5$\%, VCC 562, is also not listed. For the southern sample, of the two galaxies with values for \cii/TIR $> 0.5$\%,  VCC 737 is a ``past stripper'' while VCC 340 is not listed; while among the galaxies with values for \cii/TIR $< 0.5$\%, VCC 1437 is an ``active stripper'', VCC 324, 693, 699, 1575 and 1725 are ``past strippers'', and VCC 144 is not listed. Figure \ref{rps} gives the distribution of \cii/TIR for the sources in our sample assigned to each category by \citet{2018MNRAS.479.4367K} with the x-axis showing their $p_{loc}/p_{def}$, with their break of $p_{loc}/p_{def} = 0.5$ as the demarcation between the active and past strippers marked. It can be seen that there is no significant difference between the mean \cii/TIR for the ``active strippers'' and the ``past strippers''. Further to this, we find a Spearman's $\rho = 0.38$ giving, with 10 pairs, a significance of p = 0.28; while if the outlier VCC 737 is ignored we find $\rho = 0.40$ which, with 9 pairs, gives a significance of p = 0.29. This implies there is no significant correlation between \cii/TIR and $p_{loc}/p_{def}$.

This lack of a correlation may be partly explained by the effect of stripping on star-formation. \citet{2015AA...574A.126G}, from where the sample here is ultimately drawn, reported that the dwarfs in their sample that were undetected in HeViCS far infra-red continuum observations had a larger fraction of object with higher H\,{\sc i} deficiencies than their detected dwarfs. While this is not a very strong effect, if stripped galaxies are less likely to be forming stars then they are less likely to have been observed in this sample. This does not, however, look to be sufficient to explain the lack of correlation we see here, particularly as we do have both ``active strippers'' and ``past strippers'' in both samples.

A second effect that may lend an explanation, which is clear from \citet{2018MNRAS.479.4367K}, is that whether a galaxy is undergoing ram pressure stripping depends both on the pressure it is feeling from the ICM (their $p_{loc}$) and the pressure needed to strip its neutral hydrogen (their $p_{def}$). A galaxy where $p_{def}$ is substantially higher than $p_{loc}$ will not currently be undergoing ram pressure stripping, even if the value of $p_{loc}$ is higher than for ``active strippers''. For example, VCC 1437, identified as an ``active stripper'', has $p_{loc} = 230$ cm$^{-3}$ (km s$^{-1}$)$^2$ and $p_{def} = 280$ cm$^{-3}$ (km s$^{-1}$)$^2$ while VCC 334, identified as a ``past stripper'' at almost the same distance from M87, has $p_{loc} = 240$ cm$^{-3}$ (km s$^{-1}$)$^2$ (that is, marginally higher than for VCC 1437) but $p_{def} = 550$ cm$^{-3}$ (km s$^{-1}$)$^2$. The difference between these two galaxies is not the ram pressure they feel but the effect that that ram pressure has on their (current) H\,{\sc i} disk.\footnote{The situation may well be different where ram pressure stripping and shocks are observed to be affecting the molecular gas disk, e.g. \citet{2019ApJ...883..145J, 2020ApJ...889....9M, 2020ApJ...901...95C}, as the \cii\  and the molecular gas are likely to have similar extents, e.g. \citet{2016AJ....152...51D, 2020ApJ...903...30B}. However, such observations are currently only available for a few galaxies so do not lend themselves to inclusion in this kind of analysis.} 

\begin{figure}
\plotone{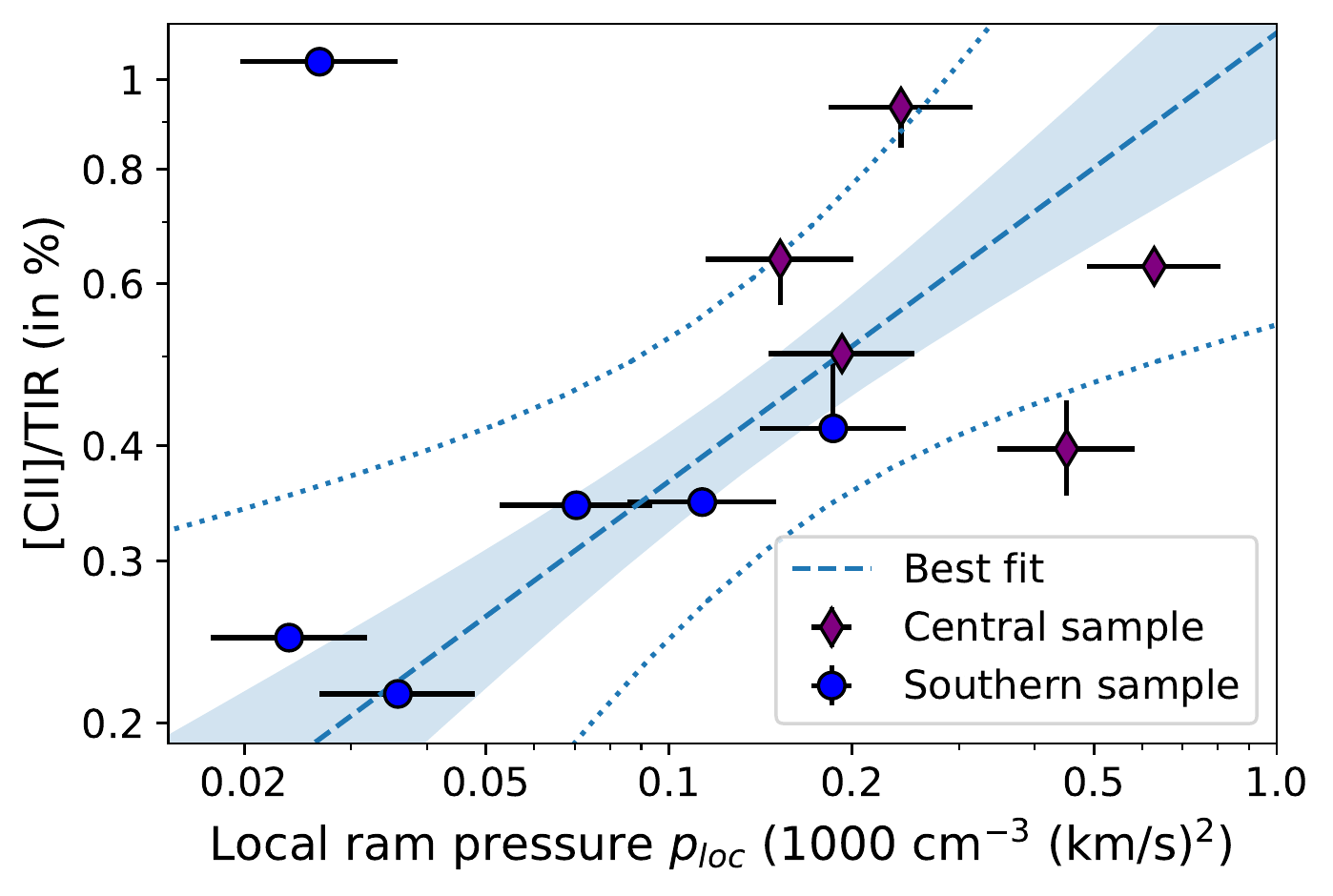}
\caption{\cii/TIR vs local ram pressure ($p_{loc}$) calculated as in Equation \ref{betafn} for our central galaxies (purple lozenges) and southern galaxies (blue circles) at 17 Mpc distance. Dashed line indicates best fit, the shaded area the 1$\sigma$ uncertainty, and the dotted line the 3$\sigma$ uncertainty.
 \label{ploc}}
\end{figure}

Following \citet{2018MNRAS.479.4367K}, we use the same $\beta$ model of the cluster and associated parameters to estimate the ICM density and the cluster mass distribution. The velocities of the galaxies at a given radius, still following \citet{2018MNRAS.479.4367K}, are estimated as being the local escape velocity, derived using the $\beta$ model for the cluster mass distribution. From this, we calculate the local ram pressure for those galaxies in our sample at the 17 Mpc distance of the main cluster (where the $\beta$ model is applicable), assuming an average deprojected distance from the cluster center of $\sqrt{4/3}$ times their projected distance from the cluster center (i.e. spherical symmetry). The results are plotted in Figure \ref{ploc}, which shows \cii/TIR versus the local ram pressure. The errorbars on $p_{loc}$ indicate fractional differences in the clustercentric distances of $\sqrt{4/3} -1$, i.e. the difference between the uncorrected and deprojected clustercentric distances.

If the outlier VCC 737 (top left of Figure \ref{ploc}) is excluded, we find a Spearman's $\rho = 0.75$ which, with 10 pairs, gives a significance of p = 0.013, indicating that there is likely to be a correlation between \cii/TIR and the local ram pressure. We fit this correlation with a power law as $\log({\rm \cii}/{\rm TIR}) = (0.49 \pm 0.13) \times \log(p_{loc}) + (0.05\pm 0.12)$, where \cii/TIR is given as a percentage and $p_{loc}$ in units of 1000 cm$^{-3}$\,(km/s)$^2$. The local ram pressure calculated here could be affected by projection effects, where the three dimensional position of a source lies in front of or behind the plane of the cluster center, result in the projected distance to the cluster center being lower than the actual three dimensional distance. As the ram pressure a galaxy feels is due to its actual distance from the cluster center but our calculation here is based on their projected distance, our estimates of the ram pressure will be high for galaxies that are projected onto the cluster center, moving them to the right on Figure \ref{ploc}. This provides a possible explanation for why VCC 562 falls clearly (including the uncertainties on its measurement) outside of the 3$\sigma$ scatter around the best-fit line -- it may lie either in front of or behind the cluster rather than near where it is seen in projection, so that its local ram pressure is lower than that calculated based on projected separation from the cluster center.

The correlation we see in Figure \ref{ploc} is unexpectedly strong, continuing as it does into the outer parts of the cluster where \cii/TIR is similar to that seen in our control sample from the DGS, which was expected to be free of environmental effects (although we do not know the local environments of the DGS galaxies), implying that ram pressure could have an effect on \cii/TIR well outside the central region of the cluster and possibly even in galaxy groups \citep[c.f.][]{2021arXiv210606315R}.

Our finding that the central sample galaxies are more likely to have high values of \cii/TIR than those in the southern sample thus appears to be linked to the higher values of $p_{loc}$ felt by the central sample, i.e. their ram pressure interaction with the ICM, without being necessarily linked to whether they are actively undergoing ram pressure stripping of their gas. It seems likely, therefore, that the excess we see in \cii/TIR is due to \cii\ formation in shocks in the ISM of these galaxies caused by the ram pressure they are feeling.

\subsection{Comparison of star formation rate indicators}

We use the calibration of \citet{2011ApJ...741..124H} to derive a star formation rate (SFR) from the GALEX far ultra-violet (FUV) luminosities combined with our TIR luminosities. We also use the \cii\ calibration of \citet{2014AA...568A..62D} for low metallicity dwarfs to derive an SFR from the luminosities. This gives us a measure of excess \cii\ based on star formation, $SFR_{\cii}/SFR_{FUV+TIR}$.\footnote{Both of these relationships are derived using a \citet{2003ApJ...598.1076K} initial mass function.}

\begin{figure}
\plotone{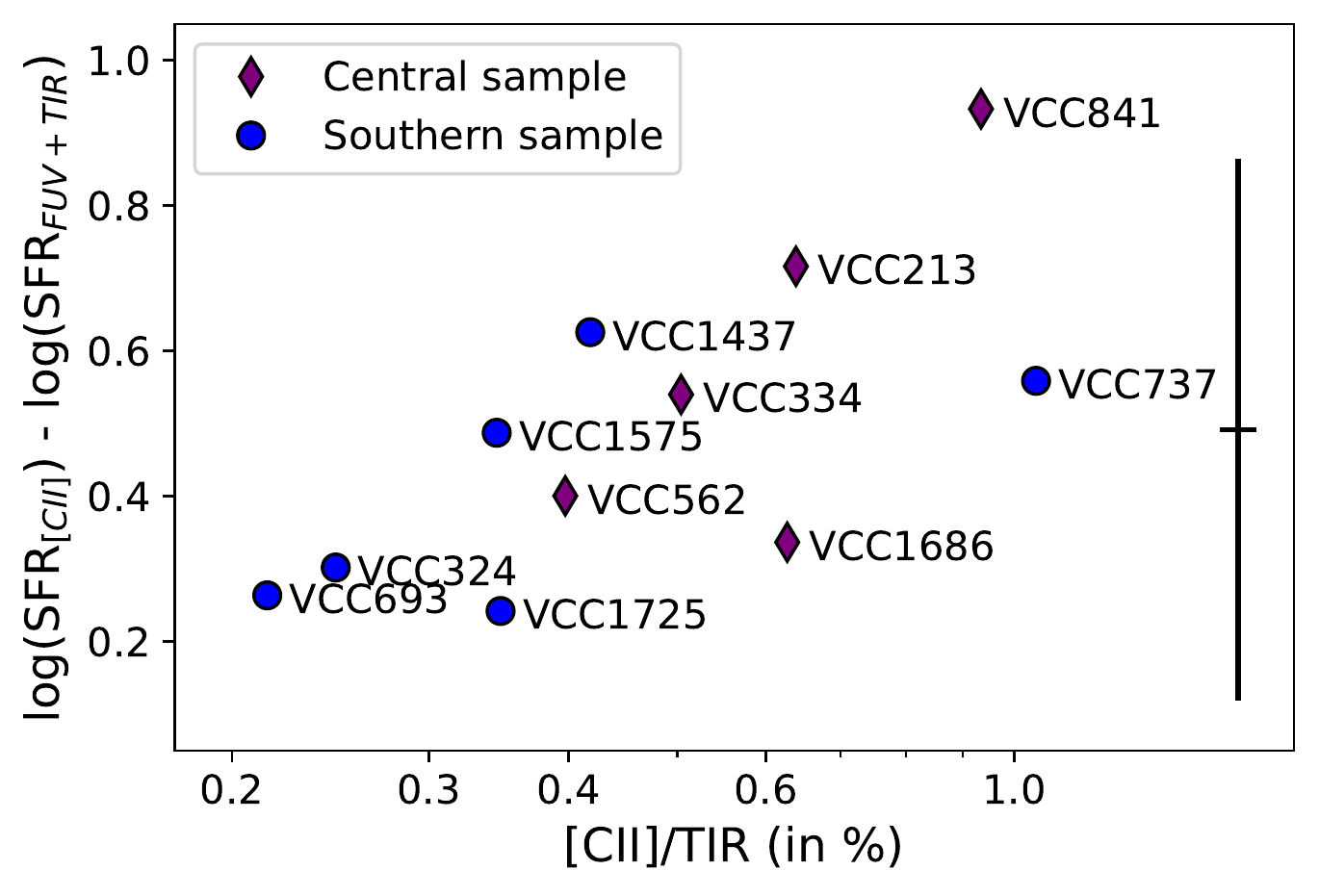}
\caption{Difference between SFR estimated from FUV+TIR and estimated from \cii, plotted against \cii/TIR. The error bar on the right hand side indicates the scatter in the SFR indicators and the average error on \cii/TIR; vertical position of the error bar marks the average difference between the SFR indicators. \label{SFRs}}
\end{figure}

In Figure \ref{SFRs} we plot this against \cii/TIR for the dwarf galaxies in the main cluster. It can be seen that this shows an increase in \cii\ excess measures via SFR indicators as \cii/TIR increases, as expected if the \cii/TIR increase is due to the creation of \cii\ via processes other than star formation. If, on the other hand, the excess in \cii/TIR were caused by a rise in unobscured star formation, i.e. the \cii\ seen here is being created by star formation that is not reflected in the TIR emission, we would not expect to see any increase in $SFR_{\cii}/SFR_{FUV+TIR}$. Two galaxies fall below this trend -- VCC 737, already identified as having an anomalously high value of \cii/TIR for its position in the cluster, and VCC 1686 which, as described in Section \ref{vcc1686section} below, has strong UV emission that is not correlated spatially with either the \cii\ or the dust and is probably due to recent star formation; this galaxy would thus be expected to show a deficit of \cii\ relative to the FUV+TIR SFR, which translates into a deficit relative to the trend here.

The trend seen here is much tighter than the scatter on the SFR indicators (shown by the error bar on the right side of the plot), which is dominated by the scatter on the \cii--SFR calibration. One possibility for our trend being tighter than the scatter is the relatively narrow range of luminosities and metallicities covered by our sample, while another is that \citet{2014AA...568A..62D} use the 24\micron\ flux as a proxy for the total infrared flux in their measurement of SFR that they compare to the \cii, whereas we measure TIR based on the whole SED here.

As can be seen in Figure \ref{SFRs}, the SFR calibrations used give an excess of \cii\ for all of the galaxies in our sample, although for four of the eleven galaxies this is within the 1$\sigma$ scatter around zero and for all but VCC 841 it is within the $2\sigma$ scatter. However, applying the \cii--SFR calibration of \citet{2015ApJ...800....1H} gives (after correcting the SFR from the \citet{1955ApJ...121..161S} initial mass function used by \citet{2015ApJ...800....1H} to the \citet{2003ApJ...598.1076K} initial mass function used by \citet{2014AA...568A..62D} and \citet{2011ApJ...741..124H} using the ratio of 0.67 found by \citet{2014ARAA..52..415M}) a similar shape while showing a deficit of $SFR_{\cii}$ relative to $SFR_{FUV+TIR}$  for all but four galaxies in the sample. Thus this appears to be an issue of calibration of the zero point, which is of secondary importance for the relationship being examined here: the clear increase in $SFR_{\cii}/SFR_{FUV+TIR}$ with increasing \cii/TIR.

\subsection{Notes on individual galaxies}

We present here notes on three of the galaxies with the highest \cii/TIR in our sample: VCC 737, VCC 841 and VCC 1686. Two of these are in the central region of the cluster, while the third (VCC 737) is on the cluster outskirts and appears anomalous in terms of its \cii/TIR compared to its cluster position.

\subsubsection{VCC 737}

\begin{figure*}
\plottwo{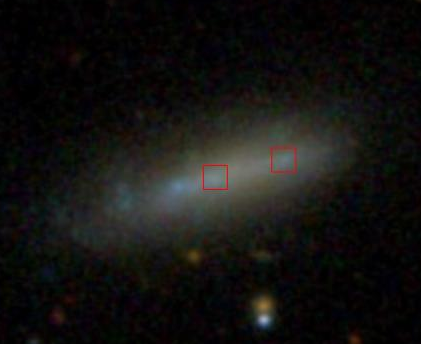}{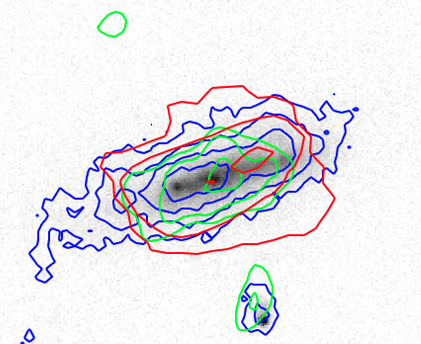}
\caption{VCC 737 optical color image from the SDSS (left) with north up and east to the left; and SDSS {\it g}-band image overlaid with \cii\ (R), WISE band 3 (G) and GALEX NUV (B) contours (right) at levels of 2.5, 5 and 10 $\times 10^{-19}$ W\,m$^{-2}$\,spaxel$^{-1}$ (4.4, 8.8 and 18 $\sigma$) for the \cii; 2, 2.8 and 4 DN\,pixel$^{-1}$ (3.5, 4.9 and 7.0 $\sigma$) above the sky value of 776.5 DN\,pixel$^{-1}$ for WISE band 3; and 0.0125, 0.025, 0.05 and 0.1 counts\,s$^{-1}$\,pixel$^{-1}$ (6.0, 12, 24 and 48 $\sigma$) for the GALEX NUV. The locations of the SDSS spectrum (Figure \ref{vcc737spectra}) are marked with red boxes on the SDSS color image.\label{vcc737maps}}
\end{figure*}

\begin{figure*}
\plottwo{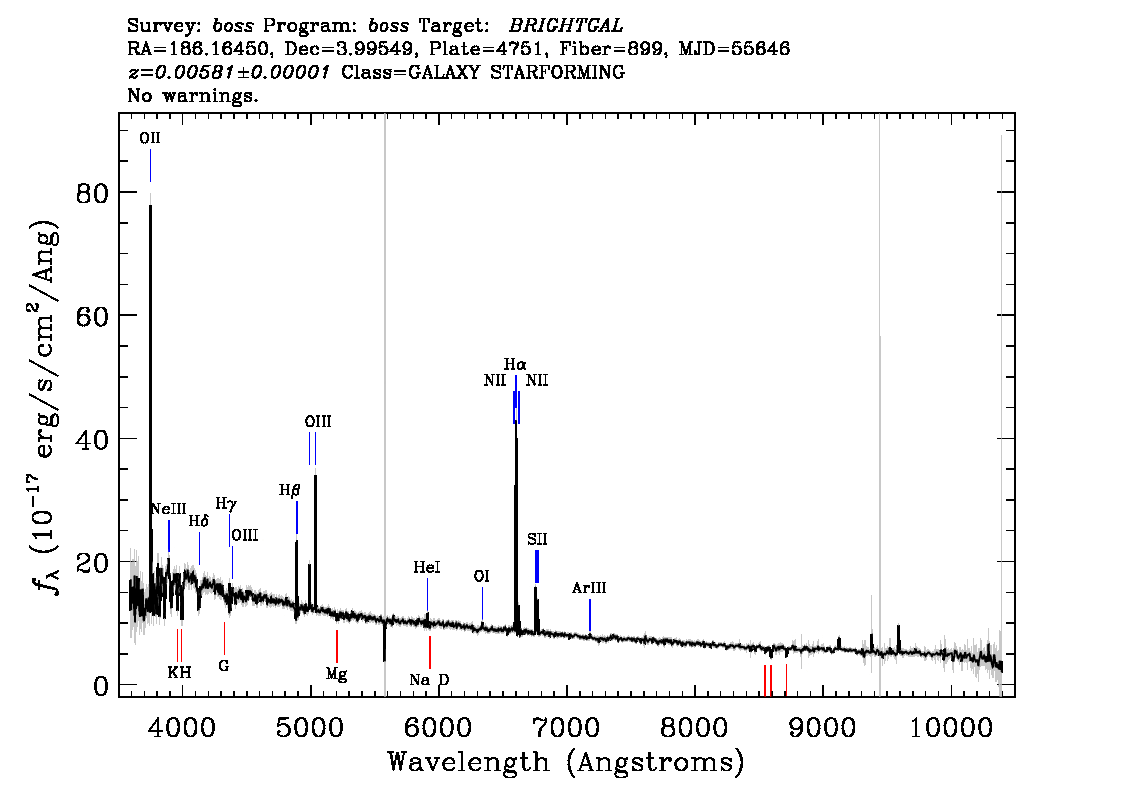}{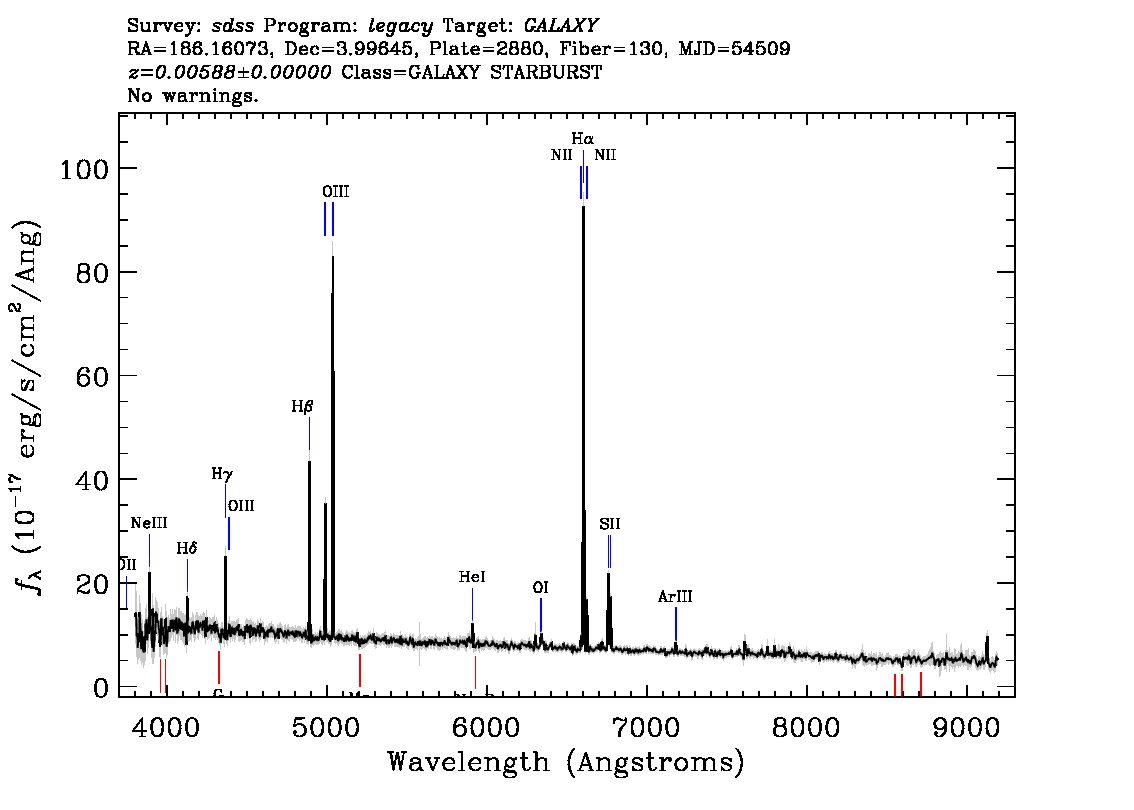}
\caption{SDSS preview spectra of VCC 737 central region (left) and western region (right). Red and blue labels indicate automatically identified absorption and emission features; grey band indicates uncertainty in the flux.\label{vcc737spectra}}
\end{figure*}

VCC 737 is far from the cluster core, at a projected distance of 2.5 Mpc, and is estimated to have a low local ram pressure, but has the highest \cii/TIR in our sample making it an outlier in these relationships. 

Figure \ref{vcc737maps} shows the SDSS \citep{2000AJ....120.1579Y} color image of VCC 737 from Data Release 13 \citep{2017ApJS..233...25A} and contour maps of the \cii, mid-IR and near UV. There are two SDSS spectra on this galaxy, shown in Figure \ref{vcc737spectra}. The eastern part of this galaxy, including the location of the central spectrum, contains coincident peaks in \cii, dust and UV. However, the western part has a stronger \cii\ peak that is not matched by peaks in either dust or UV, giving it a clear excess of \cii. The western SDSS spectrum shows a much stronger H$\alpha$ line than the central spectrum (a flux of $615.8\times10^{-17}{\rm\,erg}{\rm\,cm}^{-2}{\rm\,s}^{-1}$ in the SDSS catalog vs $200\times10^{-17}{\rm\,erg}{\rm\,cm}^{-2}{\rm\,s}^{-1}$ for the central region),\footnote{See catalog entries at \url{https://dr12.sdss.org/spectrumDetail?mjd=54509&fiber=130&plateid=2880} and \url{https://dr12.sdss.org/spectrumDetail?mjd=55646&fiber=899&plateid=4751}} and the H$\alpha$/H$\beta$ ratios of 3.6 for the western region and 3.4 for the central region (again from the SDSS catalog) imply that the two regions have similar internal absorption. However, although there is a local peak in the near UV emission near the location of the western spectrum, it is around half the strength of the near UV in the central region -- the local UV peaks at 0.088~cnt\,s$^{-1}$\,pix$^{-1}$ in the pixel on the northern edge of the western SDSS spectrum versus 0.164~cnt\,s$^{-1}$\,pix$^{-1}$ in the central pixel of the central SDSS spectrum. The eastern UV (which does not have an associated SDSS spectrum, but corresponds to the bluer eastern region in the SDSS image) peaks at 0.163~cnt\,s$^{-1}$\,pix$^{-1}$.

This pattern of \cii\ enhancement in an area with little warm dust emission is quite distinct from that expected from ram pressure shocks, where triggered star formation is expected to occur in the shocks \cite[e.g., ][]{2009AA...499...87K}, giving enhanced dust, \cii\ and UV from star formation alongside the additional \cii\ that may be formed directly from the shock (as seen in VCC 841; Figure \ref{vcc841maps}). Based on its optical diameters in GOLDMine \citep{2014arXiv1401.8123G} and its H\,{\sc i} mass from ALFALFA \citep{2018ApJ...861...49H}, we estimate an H\,{\sc i} deficiency \citep{1984AJ.....89..758H} of $-0.17$ for VCC 737, within the range of normal (unstripped) galaxies.

Clearly, the high \cii/TIR measured in VCC 737 is due to the \cii\ peak in the western part of this galaxy, but exploring the details of what might be exciting strong \cii\ and H$\alpha$ emission without the expected enhancement in the dust luminosity or UV in this galaxy is beyond the scope of this paper. 

\subsubsection{VCC 841}

\begin{figure*}
\plottwo{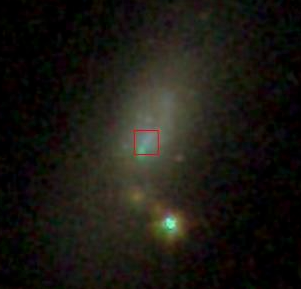}{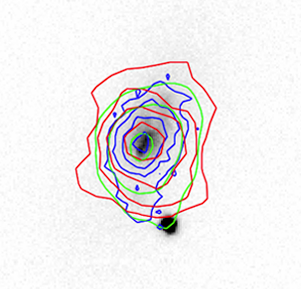}
\caption{VCC 841 optical color image from the SDSS (left) with north up and east to the left; and SDSS {\it g}-band image overlaid with \cii\ (R), WISE band 3 (G) and GALEX NUV (B) contours (right) at levels of 2.5, 5, 10 and 20 $\times 10^{-19}$ W\,m$^{-2}$\,spaxel$^{-1}$ (4.5, 9, 18 and 36 $\sigma$) for the \cii; 2.5, 5 and 10 DN\,pixel$^{-1}$ (6.0, 12 and 23 $\sigma$) above the sky value of 604 DN\,pixel$^{-1}$ for WISE band 3; and 0.0125, 0.025, 0.05 and 0.1 counts\,s$^{-1}$\,pixel$^{-1}$ (5.6, 11, 22 and 44 $\sigma$) for the GALEX NUV. The location of the SDSS spectrum (Figure \ref{vcc841spectrum}) is marked with a red box on the SDSS color image.\label{vcc841maps}}
\end{figure*}

\begin{figure}
\plotone{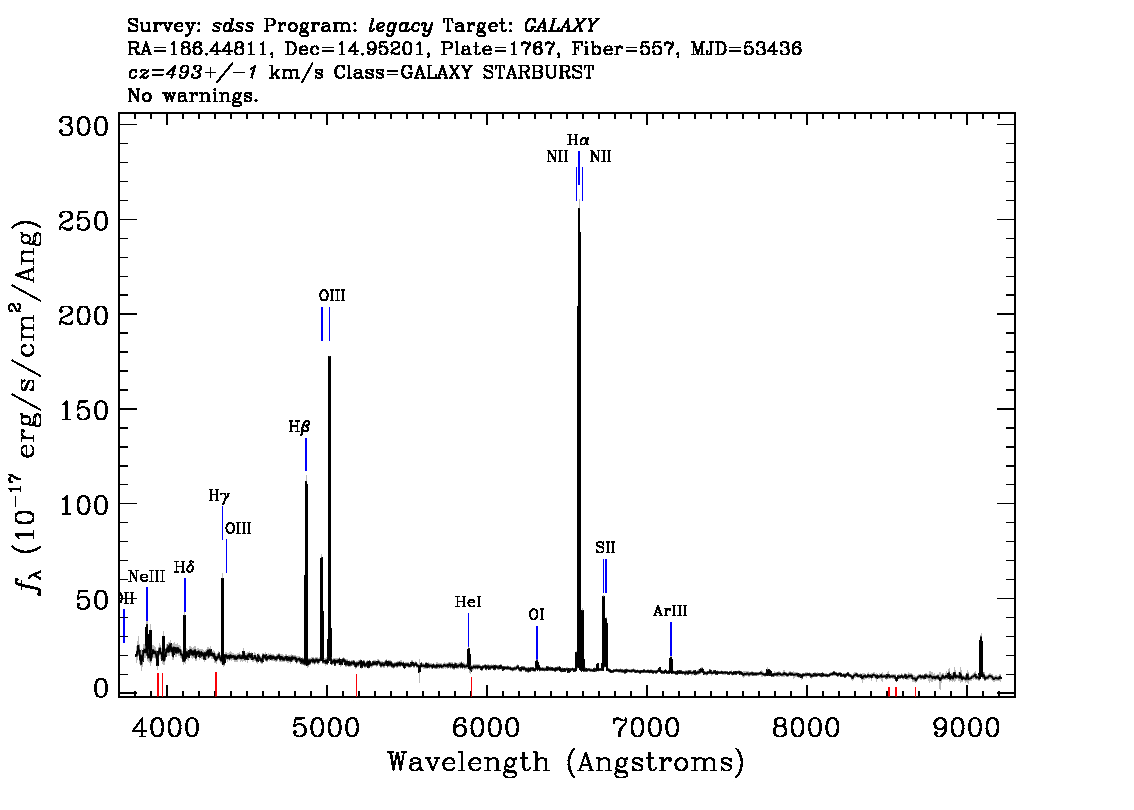}
\caption{SDSS spectrum of VCC 841. Red and blue labels indicate automatically identified absorption and emission features; grey band indicates uncertainty in the flux.\label{vcc841spectrum}}
\end{figure}

Figure \ref{vcc841maps} shows the SDSS color image of VCC 841 and contour maps of the \cii, mid-IR and near UV. The SDSS image shows an off-center nucleus, which is coincident with the \cii, dust and UV centers, with the rest of the galaxy to the northwest of this nucleus. The SDSS spectrum of the nucleus (Figure \ref{vcc841spectrum}) shows a strong H$\alpha$ line. Comparison with Figure \ref{clustermap} shows that the nucleus is on the side of the galaxy closest to M87. This is consistent with this galaxy undergoing ram pressure effects that are triggering star formation in the nucleus, which would naturally be expected to enhance both \cii\ and TIR, with the excess \cii/TIR being due to ram pressure shocks. It is not listed in  \citet{2018MNRAS.479.4367K}, but based on its optical diameters in GOLDMine and its H\,{\sc i} mass from ALFALFA we estimate an H\,{\sc i} deficiency of $0.67$, consistent with it being ram pressure stripped.

\subsubsection{VCC 1686}\label{vcc1686section}

\begin{figure*}
\plottwo{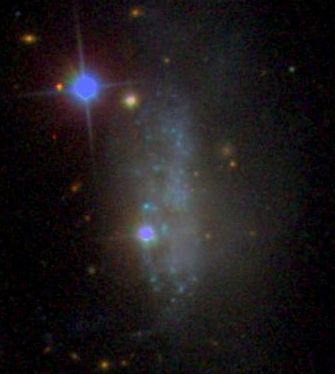}{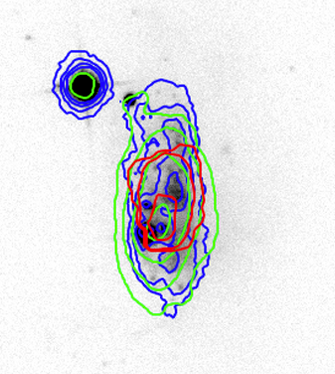}
\caption{VCC 1686 optical color image from the SDSS (left) with north up and east to the left; and SDSS {\it g}-band image overlaid with \cii\ (R), WISE band 3 (G) and GALEX NUV (B) contours (right) at levels of 1, 2 and 4 $\times 10^{-18}$ W\,m$^{-2}$\,spaxel$^{-1}$  (9.8, 19.5 and 39.0 $\sigma$) for the \cii; 2.5, 5, 10 and 20 DN\,pixel$^{-1}$ (6.3, 13, 25 and 51 $\sigma$) above the sky value of 647 DN\,pixel$^{-1}$ for WISE band 3; and 0.025, 0.05, 0.1 and 0.2 counts\,s$^{-1}$\,pixel$^{-1}$ (7.6 15, 30 and 61 $\sigma$) 
for the GALEX NUV.\label{vcc1686maps}}
\end{figure*}

Figure \ref{vcc1686maps} shows the SDSS color image of VCC 1686 and contour maps of the \cii, mid-IR and near UV. The \cii\ and dust are well aligned, with the UV being enhanced in the areas to the north west of the galaxy that also show very blue colors in the SDSS image and are probably the site of recent star formation as there is no dust or \cii\ enhancement seen in this region. It is also likely that some \cii\ flux is lost off the edge of the PACS footprint. This galaxy is considered an ``active stripper'' by \citet{2018MNRAS.479.4367K}, who calculate that it has an H\,{\sc i} deficiency of 0.45. Based on its optical diameters in GOLDMine and an updated H\,{\sc i} mass from ALFALFA we estimate a slightly lower H\,{\sc i} deficiency of $0.39$, still consistent with it undergoing stripping.

\section{Conclusions}

In recent years, the consensus that \cii\ has its origin in, and thus traces, star formation has been challenged by the discovery of \cii\ formed from the interactions of AGN jets with the disk \citep{2018ApJ...869...61A, 2019AA...626L...3S,2021ApJ...909..204F} and in galaxy-galaxy interactions \citep{2013ApJ...777...66A,2017ApJ...836...76A,2018ApJ...855..141P}. To these, as suggested by \citet{1999MNRAS.303L..29P}, we can now add the interaction of galaxies with the cluster environment. The most likely source of the \cii/TIR excess observed in our sample in galaxies near the cluster core is the formation of \cii\ in ram pressure shocks.

The dwarf galaxies studied here in the central part of the Virgo cluster have significantly higher average ratios of \cii\ to total infrared continuum than the dwarf galaxies in the southern part of the Virgo cluster (p = 0.086), with $\langle$\cii/TIR$\rangle = 0.62\pm 0.09$ in the center and $\langle$\cii/TIR$\rangle = 0.45\pm 0.08$ in the south for a $1.4\sigma$ difference. After controlling for metallicity and luminosity, the southern sample is consistent with the {\it Herschel} Dwarf Galaxy Survey (DGS) while the central sample shows a significant difference (p = 0.040), with the DGS sub-sample having $\langle$\cii/TIR$\rangle = 0.37\pm 0.05$ for a $2.4\sigma$ difference. This implies the existence of processes in the cluster environment that are injecting energy into the interstellar medium of these galaxies. The most likely candidate for such a process is an interaction between the interstellar medium of these galaxies and the intra-cluster medium, i.e. ram pressure. However, as ram pressure stripping is a combination of both the local ram pressure felt by the galaxy and how tightly bound the H\,{\sc i}  is to that galaxy, galaxies that are tightly bound (or already partially stripped) near cluster cores may suffer ram pressure shocks to their ISM, leading to the formation of \cii, without exhibiting signs of current ram pressure stripping. Similarly, galaxies that are loosely bound further from cluster cores may be undergoing stripping without showing a \cii\ excess. When we look at just the local ram pressure we find a correlation between this and \cii/TIR that can be fitted as a power law with a slope of 0.49 $\pm$ 0.13.

While the effect detected here as a difference between the central and southern galaxies and as a correlation between the ratio of \cii\ to total infrared continuum and the local ram pressure is statistically significant, increasing the number of data points would increase confidence in this result. In order to improve our understanding of the effect of ram pressure shocks on \cii\ in cluster galaxies, and whether it is indeed ram pressure that is responsible for the observed \cii\ excess, further studies will be necessary. Comparison of galaxies near the centers of galaxy clusters with suitable control samples, as done in this study, is one way forward, but other possibilities include observations of larger galaxies where the shocked regions can be identified and compared with un-shocked regions (particularly where the shocks are clearly affecting the molecular gas) and observations of galaxies in clusters, such as Coma, that have a stronger ram pressure effect than Virgo. NASA's Stratospheric Observatory for Infrared Astronomy and future proposed observatories such as the Origins Space Telescope will be vital for this effort to comprehend the origins of \cii\ in our local universe and thus better understand observations of the high redshift universe.

\section{Acknowledgments}

We thank Joachim K\"{o}ppen for useful discussions and the anonymous referee for comments that have improved this paper.

The Herschel spacecraft was designed, built, tested, and launched under a contract to ESA managed by the Herschel/Planck Project team by an industrial consortium under the overall responsibility of the prime contractor Thales Alenia Space (Cannes), and including Astrium (Friedrichshafen) responsible for the payload module and for system testing at spacecraft level, Thales Alenia Space (Turin) responsible for the service module, and Astrium (Toulouse) responsible for the telescope, with in excess of a hundred subcontractors.

PACS has been developed by a consortium of institutes led by MPE (Germany) and including UVIE (Austria); KU Leuven, CSL, IMEC (Belgium); CEA, LAM (France); MPIA (Germany); INAF-IFSI/OAA/OAP/OAT, LENS, SISSA (Italy); IAC (Spain). This development has been supported by the funding agencies BMVIT (Austria), ESA-PRODEX (Belgium), CEA/CNES (France), DLR (Germany), ASI/INAF (Italy), and CICYT/MCYT (Spain).

The observations in this paper were taken as part of guaranteed time from the SPIRE consortium. SPIRE has been developed by a consortium of institutes led by Cardiff University (UK) and including Univ. Lethbridge (Canada); NAOC (China); CEA, LAM (France); IFSI, Univ. Padua (Italy); IAC (Spain); Stockholm Observatory (Sweden); Imperial College London, RAL, UCL-MSSL, UKATC, Univ. Sussex (UK); and Caltech, JPL, NHSC, Univ. Colorado (USA). This development has been supported by national funding agencies: CSA (Canada); NAOC (China); CEA, CNES, CNRS (France); ASI (Italy); MCINN (Spain); SNSB (Sweden); STFC, UKSA (UK); and NASA (USA).

This publication makes use of data products from the Two Micron All Sky Survey, which is a joint project of the University of Massachusetts and the Infrared Processing and Analysis Center/California Institute of Technology, funded by the National Aeronautics and Space Administration and the National Science Foundation.

This publication makes use of data products from the Wide-field Infrared Survey Explorer, which is a joint project of the University of California, Los Angeles, and the Jet Propulsion Laboratory/California Institute of Technology, funded by the National Aeronautics and Space Administration.

This research has made use of the NASA/IPAC Extragalactic Database (NED), which is operated by the Jet Propulsion Laboratory, California Institute of Technology, under contract with the National Aeronautics and Space Administration.

This research has made use of the NASA/IPAC Infrared Science Archive, which is funded by the National Aeronautics and Space Administration and operated by the California Institute of Technology.

GALEX is a NASA Small Explorer, launched in April 2003. GALEX is operated for NASA by California Institute of Technology under NASA contract NAS-98034.

We acknowledge the use of NASA's {\it SkyView} facility (\url{http://skyview.gsfc.nasa.gov}) located at NASA Goddard Space Flight Center.

Funding for the SDSS and SDSS-II has been provided by the Alfred P. Sloan Foundation, the Participating Institutions, the National Science Foundation, the U.S. Department of Energy, the National Aeronautics and Space Administration, the Japanese Monbukagakusho, the Max Planck Society, and the Higher Education Funding Council for England. The SDSS Web Site is \url{http://www.sdss.org/}. 

The SDSS is managed by the Astrophysical Research Consortium for the Participating Institutions. The Participating Institutions are the American Museum of Natural History, Astrophysical Institute Potsdam, University of Basel, University of Cambridge, Case Western Reserve University, University of Chicago, Drexel University, Fermilab, the Institute for Advanced Study, the Japan Participation Group, Johns Hopkins University, the Joint Institute for Nuclear Astrophysics, the Kavli Institute for Particle Astrophysics and Cosmology, the Korean Scientist Group, the Chinese Academy of Sciences (LAMOST), Los Alamos National Laboratory, the Max-Planck-Institute for Astronomy (MPIA), the Max-Planck-Institute for Astrophysics (MPA), New Mexico State University, Ohio State University, University of Pittsburgh, University of Portsmouth, Princeton University, the United States Naval Observatory, and the University of Washington.

RT and BD acknowledge the support of the Czech Science Foundation grant 19-18647S and the institutional project RVO 67985815.

SOFIA is jointly operated by the Universities Space Research Association, Inc. (USRA), under NASA contract NNA17BF53C, and the Deutsches SOFIA Institut (DSI) under DLR contract 50 OK 0901 to the University of Stuttgart.

\facilities{IRSA, WISE, {\it Herschel}, GALEX}

\software{Astropy \citep{2013AA...558A..33A, 2018AJ....156..123A}, NumPy \citep{2020Natur.585..357H}, SAOImageDS9 \citep{2003ASPC..295..489J}, SOSPEX \citep{2018AAS...23115011F}, WebPlotDigitizer (\url{https://apps.automeris.io/wpd/})}

\appendix
\section{Input fluxes for the SED fitting}
\restartappendixnumbering

\startlongtable
\begin{deluxetable}{lR@{ $\pm$ }LR@{ $\pm$ }LR@{ $\pm$ }LR@{ $\pm$ }L}
\tabletypesize{\footnotesize}
\tablecaption{Input fluxes for the SED fitting (Jy).
\label{SEDtable}}
\tablehead{\colhead{Galaxy ID}&\multicolumn2c{GALEX FUV (0.15\micron)}&\multicolumn2c{GALEX NUV (0.23\micron)}&\multicolumn2c{SDSS $u$ (0.36\micron)}&\multicolumn2c{SDSS $g$ (0.47\micron)}\\
&\multicolumn2c{SDSS $r$ (0.62\micron)}&\multicolumn2c{SDSS $i$ (0.75\micron)}&\multicolumn2c{SDSS $z$ (0.90\micron)}&\multicolumn2c{2MASS $J$ (1.2\micron)}\\
&\multicolumn2c{2MASS $H$ (1.7\micron)}&\multicolumn2c{2MASS $K_S$ (2.2\micron)}&\multicolumn2c{WISE B1 (3.4\micron)}&\multicolumn2c{WISE B2 (4.6\micron)}\\
&\multicolumn2c{WISE B3 (12\micron)}&\multicolumn2c{WISE B4 (23\micron)}&\multicolumn2c{PACS 100 (100\micron)}&\multicolumn2c{PACS 160 (160\micron)}\\
&\multicolumn2c{PACS 250 (250\micron)}&\multicolumn2c{SPIRE 350 (350\micron)}&\multicolumn2c{SPIRE 500 (500\micron)}}
\startdata
VCC 144
&	0.00147&	0.00007&	0.00180&	0.00005&	0.00250&	0.00009&	0.00431&	0.00008\\
&	0.00510&	0.00009&	0.00490&	0.00009&	0.00510&	0.00015&	0.00656&	0.00049\\
&	0.00648&	0.00067&	0.00466&	0.00083&	0.00327&	0.00004&	0.00204&	0.00006\\
&	0.0113&	0.0003&	0.0692&	0.0041&	0.724&	0.052&	0.525&	0.040\\
&	0.182&	0.016&	0.080&	0.010&	0.036&	0.004\\
VCC 213
&	0.000581&0.000074&0.00092&	0.00007&	0.00272&	0.00010&	0.00787&	0.00014\\
& 	0.0128&	0.0002&	0.0166&	0.0003&	0.0195&	0.0006&	0.0265&	0.0008\\
& 	0.0278&	0.0011&	0.0237&   0.0013&	0.0116&	0.0001&	0.00681& 	0.00007\\
&	0.0298&	0.0004&	0.0381&	0.0035&	1.130&	0.063&	1.135&	0.064\\
&	0.516&	0.038&	0.257&	0.020&	0.093&	0.004\\
VCC 324
&	0.00127&	0.00006&	0.00179&	0.00006&	0.00392&	0.00014&	0.00874&	0.00016\\
&	0.0120&	0.0002&	0.0135&	0.0002&	0.0164&	0.0005&	0.0143&	0.0007\\
&	0.0149&	0.0012&	0.0114&	0.0011&	0.00595&	0.00006&	0.00380&	0.00006\\
&	0.0201&	0.0006&	0.104&	0.004&	0.965&	0.061&	0.717&	0.058\\
&	0.318&	0.024&	0.153&	0.014&	0.086&	0.004\\
VCC 334
&	0.000355&0.000064&0.000498&0.000058&\nodata&	\nodata&	0.00229&	0.00004\\
&	0.00307&	0.00006&	0.00377&	0.00007&	0.00412&	0.00012&	0.00434&	0.00050\\
&	0.00406&	0.00075&	0.00273&	0.00085&	0.00191&	0.00004&	0.000997&0.000047\\
&	0.00180&	0.00030&	0.00334&  0.00109&	0.137&	0.022&	0.163&	0.018\\
&	0.070&	0.009&	0.029&	0.005&	\nodata&	\nodata\\
VCC 340
&	0.000635&0.000056&0.000908&0.000053&0.00209&0.00007&	0.00525&	0.00010\\
&	0.00748&	0.00014&	0.00915&	0.00017&	0.0109&	0.0003&	0.0106&	0.0006\\
&	0.0141&	0.0010&	0.00870&	0.00089&	0.00528&	0.00005&	0.00313&	0.00006\\
&	0.00723& 0.00027&	0.0284&	0.0060&	0.455&	0.037&	0.394&	0.047\\
&	0.274&	0.018&	0.120&	0.011&	0.058&	0.004\\
VCC 562
&	0.000484&0.000043&0.000667&0.000047&0.000823&0.000029&0.00154&0.00003\\
&	0.00195&	0.00004&	0.00207&	0.00004&	0.00287&	0.00009&	0.00311&	0.00045\\
&	0.00146& 0.00083&	0.00175&	0.00070&	0.00162&	0.00003&	0.000936&0.000046\\
&	0.00177&	0.00032&	0.0115&	0.0040&	0.133&	0.021&	0.125&	0.015\\
&	0.059&	0.006&	0.045&	0.005&	0.032&	0.004\\
VCC 693
&	0.000584&0.000080&0.000882&0.000079&0.00208&0.00007&	0.00539&	0.00010\\
&	0.00714&	0.00013&	0.00826&	0.00015&	0.00742&	0.00021&	0.00675&	0.00086\\
&	0.00467&	0.00099&	0.00630&	0.00151&	0.00257&	0.00004&	0.00144&	0.00007\\
&	0.00293&	0.00036&	\nodata&	\nodata&	0.134&	0.025&	0.219&	0.039\\
&	0.153&	0.014&	0.075&	0.009&	0.030&	0.004\\
\tablebreak
VCC 699
&	0.00269&	0.00010&	0.00400&	0.00009&	0.00708&	0.00024&	0.0157&	0.0003\\
&	0.0218&	0.0004&	0.0264&	0.0005&	0.0256&	0.0007&	0.0223&	0.0009\\
&	0.0218&	0.0012&	0.0199&	0.0014&	0.0120&	0.0001&	0.00688& 	0.00007\\
&	0.0217&	0.00062&	0.0519&	0.0040&	1.427&	0.102&	1.398&	0.087\\
&	0.722&	0.054&	0.361&	0.030&	0.146&	0.016\\
VCC 737
&	0.000593&0.000107&0.000841&0.000092&0.00167&0.00006&	0.00401&	0.00007\\
&	0.00587&	0.00011&	0.00656&	0.00012&	0.00906&	0.00026&	0.00679&	0.00089\\
& 	0.00784&	0.00112&	0.00893&	0.00154&	0.00323&	0.00004&	0.00180&	0.00005\\
&	0.00293&	0.00038&	\nodata&	\nodata&	0.109&	0.019&	0.157&	0.019\\
&	0.150&	0.013&	0.093&	0.008&	0.049&	0.004\\
VCC 841
&	0.000136&0.000067&0.000248&0.000051&0.00105&0.00004&	0.00283&	0.00005\\
&	0.00460&	0.00009&	0.00591&	0.00011&	0.00762&	0.00022&	0.00554&	0.00062\\
&	0.00602&	0.00106&	0.00686&	0.00094&	0.00487&	0.00004&	0.00263&	0.00007\\
&	0.00418&	0.00032&	0.0123&	0.0033&	0.138&	0.025&	0.156&	0.017\\
&	0.101&	0.011&	0.042&	0.007&	0.011&	0.003\\
VCC 1437
&	0.000291&0.000040&0.000467&0.000049&0.00142&0.00005&	0.00404&	0.00007\\
&	0.00613&	0.00011&	0.00785&	0.00015&	0.00917&	0.00026&	0.0104&	0.0007\\
&	0.0125&	0.0012&	0.00825&	0.00110&	0.00532&	0.00004&	0.00301&	0.00006\\
&	0.0117&	0.0004&	0.0288&	0.0049&	0.454&	0.037&	0.434&	0.045\\
&	0.176&	0.017&	0.065&	0.010&	0.034&	0.004\\
VCC 1575
&  	0.000822&0.000049&0.00145&	0.00006&	0.00484&	0.00017&	0.0159&	0.0003\\
&	0.0276&	0.0005&	0.0359&	0.0007&	0.0436&	0.0012&	0.0472&	0.0016\\
&	0.0550&	0.0026&	0.0506&	0.0026&	0.0175&	0.0001&	0.0105& 	0.0001\\
&	0.0475&	0.0004&	0.136&	0.006& 	2.319&	0.141&	2.706&	0.142\\
&	1.292&	0.094&	0.542&	0.041&	0.186&	0.016\\
VCC 1686
& 	0.00303&	0.00008&	0.00435&	0.00009&	0.00550&	0.00019&	0.0168&	0.0003\\
&	0.0123&	0.0002&	0.0207&	0.0004&	0.0219&	0.0006&	0.0256&	0.0010\\
&	0.0259&   0.0018&	0.0239&	0.0015&	0.0133&	0.0002&	0.00798&	0.00015\\
&	0.0205&	0.0005&	0.0334&	0.0015&	1.061&	0.08&	1.714&	0.105\\
&	1.13&	0.106&	0.621&	0.063&	0.232&	0.026\\
VCC 1725
&	0.00169&	0.00005&	0.00205&	0.00006&	0.00295&	0.00010&	0.00788&	0.00015\\
&	0.00993&	0.00018&	0.0116&	0.0002&	0.0108&	0.0003&	0.00801&	0.00093\\
&	0.00929&	0.00157&	0.0098&	0.0016&	0.00385& 0.00005& 	0.00223&	0.00007\\
&	0.00461&	0.00029&	\nodata&	\nodata&	0.357&	0.033&	0.377&	0.034\\
&	0.287&	0.025&	0.172&	0.016&	0.082&	0.009
\enddata
\end{deluxetable}
\end{document}